\documentclass[acmsmall,nonacm]{acmart}
\setcopyright{none}
\renewcommand\footnotetextcopyrightpermission[1]{} 
\usepackage{array}
\usepackage{tabularx,multirow, booktabs}
\usepackage{hhline, makecell}
\usepackage{dsfont}   
\usepackage{threeparttable}
\usepackage{boldline}

\setcopyright{acmlicensed}
\copyrightyear{2025}
\acmYear{2025}
\acmDOI{10.1145/3714458}

\acmJournal{csur}
\acmVolume{57}
\acmNumber{7}
\acmArticle{165}
\acmMonth{February}

\begin{document}

\title{A Survey on Speech Deepfake Detection}

\author{Menglu Li}
\email{menglu.li@torontomu.ca}
\affiliation{%
    \institution{Department of Electrical, Computer and Biomedical Engineering, Toronto Metropolitan University}
  \city{Toronto}
  \state{ON}
  \country{Canada}
  \postcode{M5B 2K3}
}
\author{Yasaman Ahmadiadli} 
\email{yahmadiadli@torontomu.ca}
\affiliation{%
    \institution{Department of Electrical, Computer and Biomedical Engineering, Toronto Metropolitan University}
  \city{Toronto}
  \state{ON}
  \country{Canada}
  \postcode{M5B 2K3}
}

\author{Xiao-Ping Zhang}
\authornote{Corresponding Author}
\email{xpzhang@ieee.org}
\affiliation{%
    \institution{Shenzhen Key Laboratory of Ubiquitous Data Enabling, Tsinghua Shenzhen International Graduate School, Tsinghua University, China, and with the Department of Electrical, Computer and Biomedical Engineering, Toronto Metropolitan University}
  \city{Toronto}
  \state{ON}
  \country{Canada}
  \postcode{M5B 2K3}
}
\thanks{This work was supported in part by Shenzhen Key Laboratory of Ubiquitous Data Enabling (No. ZDSYS20220527171406015), and by Tsinghua Shenzhen International Graduate School-Shenzhen Pengrui Endowed Professorship Scheme of Shenzhen Pengrui Foundation.}

\renewcommand{\shortauthors}{Li et al.}

\begin{abstract}
The availability of smart devices leads to an exponential increase in multimedia content. However, advancements in deep learning have also enabled the creation of highly sophisticated Deepfake content, including speech Deepfakes, which pose a serious threat by generating realistic voices and spreading misinformation. To combat this, numerous challenges have been organized to advance speech Deepfake detection techniques. In this survey, we systematically analyze more than 200 papers published up to March 2024. We provide a comprehensive review of each component in the detection pipeline, including model architectures, optimization techniques, generalizability, evaluation metrics, performance comparisons, available datasets, and open source availability. For each aspect, we assess recent progress and discuss ongoing challenges. In addition, we explore emerging topics such as partial Deepfake detection, cross-dataset evaluation, and defenses against adversarial attacks, while suggesting promising research directions. This survey not only identifies the current state of the art to establish strong baselines for future experiments but also offers clear guidance for researchers aiming to enhance speech Deepfake detection systems.

\end{abstract}





\maketitle

\section{Introduction}
Deep learning (DL) techniques have significantly advanced the creation of fake media, commonly referred to as "Deepfake". The rise of Deepfakes has introduced significant challenges across various domains, as it can manipulate visual, or audio content, leading to the spread of misinformation. This includes defaming the credibility of prominent figures, leading to political insecurity, fake news, and manipulation of public opinion \cite{chintha2020recurrent}. The Deepfake landscape can be broadly categorized into visual, audio, and multimedia Deepfake, each requiring specialized detection methods. While a substantial portion of research efforts has focused on detecting visually related deepfakes, this survey specifically explores advanced techniques for detecting speech Deepfakes. Speech Deepfakes involve the manipulation of speech content or the creation of synthetic voices through the application of machine learning algorithms, particularly deep learning. For instance, in 2019, a YouTube channel released several Deepfake speeches mimicking Canadian politicians \cite{misinfo}, highlighting the potential of speech Deepfake to influence the outcome of elections by misleading voters. The rapid advancement of speech Deepfake synthesis algorithms also can be weaponized for phone scams or compromising voice-enabled devices, where synthesized voices could maliciously take control. Given these challenges, speech Deepfake detection demands focused attention, distinct from other forms of Deepfake detection. At the same time, developments in this area contribute valuable insights to other modalities. For example, advanced techniques for detecting speech Deepfakes could be integrated into audio-visual Deepfake detection systems to assess the synchronization between speech and video, or to enhance interpretability by identifying discrepancies in the speech domain.

The primary techniques to synthesize speech Deepfake are Text-to-Speech (TTS) and Voice Conversion (VC). TTS models take text as input and utilize vocoders to generate natural-sounding speech that adheres to the linguistic rules of the input \cite{renfastspeech}. VC, on the other hand, modifies original speech to mimic the voice of a specified target speaker while preserving the linguistic content \cite{pasini2019melgan}. Notably, the source speech for VC models can originate from a TTS algorithm. Beyond traditional vocoder-based TTS and VC approaches, Large Language Model (LLM)-based audio generation methods, leveraging neural audio codecs, have recently gained prominence, producing highly realistic synthesis results \cite{wang2024speechx}. With the rise of speech Deepfake threats, initiatives like the ASVspoof (Automatic Speaker Verification Spoofing and Countermeasures) \cite{yamagishi2019asvspoof} and ADD (Audio Deep Synthesis Detection) \cite{yi2022add} challenges have played a pivotal role in fostering the development of advanced algorithms to combat speech Deepfake attacks. In addition to speech synthesis techniques, physical attacks also pose a significant threat to audio security by physically manipulating or re-recording audio signals \cite{yamagishi2019asvspoof}. A common example is a replay attack, where an attacker records a legitimate audio signal and replays it to deceive a voice-recognition biometric system. Unlike Deepfake attacks that involve DL-based alteration or synthesis of audio, replay attacks involve physical devices such as microphones and speakers to capture and replay sound, thus bypassing digital security measures like Automatic Speaker Verification (ASV) systems.

In this survey, we focus on reviewing and analyzing recent advances in speech Deepfake countermeasures (CMs) targeting TTS and VC attacks across diverse application scenarios, extending beyond the scope of ASV systems. We do not include physical attacks in this survey because they fundamentally differ from Deepfake attacks in their method of execution and the challenges they present. While physical attacks are critical to consider in the broader context of audio security, our emphasis is on detecting speech Deepfake that involves speech synthesis and voice conversion, where tremendous advances have been made in recent years.

Several surveys on speech Deepfake detection have been published. The main differences between our survey and the existing ones are summarized as follows.
\begin{enumerate}
\item[-] \textbf{Broader definition of speech Deepfake.} We adopt a more inclusive definition of speech Deepfake, to include both the fully fake speech clips and the partial Deepfake speech. We evaluate the detection models that specifically target partially fake portions within a speech clip, which has not been addressed in the previous surveys. We also analyze speaker-aware Deepfake detection models, exploring the integration of speaker verification and Deepfake CMs to protect the ASV systems.
\item[-] \textbf{Optimization and performance enhancement techniques during the training process.} While the existing surveys primarily concentrate on the architecture of detection algorithms, our survey extends the analysis to include optimization and training techniques. We highlight the significance of factors like data augmentation and the choice of loss function, which can substantially impact the detection performance. By examining these techniques collectively, we equip researchers with a comprehensive toolkit for developing enhanced algorithms.
\item[-] \textbf{Emphasis on open-source availability.} Open source plays a pivotal role as it fosters technological advancement. Our survey places significant emphasis on providing open-source information for both models and datasets.
\item[-] \textbf{Explorations of emerging research topics.} The publications in the field of speech Deepfake have increased significantly in recent years, leading to the emergence of advanced topics like adversarial attack defense and cross-dataset evaluation. Our survey provides a thorough review and evaluation of newly published articles, ensuring up-to-date coverage of these topics.
\end{enumerate}

More specifically, several existing surveys \cite{zhang2022deepfake,kandari2023comprehensive,masood2023deepfakes,khanjani2021deep} provide literature reviews on Deepfake media content, with an emphasis on image and video aspects rather than audio/speech. Besides, specific surveys addressing speech Deepfake differ from this paper in several aspects. For example, 
Almutairi et al. \cite{almutairi2022review} and Cuccovillo et al. \cite{cuccovillo2022open} highlight the open challenges in the current solutions rather than providing a detailed comparison of state-of-the-art (SOTA). \cite{wang2022practical, dixit2023review, ijcai2023p756} provide evaluation of the detection methods specifically targeting TTS/VC-generated fully Deepfake speech clips. More recently,  Khan et al. \cite{khan2023battling} and Yi et al. \cite{yi2023audio} briefly touch upon partially Deepfake detection in their surveys but do not offer a systematic discussion and evaluation of the specific algorithms involved. Furthermore, the existing work also lacks reviews on the optimization techniques that can be applied to the training process for performance enhancement.

Therefore, we provide an in-depth review of various aspects of speech Deepfake technology, including the elements of detecting architecture, prevailing training techniques, methodologies embracing diverse application scenarios, and the latest available datasets. For each aspect, we will discuss the detailed design, evaluate the performance, address current limitations and explore future directions. We aim to provide a thorough understanding of the broader picture for preventing malicious speech Deepfake, serving as a valuable reference guide for future researchers. Specifically, the contribution of our survey can be summarized as follows:
\begin{enumerate}
\item[-] We present a comprehensive review of each building block of developing speech Deepfake detection algorithms and provide the evaluation of performance on both fully and partially Deepfake scenarios.
\item[-] We are the first to evaluate the effectiveness of optimization techniques applied in the model training process, such as data augmentation, activation functions and loss functions. We also address the current stage of emerging research tasks, including transfer learning and the explainability of the detection architecture.
\item[-] We provide open-source information regarding SOTA and benchmarking datasets, which give a feasible guide to achieve reproductions. 
\item[-] We review the existing challenges faced by SOTA models and propose future research directions for advancing speech Deepfake detection.
\end{enumerate}

\section{Datasets and evaluation metrics for speech Deepfake detection}

\subsection{Datasets}
We discuss the characteristics of the most recent widely utilized datasets in speech Deepfake detection, categorized into two main types: fully Deepfake, and partially Deepfake datasets. 
Detailed information of all mentioned datasets is presented in Table \ref{TABLE:DATASET}.

\subsubsection{Fully Deepfake speech datasets}  \hfill\\
\textbf{ASVspoof2015}\cite{wu15e_interspeech} This dataset is based on the Voice cloning toolkit (VCTK) database \cite{yamagishi2019cstr}, which has been recorded in a studio environment, with eight different accents, including American, Australian, English, Canadian, Indian, Irish, Scottish and Northern Irish. 
No channel or background noise effects are added. 

\textbf{FakeorReal-orginal (FoR)}\cite{reimao2019dataset} FoR-orginal is an English speech dataset that contains both bona fide in a variety of accents and Deepfake speech generated by diverse TTS algorithms. The FoR dataset provides three publicly available versions, each with different pre-processing methods.

\textbf{ASVspoof2019-LA}\cite{yamagishi2019asvspoof} This dataset is divided into two subsets, logical access (LA) and physical access (PA), with PA featuring the replay-spoof speech and LA containing TTS and VC-generated Deepfake speech, all derived from the VCTK database. This survey mainly focuses on the LA subset. 
It's important to note that all data in the dataset are also clean, without noise, or channel variation, which may lead to a detachment from real-world conditions. 

\textbf{ASVspoof2021-LA}\cite{liu2023asvspoof} This dataset is an extended version of the ASVspoof2019-LA, aiming to bridge the gap between ideal experimental and real-world conditions. The evaluation data are across the real telephone systems, incorporating various codecs, transmission channels, bitrates and sample rates. 

\textbf{ASVspoof2021-DF}\cite{liu2023asvspoof} The Deepfake Speech (DF) subset is newly introduced in ASVspoof2021, distinct from ASV systems. Both bona fide and Deepfake speech utterances in the DF track are processed with different lossy codecs, potentially introducing distortion. This DF subset contains the audio clips from the ASVspoof2019-LA evaluation set, along with data from Voice Conversion Challenge (VCC) 2018 \cite{lorenzo2018voice} and 2020 \cite{yi2020voice} databases. VCC 2018 includes a collection of clips of professional US English speakers recorded in a studio setting. VCC2020 contains English utterances recorded from both English and bilingual speakers. 

\textbf{FMFCC-A}\cite{zhang2021fmfcc} FMFCC-A is a publicly available Mandarin speech Deepfake dataset. 
Half of the evaluation dataset is randomly selected to process a compression-decompression operation or add Gaussian noise to enhance its diversity.

\textbf{WaveFake}\cite{frank2021wavefake} WaveFake is a speech Deepfake dataset, generated by six different GAN-based TTS algorithms across two languages, English and Japanese. There is no additive noise in this dataset and it only contains two speakers. 

\textbf{ADD2022-LF}\cite{yi2022add} The low-quality (LF) track of the ADD2022 challenge focuses on utterances with various real-world noises and background music effects. However, this dataset is inaccessible online.

\textbf{In-the-Wild (ITW)}\cite{muller22_interspeech} ITW is a dataset of speech Deepfakes with corresponding bona fide speech for English-speaking celebrities and politicians. Both bona fide and Deepfake speech samples are collected from publicly available sources such as social networks and video streaming platforms, potentially containing background noises. There is no information on accent variation provided. This dataset is designed to assess the generalizability of detection models, including their performance across different datasets.

\begin{table*}
  \caption{Statistics of Datasets for speech Deepfake Detection}
  \label{TABLE:DATASET}
  \scriptsize
  \setlength{\tabcolsep}{1.5pt} 
  \begin{tabularx}{\textwidth} {
  >{\centering\arraybackslash}m{0.07\textwidth}
  >{\centering\arraybackslash}m{0.04\textwidth}
  >{\centering\arraybackslash}m{0.09\textwidth}
  >{\centering\arraybackslash}m{0.07\textwidth}
  >{\centering\arraybackslash}m{0.07\textwidth}
  >{\centering\arraybackslash}m{0.06\textwidth}
  >{\centering\arraybackslash}m{0.07\textwidth}
  >{\centering\arraybackslash}m{0.07\textwidth}
  >{\centering\arraybackslash}m{0.04\textwidth}
  >{\centering\arraybackslash}m{0.06\textwidth}
  >{\centering\arraybackslash}m{0.045\textwidth}
  >{\centering\arraybackslash}m{0.06\textwidth}
  >{\centering\arraybackslash}m{0.05\textwidth}
  >{\centering\arraybackslash}m{0.05\textwidth}
  >{\centering\arraybackslash}m{0.055\textwidth}}
    \toprule[1.2pt]
     Dataset & Year & Accessibility & Language & Deepfake type & Unseen attacks* & \#Deepfake methods & Condition & Format & Sample rate & \# real & \# Deepfake & \# Male speaker & \#Female speaker & Accent variation\\
    \midrule
    \multicolumn{14}{c}{Fully Deepfake datasets}\\
    \midrule
    ASVspoof 2015 &2015.09 &Yes &English &TTS, VC &Yes &10 &Clean &Flac &16kHz &16651 &246500 &45 &61 & Yes\\[3pt]
    FoR-original&2019.10 &Yes  &English &TTS &No &7 &Clean &WAV &Multiple &108256 &87285 & \multicolumn{2}{c}{173}& Yes\\[1pt]
    ASVspoof 2019-LA& 2019.11 &Yes &English &TTS, VC &Yes &19 &Clean &Flac &16kHz &10256 &90192 &45  &61 & Yes\\[5pt]    
    ASVspoof 2021-LA &2021.09 &Yes &English &TTS, VC &Yes &19 &Codec &Flac &Multiple &14816 &133360 &60 &74& Yes\\[3pt]
    ASVspoof 2021-DF &2021.09 &Yes &English &TTS, VC &Yes &100+ & Codec &Flac &Multiple &14869 &519059 &73 &87& Yes\\[1pt]
    FMFCC-A &2021.10 &Yes &Chinese &TTS, VC &Yes &13 &Noisy, Codec &WAV &16kHz &10000 &40000 &58 &73& No\\[1pt]
    WaveFake  &2021.11 &Yes &English, Japanese &TTS &No &7 &Clean &WAV &16kHz & 0 &117985 &0 &2& No\\[6pt]
    ADD2022-LF &2022.02 &Restricted &Chinese &TTS, VC &Yes &Unknown &Noisy &WAV &16kHz&36953&123932 &40 &40& No\\[4pt]
    ITW & 2022.09 &Yes &English & \multicolumn{3}{c}{Not provided} & Noisy &WAV &16kHz &19963 &11816& \multicolumn{2}{c}{58}& Unknown\\[2pt]
    Latin-American &2022.11 &Yes &Spanish &TTS, VC &No &6 &Clean &WAV &48kHz &22816 &758000 &78 &84& Yes\\[1pt]
    TIMIT-TTS &2023.04 &Yes &English  &TTS &No &12 & Noisy, Codec &WAV &16kHz &0 &5160&\multicolumn{2}{c}{46}& Yes\\[3pt]
    DeepVoice &2023.08 &Yes & English &VC &No &1 & Noisy &WAV &16kHz &0&8 &\multicolumn{2}{c}{8}& Unknown\\[3pt]
    MLAAD &2024.01 &Yes &23 &TTS &No &54 & Clean &WAV &22kHz &0&76000 &\multicolumn{2}{c}{Not provided}& Unknown\\[3pt]
    
    CodecFake &2024.06 &Yes &English &codec &Yes &6 & Clean &WAV &Multiple &44242&44242 &\multicolumn{2}{c}{107}& Unknown\\[1pt]
    Codecfake &2024.09 &Yes &English, Chinese &codec &Yes &7 & Clean &WAV &Multiple &132277&955939 &\multicolumn{2}{c}{Not provided}& Unknown\\[4pt]
    
    CFAD &2024.10 &Yes &Chinese &TTS, VC &Yes &12 & Noisy, Codec &WAV &16kHz &38600&77200 &\multicolumn{2}{c}{1212}& No\\[4pt]
    \midrule
    \multicolumn{14}{c}{Partially Deepfake datasets}\\
    \midrule
    Partial Spoof**  &2021.04 &Yes &English &TTS, VC &Yes &19 &Clean &Flac &16kHz &12483 &108978&45&61& Yes\\[5pt]
    HAD** &2021.04 &Yes &Chinese &TTS &Yes &Unknown &Clean &WAV &44.1kHz &53612 &753612 &43 &175& No\\[2pt]
    ADD2022-PF &2022.02 & Restricted &Chinese &TTS, VC &No &Unknown &Clean &WAV &16kHz &23897 &127414&\multicolumn{2}{c}{Not provided}& No\\[1pt]
    Psynd &2022.08 &Restricted &English &TTS & No &1 &Codec &WAV &24kHz &30 &2371 &537 &507& No\\[1pt]
    ADD2023-PF** &2023.05 & Restricted &Chinese & TTS, VC  &Yes &Unknown &Noisy, Codec &WAV &16kHz &55468 &65449 &\multicolumn{2}{c}{Not provided}& No\\
  \bottomrule[1.2pt]
          \begin{minipage}{0.9\textwidth} 
          
          \vspace{2pt}
        \linespread{0.6}\selectfont
        {\scriptsize  $^{*}$Unseen attacks refer to Deepfake attacks present in the evaluation set but not included in training set.\\
         $^{**}$These datasets also offer fine-grain labels at segment-level for partial Deepfake.}
        \end{minipage}
\end{tabularx}
\end{table*}

\textbf{Latin-American Voice Anti-spoofing}\cite{tamayo2022voice} The dataset utilizes TTS and VC algorithms to generate Deepfake speeches with five different accents of Latin-American Spanish. 

\textbf{TIMIT-TTS}\cite{salvi2023timit} TIMIT-TTS is a synthetic speech dataset generated by 12 TTS algorithms. 
Various post-processing techniques, including adding Gaussian noises, applying MP3 codecs and adding reverberation compression, are applied to reduce the speech quality and hide some artifacts. 

\textbf{DeepVoice}\cite{bird2023real} This dataset contains one hour of fake speech created by a retrieval-based VC process using real speech from eight well-known figures. Different recording conditions of the speech provide variation within the dataset.

\textbf{Multi-Language Audio Anti-spoofing (MLAAD)}\cite{muller2024mlaad} MLAAD is a newly published speech Deepfake dataset created using 54 TTS models across 23 languages. It can be utilized either as new out-of-domain test data for existing Deepfake detection models or as an additional training resource.

\textbf{CodecFake}\cite{wu2024codecfake} is a dataset featuring Deepfake speech generated by LLM-based methods, as opposed to traditional vocoder-based synthesis techniques. It includes speech synthesized using six representative neural codec models under varying sampling rates and bit rates. This dataset is particularly useful for evaluating the out-of-domain performance of detection models trained on vocoder-based samples.

\textbf{Codecfake}\cite{xie2024codecfake} This dataset contains seven codec models and extends codec-based Deepfake speech to Chinese. This dataset provides another valuable benchmark for codec-based Deepfake detection tasks.

\textbf{Chinese Fake Audio Detection (CFAD)}\cite{ma2024cfad} CFAD is another publicly available Mandarin speech Deepfake dataset consisting of 12 different Deepfake algorithms. 
CFAD provides detailed labelling, including information on the Deepfake type, real data source, noise type, signal-to-noise ratio, and media codec types.

\subsubsection{Partially Deepfake speech datasets}  \hfill\\

\textbf{PartialSpoof}\cite{zhang2022partialspoof} The PartialSpoof dataset contains speech Deepfake with varying proportions of speech Deepfake segments within a single utterance. This is achieved by pairing speech utterances, with one entirely generated using TTS or VC, and another being an original bonafide speech utterance. Short segments within the pairs are randomly substituted with different lengths. This dataset is constructed using utterances in the ASVspoof2019-LA database. Notably, this dataset provides fine-grained labels, including segment-level annotations at different temporal resolutions, as well as detailed labels indicating real segments, spoofing methods, non-speech segments, and concatenated parts.

\textbf{Half-Truth (HAD)}\cite{yi21_interspeech} The HAD dataset features partially fake speech where a few words in an utterance are altered using TTS generation techniques. This dataset is designed to evaluate Deepfake detection methods and localize partially fake speech. The replaced keywords include entities, such as person, location, organization, and time.

\textbf{ADD2022-PF}\cite{yi2022add} The Partially Fake audio detection (PF) track in the ADD2022 challenge dataset contains fake utterances generated by replacing the partial segments of the original genuine utterances with real as well as synthesized speech. The details of generation algorithms for the Deepfake segments are not provided.

\textbf{Partial Synthetic Detection (Psynd)}\cite{zhang2022localizing} The data samples in this dataset are bona fide utterances injected with synthetic speech segments closely resembling the target speakers, generated by multi-speaker TTS algorithms. The bona fide samples are from LibriTTS \cite{
panayotov2015librispeech}, a reading English speech corpus with accents closer to US English. In training, validation, and preliminary test data, each partially Deepfake utterance incorporates one single fake segment. Special cases, such as fully fake, fully real, and multi-fake segments, are stored in the special test set.

\textbf{ADD2023-PF}\cite{yi2023add} The PF track is extended in the ADD2023 challenge, which focuses on locating the manipulated regions in partially fake speech in addition to Deepfake detection. Additive noise and format conversions are also applied to the utterances.






\subsection{Evaluation metrics}
We examine the evaluation metrics utilized in the speech Deepfake detection literature. Furthermore, we emphasize metrics tailored to address the specific challenges of detecting partially Deepfake content.

\textbf{Equal Error Rate (EER)} The EER is one of most widely used evaluation metrics for speech Deepfake CMs. It represents the CM threshold where the false acceptance rate equals the false rejection rate. This metric offers a more comprehensive and objective assessment compared to accuracy, particularly in scenarios with unbalanced evaluation datasets. Notably, EER serves as the evaluation metric in the ASVspoof and ADD challenge series.



\textbf{Tandem Detection Cost Function (t-DCF)} \cite{kinnunen2020tandem} The t-DCF metric is developed during the ASVspoof2019 challenge as an ASV-centric evaluation method. It shifts the focus from the Deepfake CMs alone to offer a more comprehensive assessment for ASV attack detection. Recognizing that Deepfake CMs and ASV systems operate under different hypotheses and objectives, the t-DCF considers both systems combined in cascaded order. It reflects the cost of detection decisions made by the combination of ASV and CM in a Bayesian sense, enabling the integration of Deepfake detector outputs with speaker detection results to assess the robustness of ASV systems against Deepfake spoofing attacks. The t-DCF can be computed through:
\begin{equation}   \label{equation:tdcf} 
\begin{aligned}
     \text{t-DCF}(t_{\text{cm}},t_{\text{asv}})&:=C_{\text{miss}}\cdot \pi_{\text{tar}} \cdot P_{\text{miss,tar}}(t_{\text{cm}},t_{\text{asv}}) + C_{\text{fa,non}}\cdot \pi_{\text{non}}\cdot P_{\text{fa,non}}(t_{\text{cm}},t_{\text{asv}})\\
     &+ C_{\text{fa,df}}\cdot \pi_{\text{df}}\cdot P_{\text{fa,df}}(t_{\text{cm}},t_{\text{asv}}).\\
\end{aligned}
\end{equation}
In Eq. \ref{equation:tdcf}, $\pi_{\text{tar}}$, $\pi_{\text{non}}$, and $ \pi_{\text{df}}$ are the class priors for ASV target, ASV non-target and Deepfake attacks, where $\pi_{\text{tar}} + \pi_{\text{non}} + \pi_{\text{df}} = 1$. $C_{\text{miss}}$, $C_{\text{fa,non}}$ and $C_{\text{fa,df}}$ are
non-negative costs assigned to rejecting an ASV target trial, accepting an ASV non-target trial and accepting a Deepfake trial, respectively. $P_{\text{miss,tar}}(t_{\text{cm}},t_{\text{asv}})$,  $P_{\text{fa,non}}(t_{\text{cm}},t_{\text{asv}})$ and $P_{\text{fa,df}}(t_{\text{cm}},t_{\text{asv}})$ are the three detection error rates as a function of two detection thresholds, where $t_{\text{cm}}$ is for Deepfake detector and $t_{\text{asv}}$ is for speaker verification system. These two sub-systems operate independently and make separate decisions. 

\textbf{Architecture-agnostic Detection Cost Function (a-DCF)} \cite{shim2024dcf} Shim et al. \cite{shim2024dcf} argue that t-DCF is limited to evaluating cascaded ASV and CM systems; however, integrating the ASV and CM components with joint optimization could enhance spoofing-robust ASV performance, as the strengths of one subsystem can compensate for the weaknesses of the other.  Therefore, the a-DCF is proposed, which offers greater flexibility than t-DCF regarding model architectures. Unlike t-DCF, which requires two detection thresholds, $t_{\text{cm}}$ and $t_{\text{asv}}$, a-DCF employs a single detection threshold, $t$. The decision remains binary, classifying inputs as either bona fide speech with a verified speaker (positive class) or as part of the negative class, which can be non-targets of Deepfake spoofs. The a-DCF is defined as
\begin{equation}   \label{equation:adcf} 
\begin{aligned}
     \text{a-DCF}(t_{\text{cm}},t_{\text{asv}})&:=C_{\text{miss}}\cdot \pi_{\text{tar}} \cdot P_{\text{miss}}(t) + C_{\text{fa,non}}\cdot \pi_{\text{non}}\cdot P_{\text{fa,non}}(t)\\
     &+ C_{\text{fa,df}}\cdot \pi_{\text{df}}\cdot P_{\text{fa,df}} (t) ,\\
\end{aligned}
\end{equation}
where $P_{\text{miss}}(t)$ is the miss rate, $P_{\text{fa,non}}(t)$ and $P_{\text{fa,df}}(t)$ are the false alarm rates for non-target speaker verification and Deepfake spoofs, respectively. The a-DCF also have been served as primary metric for the evaluation of spoofing-aware speaker verification systems in ASVspoof5 challenge \cite{wang2024asvspoof}.

\textbf{Range-based EER} \cite{zhang2023Range} This metric is specifically proposed to evaluate segment-level Deepfake detection in the partially Deepfake task. The point-based EER is commonly used for this purpose, requiring the speech to be divided into uniform segments at a fixed resolution, with detection results compared segment by segment. However, using a fixed resolution can lead to mixed segments containing both bona fide and fake speech, which reduces detection precision. To address this issue, Zhang et al. \cite{zhang2023Range} introduce the range-based EER. Instead of pre-segmenting the speech or pre-defining the resolution, the range-based EER focuses on the boundaries between bona fide and Deepfake regions, measuring the duration of misclassified regions between the reference and hypothesis for each trial, thus avoiding resolution-related limitations.

\section{Fully Deepfake detection}
In this section, we evaluate key components of the deep learning-based pipeline for speech Deepfake detection, including model architectures and training optimizations. Most recent models are structured with two modules: front-end feature extraction and back-end classification. However, end-to-end (E2E) architectures have gained attention for avoiding the information loss associated with pre-defined feature extraction. We also assess various training optimizations from the literature, such as data augmentation, loss functions, and activation functions, focusing on their effectiveness in improving Deepfake detection performance. Lastly, we address non-machine learning-based detection methods, highlighting their unique approaches and advantages.

\begin{table*}[t]
  \begin{threeparttable}[b]
  \caption{The Performance of Single-System State-of-the-art Models on the Series of ASVspoof Evaluation Sets} 
  \label{TABLE:PERFORMANCE}
  \scriptsize
  \setlength{\tabcolsep}{1.2pt} 
  \renewcommand{\arraystretch}{1.2}
  \begin{tabularx}{\textwidth} {
   >{\raggedleft\arraybackslash}m{0.04\textwidth}
   |>{\centering\arraybackslash}m{0.13\textwidth}
  |>{\centering\arraybackslash}m{0.145\textwidth}
  |>{\raggedright\arraybackslash}m{0.15\textwidth}
  |>{\raggedright\arraybackslash}m{0.13\textwidth}
  |>{\centering\arraybackslash}m{0.1\textwidth}
  |>{\centering\arraybackslash}m{0.08\textwidth}
  |>{\centering\arraybackslash}m{0.035\textwidth}
  |>{\centering\arraybackslash}m{0.035\textwidth}
  |>{\centering\arraybackslash}m{0.035\textwidth}
  | >{\raggedright\arraybackslash}m{0.04\textwidth}}  \clineB{1-11}{2.5}
    
    \multicolumn{2}{c|}{\multirow{2}{*}{Publication}} & 
    \multicolumn{1}{c|}{\multirow{2}{*}{\shortstack{Data \\augmentation}}} &
    \multicolumn{1}{c|}{\multirow{2}{*}{Feature}} & 
    \multicolumn{1}{c|}{\multirow{2}{*}{Classifier}} & 
    \multicolumn{1}{c|}{\multirow{2}{*}{\shortstack{Loss \\funcion}}} & 
    \multicolumn{1}{c|}{\multirow{2}{*}{\shortstack{\# \\Params}}} &
    \multicolumn{3}{c|}{EER (\%) $\downarrow$} &
    \multicolumn{1}{c}{\multirow{2}{*}{\shortstack{Access-\\ibility}}}\\ \cline{8-10}

    \multicolumn{2}{c|}{}& \multicolumn{1}{c|}{}& \multicolumn{1}{c|}{} & \multicolumn{1}{c|}{}&\multicolumn{1}{c|}{}&\multicolumn{1}{c|}{}&19-LA&21-LA&21-DF& \\ \cline{1-11}

    \cite{Zhang2021silence}&INTERSPEECH'21&w/o&Mel-Spec on 0-4kHz&SE-ResNet-18&AM-Softmax&1.1M&1.14&-&-&No \\ \cline{1-11}
    \cite{tak21_asvspoof}&INTERSPEECH'21&channel masking&RawNet2*&GAT&CE&440K&1.06&6.92&-&Yes\tnote{1} \\ \cline{1-11}
    \cite{hua2021towards}&SPL'21&mix-up&\multicolumn{2}{l|}{E2E: CNN\textrightarrow ResNet\textrightarrow MLP}&CE&350M&1.64&-&-&Yes\tnote{2} \\ \cline{1-11}
    
    \cite{teng2022arawnet}&ICPR'22&w/o&RawNet2+(CQT \textrightarrow ECAPA-TDNN)&CNN\textrightarrow MLP&CE&7.19M&1.11&-&-&Yes\tnote{3} \\ \cline{1-11}
    \cite{Lee2022sasv}&INTERSPEECH'22&w/o&wav2vec2.0-XLSR&MLP&CE&317M&0.31&-&-&No \\ \cline{1-11}
    \cite{Choi2022overlapped}&INTERSPEECH'22&frequency masking&CQT-Spec&LCNN&CE&135K&1.35&-&-&No \\ \cline{1-11}
    \cite{Wang2022SSL}&ODYSSEY'22&w/o&wav2vec2.0-XLSR&Bi-LSTM \textrightarrow MLP&CE&317M&1.28&6.53&4.75&No \\ \cline{1-11}
    \cite{Tak2022autometic}&ODYSSEY'22&RawBoost&wav2vec2.0-XLSR&AASIST&CE&Unknown&-&\textbf{0.82}&\textbf{2.85}&Yes\tnote{4} \\ \cline{1-11}
    \cite{wang2022fully}&DDAM'22&w/o&wav2vec2.0-Large&DARTS&Unknown&Unknown&1.08&-&7.89&No \\ \cline{1-11}
    \cite{jung2022aasist}&ICASSP'22&w/o&RawNet2&GAT&CE&297K&0.83&5.59&-&Yes\tnote{5} \\ \cline{1-11}
    \cite{li2022role}&SPL'22&w/o&LFCC&OCT&Focal loss&250K&1.06&-&-&No \\ \cline{1-11}
    
    \cite{ma2023dualbranch}&SPL'23&w/o&(LFCC \textrightarrow ResNet) + (CQT-Spec \textrightarrow ResNet)& GRL \textrightarrow MLP &CE&Unknown&0.80&-&-&Yes\tnote{6} \\ \cline{1-11} \cite{liu2023leveraging}&ICASSP'23&w/o&RawNet2&Rawformer&CE&370K&0.59&4.98&4.53&Yes\tnote{7} \\ \cline{1-11}
    \cite{chen2023graph}&ICASSP'23&time \& frequency masking&LFB-Spec&GCN&CE&Unknown&0.58&-&-&No \\ \cline{1-11}    
    \cite{zhang2023audio}&ALGORITHM'23&RawRoost&wav2vec 2.0&Transformer&CE&Unknown&-&1.18&4.72&No \\ \cline{1-11}
    \cite{khan2024frame}&ICASSP'24&FIR filter, codec, noises, shift&SDC + Bi-LSTM&Auto-encoder\ \textrightarrow SE-ResNeXT&CE&Unknown&\textbf{0.22}&3.50&3.41&No \\ \clineB{1-11}{2.5}

\end{tabularx}
     \begin{tablenotes}
       \item [1] \url{https://github.com/eurecom-asp/RawGAT-ST-antispoofing}
       \item [2] \url{https://github.com/ghua-ac/end-to-end-synthetic-speech-detection}
       \item [3] \url{https://github.com/magnumresearchgroup/AuxiliaryRawNet}
       \item [4] \url{https://github.com/TakHemlata/SSL_Anti-spoofing}
       \item [5] \url{https://github.com/clovaai/aasist}
       \item [6] \url{https://github.com/imagecbj/End-to-End-Dual-Branch-Network-Towards-Synthetic-Speech-Detection}
       \item [7] \url{https://github.com/rst0070/Rawformer-implementation-anti-spoofing}
       
       \item[*] RawNet2 consists of a learnable SincNet filter and six ResNet blocks 

     \end{tablenotes}
  \end{threeparttable}
          \begin{minipage}{\textwidth}         
          \vspace{2pt}
        \linespread{0.6}\selectfont
        {\scriptsize  The evaluation metric is EER (\%). “$\downarrow$” indicates that a lower score corresponds to better detection performance. "-” indicates that the performance with the corresponding dataset is not reported. The bold values refer to the best performance on the same dataset. "+” indicates multiple techniques processed in parallel, while "\textrightarrow" denotes sequential order.}
        \end{minipage}
\end{table*}

\subsection{Feature Engineering}
We categorize the current methodologies of feature extraction into three groups: hand-crafted traditional spectral features, deep-learning features, and other analysis-oriented approaches, as summarized in TABLE \ref{TABLE:FEATURE}.

\subsubsection{Hand-crafted spectral features} Hand-crafted features have been demonstrated as a strong baseline for speech Deepfake detection, providing a reliable foundation for capturing discriminative patterns of artifacts.

\textbf{Magnitude-based spectral coefficient} The literature shows that the majority of front-end features are derived from the magnitude/power spectrum, where the power spectrum is the square of the magnitude spectrum. Short-term magnitude spectral features are commonly obtained from discrete Fourier transforms (DFT), such as Mel frequency cepstral coefficient (MFCC) \cite{davis1980comparison}, and linear frequency cepstral coefficients (LFCC) \cite{alegre2013one}. Sahidullah et al. \cite{sahidullah15_interspeech} compare nine short-term magnitude-based spectral features alongside their first and second-order derivatives, highlighting the benefits of incorporating dynamic features in capturing temporal changes. 

In contrast to short-term spectral features derived from fixed short window lengths, long-term window transforms have been proposed to capture long-range information and achieve higher frequency resolution. Constant-Q cepstral coefficient (CQCC) \cite{Tak2020CQCC} has shown its effectiveness in speech Deepfake detection by providing higher frequency resolutions at lower frequencies and higher temporal resolution at higher frequencies. Variants of CQCC are continuously proposed \cite{li2023investigation, yang2018extended, yang2020long}. 
In addition to Constant-Q transform (CQT), Li et al. \cite{li2022long} introduces a long-term variable Q transform (L-VQT), where the frequencies vary as a power function rather than exponential as in CQT, aiming to capture better high-frequency information and detect artifacts created by commonly-used vocoders like WaveNet, even in noisy conditions. Apart from CQT-based features, Gao et al. \cite{Gao2021} utilizes global 2D-DCT on Mel-scale magnitude spectrum across both temporal and frequency dimensions to capture long-term modulation artifacts. 

\begin{table*}
  \caption{The Categorization of Feature Extraction Methods}
  \label{TABLE:FEATURE}
  \scriptsize
  \setlength{\tabcolsep}{3pt} 
  \renewcommand{\arraystretch}{1.3}
  \begin{tabularx}{\textwidth} {
  >{\centering\arraybackslash}m{0.08\textwidth}
  |m{0.1\textwidth}
  |m{0.4\textwidth}
  |m{0.35\textwidth}}  \clineB{1-4}{2.5}
    \multicolumn{2}{c|}{Category}&\multicolumn{1}{c|}{Description}&\multicolumn{1}{c}{Methods} \\ \cline{1-4}
    
    \multirow{5}{*}{\shortstack{Hand\\-crafted\\spectral\\features}} & \multirow{2}{*}{\shortstack[l]{Magnitude / \\power-based \\spectral\\ coefficients}} & Short-term: Computing a frequency domain transform on each temporal window of the speech signal to enhance time resolution at lower frequencies. & MFCC \cite{davis1980comparison}, IMFCC \cite{sahidullah15_interspeech}, LFCC \cite{alegre2013one}, RFCC \cite{hasan2013crss}, LPCC \cite{sahidullah15_interspeech}, SSFC \cite{sahidullah15_interspeech}, SCFC \cite{sahidullah15_interspeech}, SCMC \cite{sahidullah15_interspeech}, WPT \cite{gasenzer2023towards}, STFT \cite{li2023robust}, PLP \cite{li2022comparative}, Spec-Env \cite{khan2023spotnet}, Spec-Contrast \cite{khan2023spotnet} \\ \cline{3-4} 
    
    & & Long-term: Longer temporal window. & CQCC \cite{Tak2020CQCC}, CQBC 
    \cite{li2023investigation}, eCQCC \cite{yang2018extended}, CLBC \cite{yang2020long}, L-VQT \cite{li2022long}, CQMOC \cite{yang2021modified}, CFCC \cite{patil2022effectiveness}, Global M \cite{Gao2021}\\ \cline{2-4}
    
    & \shortstack[l]{ \\Phase-based \\spectral \\coefficients} & Working effectively as a complement to the magnitude features. However, it may not be helpful for unknown attacks, especially for VC attacks. & MGDCC \cite{wu2013synthetic}, Relative phase \cite{wang2015relative}, Quadrature phase \cite{gupta2022significance},  MMPS \cite{yang2021modified}, IF \cite{patil2022effectiveness} \\ \cline{2-4}
    
    & Bispectrums & Describing the higher-order spectral correlation in the Fourier domain; is useful for the known attacks. & Statistics of bispectral correlations \cite{albadawy2019detecting} \\ \cline{2-4}

    & Spectrograms & Propagating in time to show the variations in frequencies and intensities of a speech signal in a 2D feature, which are interpreted as images. & STFT-spec \cite{ranjan2023sv}, Mel-spec \cite{ray2021feature}, CQT-spec \cite{abdzadeh2023comparison}, E-Spect \cite{xiang2023extracting}, C-CQT spectrogram \cite{muller23_interspeech}, LBP \cite{gonzalez2020texture}, MLTP \cite{ibrar2023voice}, SDC \cite{khan2024frame} \\ \cline{1-4}
    
    \multirow{3}{*}{\shortstack{DL \\features}} & Filter-learning features & Utilizing DL techniques to construct learnable filterbanks or approximate the standard filtering process. & nnAudio \cite{cheuk2020nnaudio}, FastAudio \cite{fu2022fastaudio}, SincNet \cite{Zeinali2019Detecting} \\ \cline{2-4}
    
    & Deep embeddings & Constructing deep embeddings using DL models through supervised training & ResNet \cite{shim2022graph}, X-vector \cite{Chang2019TransferRepresentation}, auto-encoder \cite{balamurali2019toward}, Bi-LSTM \cite{khan2024frame}, U-net \cite{chen2023twice} \\ \cline{2-4}

    & Pre-trained embeddings & Utilizing SSL models or other DL models pre-trained by external large datasets to extract latent representations of the raw speech waveform. & wav2vec2.0 \cite{Wang2022SSL}, WavLM \cite{zhu2023characterizing}, HuBERT \cite{li2023voice}, TDNN \cite{Ma2023xvector}, ImageNet \cite{lu2022acoustic} \\ \cline{1-4}
    
    \multirow{4}{*}{\shortstack{Analysis\\-oriented \\features}} & Prosody/ semantic features & Focusing on perceptual features, such as prosody and emotion of the speech sounds, which works effectively on TTS-based speech Deepfake, not the VC. & shimmer \cite{li2023contributions}, phoneme duration \cite{Wang2023prodoscic}, pronunciation \cite{Wang2023prodoscic}, prosody \cite{attorresi2022combining}, emotion \cite{conti2022deepfake}, VOT \cite{dhamyal2021fake}, coarticulation \cite{dhamyal2021fake}, breath \cite{layton2024every}\\ \cline{2-4}
    
    & The impact of silence & Contributing effectively to the current speech Deepfake detection models. & Silence portion \cite{Muller2021silence},  BTS-Encoder \cite{doan2023bts}\\ \cline{2-4}
    
    & Frequency sub-band features  & Focusing on one or more specific portions of the frequency band, rather than the entire frequency range.   & F0 \cite{Fan2023F0}, 0-4kHz 
    \cite{Zhang2021silence}, 4-8kHz \cite{Sankar2022voicedforVC} \\ \cline{2-4}
    
    & \shortstack[l]{\\Other\\directions} & Including recent efforts in the development of features for detecting speech Deepfake. & Varied input length \cite{Wang2021Comparative}, energy loss \cite{deng2022detection}, face embedding \cite{xue2023cross}, dual channel stereo feature \cite{Liu2023betray}, Compressed coding metadata \cite{yadav2023assd}\\ \clineB{1-4}{2.5}
  
\end{tabularx}
\end{table*}

\textbf{Phase-based spectral coefficient} The speech Deepfake often lacks natural phase information, as the human auditory system tends to be less sensitive to phase spectrum characteristics compared to magnitude spectrum features. Therefore, investigating the phase information can be effective in capturing these artifacts in speech Deepfake. Wang et al. \cite{wang2015relative} propose a relative phase extraction method aimed to reduce the phase variation, where the peaks of the utterance waveform serve as the center of each window section.  Furthermore, Gupta et al. \cite{gupta2022significance} demonstrate the significance of a quadrature phase over other phase angles by performing Mutual Information-based analysis. 

\textbf{Bispectrum} The magnitude/power spectrum lacks sensitivity to higher-order spectral correlations revealed by bispectral analysis. A qualitative assessment \cite{albadawy2019detecting} reveals the difference in the first four moments of bispectral correlations between real and synthesized speech, particularly under a high Signal-to-Noise Ratio (SNR) condition. 

\textbf{Spectrogram-based feature} In speech Deepfake detection, magnitude-based spectral coefficients are typically integrated with the magnitude of the speech signal over time to form a spectrogram, a two-dimensional (2D) feature. The spectrogram includes information regarding frequencies and intensities of the speech signal as it propagates in time. Front-end features, such as Mel-spectrogram (Mel-Spec), and CQT-spectrogram (CQT-Spec) are always treated as images and passed to CNN-based back-end classifiers  \cite{ray2021feature}. The spectrogram requires less computation power than extracting the spectral coefficients through DCT while promising detection accuracy \cite{abdzadeh2023comparison}. Xiang et al. \cite{xiang2023extracting} propose an Efficient Spectrogram (E-Spec), which applies STFT directly to the decoded speech signal along the frequency axis after the compression filterbank. Phase information is also integrated with the magnitude spectrogram \cite{pilia2021time, Jung_Shim_Heo_Yu_2019}. Muller et al. \cite{muller23_interspeech} introduce a complex-valued CQT (C-CQT) log-spectrogram to embed phase information, which requests modifications to classifier networks, activation functions, and Batch Normalization (BN) to handle complex-valued input and weights. Furthermore, some research applies texture analysis tools to spectrogram-based features, including local binary pattern (LBP), and Modified local ternary patterns (MLTP). Khan et al. \cite{khan2024frame} propose a method to slice the log-mel spectrogram into square segments, forming local spectral deviation coefficients (SDC) for detecting the frame-level inconsistencies. 

\subsubsection{Deep-learning (DL) features} With the advancement of deep learning approaches, DL-based structures have been adopted to extract learnable embeddings to describe the underlying characteristics of the raw speech, alongside traditional hand-crafted features. Various types of the recent DL-based features are discussed below.

\textbf{Filter-learning feature} DL techniques are involved in approximating the standard filtering process for both STFT-based and First-order Scattering Transform (FST)-based front-ends. Learnable STFT-based features like nnAudio \cite{cheuk2020nnaudio} are implemented without constraining the shape of the filter, leading to a larger number of training parameters and potential overfitting. Fu et al. \cite{fu2022fastaudio} enhance nnAudio by restricting the filter shape to triangular and making only the filterbanks learnable,  resulting in improved performance. On the other hand, learnable FST-based front-ends, such as SincNet \cite{Zeinali2019Detecting}, utilize a convolutional layer to parameterize the sinc function, effectively acting as a customized filter bank. However, it also may suffer from overfitting by learning the low and high cut-off frequencies during training. In the work of RawNet2 \cite{tak2021end}, SincNet is utilized to extract the front-end feature directly from the raw speech while fixing the cut-off frequencies to reduce overfitting. The success of RawNet2 makes it one of the most well-known and reproducible models in speech Deepfake detection, serving as an official baseline in the ASVspoof challenge series. Furthermore, an additional channel dimension is added to the output of the SincNet front-end to form a time-frequency representation \cite{tak21_asvspoof} .

\textbf{Deep embedding derived from hand-crafted features} In this category, DL models, such as Deep Neural Networks (DNN), Residual Networks (ResNet), and Recurrent Neural Networks (RNN) are applied after extracting hand-crafted features to construct deep embeddings \cite{shim2022graph, khan2024frame}. Teng et al. \cite{teng2022arawnet} choose to use the whole structure of RawNet2 as an encoder for raw waveforms, combined with ECAPA-TDNN \cite{Chen2021urchannel} as the encoder for the CQT feature. Two embeddings from both encoders are concatenated and encoded by a convolutional layer with BN. Most recently, a U-net network is utilized with attention mechanisms and skip connections, termed Twice Attention U-net (TA-Unet), to process CQT-spectrograms, aiming to prevent overfitting while emphasizing artifact locations on the spectrogram feature \cite{chen2023twice}. 

\textbf{Pre-trained embedding} Self-supervised learning (SSL) models have proven effective in generating latent representations from raw speech waveforms. These SSL-based features outperform traditional hand-crafted acoustic features and other learnable features across various tasks, such as speech and emotion recognition. In the context of speech Deepfake detection, research indicates that using pre-trained wav2vec 2.0 features instead of hand-crafted features or SincNet front-ends, significantly improves detection performance when the same back-end classifier and data augmentation techniques are used \cite{lee2023experimental, Tak2022autometic, lv2022fake}. Furthermore, fine-tuning SSL features alongside the classifier during training accelerates convergence and enhances detection accuracy for both known and unknown attacks \cite{Wang2022SSL}. Martin-Donas et al. \cite{donas2022vicomtech} apply temporal normalization to the hidden states of each transformer layer in the wav2vec 2.0 model, where each normalized representation is scaled with a trainable weight during fine-tuning. Moreover, integrating attention mechanisms into SSL-based models further enhances their detection capabilities \cite{zhu2023characterizing}.

Experiments have also shown that pre-trained models trained on diverse speech data sources achieve better results on out-of-domain samples. For example, Phukan et al. \cite{phukan2024heterogeneity} demonstrate that models like wav2vec2.0-XLS-R \cite{babu22_interspeech}, trained on 436k hours of speech across 128 languages, Whisper \cite{radford2023robust}, trained on 96 languages, and MMS \cite{pratap2024scaling}, trained on over 1400 languages, perform better in detecting non-English Deepfakes than SSL models trained solely on English data.


\subsubsection{Other analysis-oriented features} In addition to extracting hand-crafted features or high-level embeddings, various other specific directions for feature development have been explored to improve the robustness of detection systems.

\textbf{Prosody and semantic feature} Dhamyal et al. \cite{dhamyal2021fake} investigate microfeatures, including Voicing Onset Time (VOT) and coarticulation, which are associated with voice production mechanisms in humans. The continuous shimmer is proven to reflect the stability of amplitude and frequency perturbation in the voice \cite{li2022analysis}. DL-based prosody and semantic features have also been proposed. Wang et al. \cite{Wang2023prodoscic} incorporate two types of prosodic features with wav2vec 2.0 embeddings, including phoneme duration feature and pronunciation features. Attorresi et al. \cite{attorresi2022combining} train the same structure of the prosody encoder used by Tacotron to enhance prosody in TTS-synthesized speech as the prosodic embedding extractor in the detection process. A pre-trained Speech Emotion Recognition (SER) system is also used to extract emotion embeddings based on the semantics for speech Deepfake detection \cite{conti2022deepfake}. Like other mentioned prosody and semantic features, this emotion embedding is effective for TTS-generated fake speeches, rather than VC-generated.

\textbf{The impact of silence} Some researchers argue that the silence portion works as a significant feature for the current speech Deepfake detectors \cite{Muller2021silence}. Specifically, the duration proportion of silence plays a significant role in detecting TTS Deepfakes, while the content of silence is an important factor in detecting VC attacks. It is because TTS algorithms lack the ability to model diverse and accurate pauses, whereas silence portions in VC-based speech Deepfake have signal discontinuity compared to bona fide speech. Utilizing voice activity detection (VAD) to detect and remove silent portions leads to performance degradation. To utilize the effectiveness of the silent part, Doan et al. \cite{doan2023bts} encode the correlation between breathing, talking and silence sounds in speech clips as the front-end feature. It is important to note that all existing research on the impact of silence is conducted using clean data. These findings may not necessarily hold in noisy conditions or when using various codecs.

\textbf{Frequency sub-band feature} Studies have investigated sub-band spectral information to identify specific frequency ranges containing more discriminative information relevant to detecting speech Deepfake \cite{Sriskandaraja2016}. Research shows that sub-band features, particularly within the low-frequency band of 0-4kHz, outperform the full-band features against channel effects, codecs and noisy conditions, while the high-frequency part of spectrograms may lead to overfitting \cite{li2023robust}. This finding holds significance for the development of resource-constrained Deepfake detectors. Works in \cite{Fan2023F0} and \cite{xue2022audio} further narrow down the low-frequency band to 0-400Hz, focusing on the fundamental (F0) frequency. By only including the F0 sub-band of the log-power spectrogram as the feature, it can still achieve a satisfactory detection result on the ASVspoof2019-LA set  with an EER of 1.15\%. Sankar et al. \cite{Sankar2022voicedforVC} point out that, in voiced segments, most spectral differences lie within the 0-4kHz frequency band. Conversely, for silence and unvoiced segments, the spectral discriminating features predominantly reside in the 4-8kHz range. This observation may explain performance degradation after silence removal, particularly when the feature emphasizes high frequencies.

\textbf{Varied input length} To pass inputs into DL architectures in batch, speech inputs are often set to be fixed-size through trimming or padding. However, this approach may lead to information loss or the propagation of irrelevant information \cite{chettri2020dataset}. Wang et al. \cite{Wang2021Comparative} propose to add a pooling layer before DL models to handle varied input lengths, which outperforms fixed-length inputs with the same back-end classifiers. 

\textbf{Other directions} Other research endeavours have aimed to develop alternative types of features contributing to speech Deepfake detection. Deng et al. \cite{deng2022detection} investigate the energy loss in pauses between words and the high-frequency range caused by Deepfake algorithms. Yadav et al. \cite{yadav2023assd} explore the utilization of compression coding metadata information solely from the compressed bit-stream as a feature, which overcomes detection problems for compressed speech.  Xue et al. \cite{xue2023cross} construct face embeddings from spectrograms to describe speaker information and combine face features and speech features as the front-end. Liu et al. \cite{Liu2023betray} convert mono audio signals to dual channels, encoding and processing each channel signal separately to capture different detail cues for speech Deepfake detection.

\subsubsection{Performance discussion on feature selection}
The trend in feature engineering for speech Deepfake detection is moving away from hand-crafted features and towards deep embedding representations, especially those derived from pre-trained SSL-based models. This shift is driven by two main factors. First, traditional acoustic features, which are based on mathematical principles, may not capture hidden information from unknown attacks as effectively as learnable features. Second, hand-crafted perceptual features, such as pitch, formant frequencies, and articulatory patterns, can be influenced by language. Even within the same language, different accents introduce variations in these features due to differences in pronunciation, vowel length, and stress. As a result, the effectiveness of hand-crafted features can vary significantly depending on the diversity of both bona fide data and Deepfake types in the training set. Learnable features, on the other hand, excel at extracting high-level representations from raw speech data, leading to better performance in cross-dataset testing scenarios. Additionally, since SSL-based models are pre-trained on large datasets in multiple languages, they are better equipped to handle multilingual deepfake detection tasks.

Despite these advantages, hand-crafted features still have their merits and should not be overlooked. They require fewer computational resources and offer a high degree of interpretability, whereas learnable features typically need longer training times and larger model parameters. In situations where training data is limited in quantity or quality, hand-crafted features and SSL-based features using transfer learning often outperform learnable features. Therefore, a promising approach is to integrate both hand-crafted and learnable features to build a robust detection system. Furthermore, while prosodic-based and phase-based features alone may not provide the most competitive detection results, they can be valuable when used in combination with other features, such as magnitude-based spectral features and learnable deep embeddings.


\subsection{Classifier Architecture}
In addition to traditional machine learning classifiers, DL-based approaches have come to dominate speech Deepfake detection. Building on advancements made in detecting visual Deepfake artifacts in images and videos, the development of speech Deepfake detection techniques has similarly evolved, with early models based on CNNs and ResNets giving way to more sophisticated architectures such as GNNs, Transformers, and other DL-based structures. We assess the strengths and limitations of different classifier structures, as summarized in TABLE \ref{TABLE:CLASIFIER}. Certain models incorporate multiple DL architectures. For instance, RawNet2 \cite{tak2021end} integrates a gated recurrent unit (GRU) layer after ResNet blocks. Here, we categorize these models based on their primary structure.

\subsubsection{Traditional Machine Learning (ML) classifiers} The ML-based classifiers are widely used in the early years of Deepfake detection, including support vector machines (SVM), Gaussian mixture models (GMM), and random forest (RF) \cite{yu2016effect, Chang2019TransferRepresentation, ji17_interspeech}. In particular, GMM-based classifiers serve as a fundamental baseline in the ASVspoof challenge series.

\subsubsection{Convolutional Neural Network (CNN)} CNN architecture is well-known for its effectiveness in capturing local and hierarchical features. Lavrentyeva et al. \cite{Lavrentyeva2019} apply the CNN architecture to address the Deepfake detection problem while reducing model size by implementing a Light-CNN (LCNN). LCNN mainly replaces ReLU with Max-Feature-Map (MFM) activation, which selects the maximum value of each of the two feature channels as output, effectively halving the LCNN architecture. The MFM layer also performs feature selection. However, the translation invariance property in CNN may lead to information loss in the frequency domain, particularly because different sub-band frequencies contain diverse information. Choi et al. \cite{Choi2022overlapped} suggest splitting spectrogram inputs along the frequency axis and processing the high-, mid- and low-frequency bands by the LCNN separately. 
Luo et al. \cite{Luo2021capsule} refine the dynamic routing strategies in capsule networks to enhance speech deepfake detection by focusing on hierarchical feature structures and the spatial characteristics of artifacts. 

\subsubsection{Residual Network (ResNet)} ResNet is one of the significant variants of CNN, addressing the vanishing gradient problem in a deep network by incorporating skip connections. Recent works have focused on enhancing the fundamental structure of ResNet. Lai et al. \cite{Lai2019ASSERT} integrate squeeze-and-excitation (SE) blocks with ResNet, forming SE-Nets, to perform dynamic channel-wise feature recalibration. In \cite{kwak2021resmax} and \cite{lee2023experimental}, the MFM activation layer is adapted to each ResNet block, termed ResMax. 
To improve the information flow in ResNet, DenseNet \cite{ wang2020densely} is proposed to skip connections linking each layer to all layers within the same dense block. DenseNets also have fewer parameters compared to the conventional ResNets.

In addition to the previously discussed techniques, Res2Net \cite{li2021replay} stands out as another significant variant of ResNet. Res2Net modifies the bottleneck block to incorporate a hierarchical residual-like connection. Instead of passing the entire feature map to the convolutional layer as a whole, Res2Net divides the input feature map along the channel dimension into several feature segments of equal size. Before conducting the convolutional operation on each subsegment, the convolutional result from the previous subsegment is added, creating a multi-scale feature representation. Furthermore, adjustments are made to the addition operation of feature segments in the basic Res2Net structure \cite{yang2023comparative}. 

\begin{table*}[t]
  \caption{The Categorization of Classifiers}
  \label{TABLE:CLASIFIER}
  \scriptsize
  \setlength{\tabcolsep}{3pt} 
  \renewcommand{\arraystretch}{1.3}
  \begin{tabularx}{\textwidth} {
  >{\centering\arraybackslash}m{0.1\textwidth}
  |m{0.3\textwidth}
  |m{0.3\textwidth}
  |m{0.24\textwidth}}  \clineB{1-4}{2.5}
    \multicolumn{1}{c|}{Category}&\multicolumn{1}{c|}{Advantages}&\multicolumn{1}{c|}{Disadvantages} &\multicolumn{1}{c}{Methods}\\ \cline{1-4}
    
    Traditional ML & Light-weight; facilitating easier interpretation of the distribution outcomes & Poor generalization performance on unseen attacks   & GMM \cite{yu2016effect}, RF \cite{ji17_interspeech}, SVM \cite{Chang2019TransferRepresentation}\\ \cline{1-4} 

    CNN & Light-weight; Producing promising detection performance & Causing information loss in the frequency domain due to the translation invariant property& LCNN \cite{Lavrentyeva2019}, Non-OFD \cite{Choi2022overlapped}, CapsuleNet \cite{Luo2021capsule} \\ \cline{1-4} 

    ResNet & Enabling architectural adjustments for modifying receptive fields; enhancing generalizability to unseen attacks; accommodating deeper networks & High computational cost; The performance can be highly varied by feature selection & ResNet \cite{alzantot19_interspeech}, SE-Net \cite{Lai2019ASSERT}, ResMax \cite{kwak2021resmax}, 
    Res2Net \cite{li2021replay}, DenseNet \cite{wang2020densely}, xResNet \cite{Castan2022speaker} \\ \cline{1-4} 

    GNN & Aggregating all note features for message passing; enhancing the formulation of inter-relationships among frame-level features & Challenging to construct a deep network; high time and space complexity & RawGAT \cite{tak21_asvspoof}, AASIST \cite{jung2022aasist}, GCN \cite{chen2023graph} \\ \cline{1-4} 

    Transformer & Effectively capturing long-term dependencies & Potential for overfitting; high computational costs & CCT \cite{bartusiak2021synthesized}, OCT \cite{li2022role}, TFT \cite{wang2022npu}, Rawformer \cite{liu2023leveraging} \\ \cline{1-4} 

    TDNN & Lightweight; allowing varying input lengths & Unsatisfactory detection performance & ECAPA-TDNN \cite{Chen2021urchannel}\\ \cline{1-4} 

    DART & Enabling architecture optimization during back-propagation & Performance may be influenced by pre-defined hyperparameters & PC-PARTS \cite{Ge2021pcdart}, light-DARTS \cite{wang2022fully} \\ \clineB{1-4}{2.5}
  
\end{tabularx}
\end{table*}

\subsubsection{Graph Neural Network (GNN)}
To apply GNN in speech Deepfake detection, the frequency bins and time frames can be utilized as nodes to form a fully connected graph. One common variant of GNN is the graph convolutional network (GCN). Chen et al. \cite{chen2023graph} divide the spectrogram into grid patches and extracts each patch embedding using a CNN, emphasizing the network's focus on the relationship between patches. Subsequently, each patch embedding passes through several layers of GCN to aggregate node features within the same time frames or frequency bands. 
      
Graph attention network (GAT) introduces the attention mechanism during node feature aggregation. Tak et al. \cite{Tak2021Graph} conduct experiments utilizing several GATs with the conventional handcrafted feature and log-linear filterbank feature (LFB). Then, the LFB features is replaced by the learnable SincNet front-end, named RawGAT \cite{tak21_asvspoof}. The GAT-based classifier consists of three components. Initially, the first two GATs individually model the relationships within spectral and temporal domains. Then, these two sub-graphs are fused to facilitate the processing of the third GAT, thereby leveraging complementary information. AASIST \cite{chen2023graph} is introduced as an enhancement to RawGAT by incorporating a heterogeneity-aware technique to integrate spectral and temporal sub-graphs. In AASIST, each node aggregates information from all other spectral nodes and temporal nodes in the graph, whereas nodes in RawGAT only aggregate information within the spectral and temporal sub-graphs individually. Huang et al. \cite{huang2023frequency} make two adjustments to RawGAT by adding a pre-emphasis module before the SincNet filter to enhance the high-frequency components and replacing BatchNorm with LayerNorm to reduce the impact caused by uneven samples. These two adjustments lead to a 51\% improvement in the ASVspoof2019-LA set.

\subsubsection{Transformer} In the context of speech Deepfake detection, Transformer encoders are often integrated with other DL architectures, such as ResNet \cite{zhang2021fake} or CNN \cite{ li2023bilevel}. The compact convolutional Transformer (CCT) \cite{bartusiak2021synthesized} is proposed by incorporating two 2D convolutional layers before Transformer encoders to enhance generalization ability. This strategy aims to extract high-level embeddings from the input spectrogram feature, rather than directly dividing the spectrogram into patches and feeding them to the Transformer. Li et al. \cite{li2022role} modify the 2D convolutional layer into 1D, accompanied by a smaller number of Transformer encoders, named OCT,  to reduce overfitting. Rawformer, proposed by \cite{liu2023leveraging} combines SE-Res2Net with a positional aggregator before passing to Transformer encoders. This integration of Res2Net and Transformer architectures aims to effectively capture both local and global dependencies.

The conventional Transformer architecture typically focuses on the temporal domain exclusively. However, recent advancements have been made to adapt the Transformer to treat the temporal and frequency dimensions equally.  Zhang et al. \cite{zhang2023audio} leverage both the feature matrix and its transposed version to facilitate self-attention mechanism across both temporal and frequency domains. 

\subsubsection{Time-Delay Neural Network (TDNN)} TDNN is widely recognized in tasks like speech recognition by converting acoustic signals into phonetic representations. Various efforts have been made to utilize TDNN variants, such as ECAPA-TDNN, for speech Deepfake detection \cite{Chang2019TransferRepresentation, Caceres2021Vox, Chen2021urchannel}. However, the performance of TDNN in Deepfake countermeasures remains suboptimal compared to their effectiveness in speaker verification.

\subsubsection{Differentiable Architecture Search (DART)} DART introduces a dynamic detection model, enabling optimization of both architectures and parameter values based on performance on the validation set using gradient descent. The candidate operations for network building blocks include convolutional, pooling, and residual layers. To reduce computation power and memory usage,  Ge et al. \cite{Ge2021pcdart} propose a partially-connected DARTS (PC-DARTS) by adding a random mask to some partial channels during the architecture search stage. With random masking, PC-DARTS ensures complex architecture learning while reducing training time by 50\% compared to standard DARTS. In \cite{wang2022fully}, Light-DARTS is introduced, incorporating the MFM module as one of the potential candidate operations within the architecture search space, where the MFM module functions as a feature selection mechanism.

\subsubsection{Other Architectures} In addition to architectures mentioned above, advanced techniques like quantum neural networks are also being explored to address the problem of speech Deepfake. Wang et al. \cite{wang2023quantum} propose employing a 4-qubit variational quantum circuit network as the back-end classifier. They design a pipeline that utilizes Bi-LSTM to convert the feature maps into low-dimensional embedding vectors, which can be processed by the quantum circuit.

\subsubsection{Pooling and Attention mechanism} Pooling layers and attention mechanisms serve critical functions in back-end classifiers by highlighting discriminative information.

\textbf{Statistical pooling} Common pooling methods include max pooling and average pooling, which respectively select the maximum value or compute the average value along a dimension. Introduced by \cite{snyder17_interspeech}, statistics pooling, employed in x-vector architectures with TDNN, calculates and concatenates the mean and standard deviation of frame-level embeddings to generate an utterance-level representation. An enhanced version, attentive statistics pooling (ASP) \cite{Lee2022sasv, donas2022vicomtech, ma2023dualbranch} incorporates an attention mechanism into statistics pooling. ASP assigns channel-dependent weights to each frame, dynamically emphasizing the most informative frames during pooling. 

\textbf{Attention} In recent research, attention mechanisms have been integrated into classifier architectures such as LCNN and ResNet \cite{ma2023dualbranch, wang2022singleattention}. Well-known attention mechanisms include convolutional block attention modules (CBAM) and dual attention network (DANet), which address both channel attention and spatial attention. CBAM comprises two sub-attention modules in sequence, while DANet incorporates them in parallel. Zhou et al. \cite{zhou2022spoof} employ topK pooling before computing attention weights across temporal and spatial dimensions, aimed at preserving a lightweight architecture. 

\subsubsection{Performance discussion on classifier selection}While the existing countermeasures have shown effectiveness in detecting fully Deepfake speech, each classifier type comes with its own set of limitations that require attention, as outlined in TABLE \ref{TABLE:CLASIFIER}. Employing ensemble methods can be advantageous in addressing these limitations. For instance, CNN-based detectors are widely used for their proficiency in extracting local patterns, but they may struggle to capture long-term information. In such cases, ensembling with Transformer-based models can help in capturing global temporal features. GATs excel in formulating inter-relationships among frame-level features, which can lead to improved performance when combined with other CNN-based detectors. Additionally, DL-based classifiers may face challenges related to overfitting, resulting in poor performance if the training dataset is small or lacks diversity.

\subsection{End-to-End (E2E) Architecture}
One limitation of detection models with a two-stage architecture is their high dependency on extracted features.  The performance of classifiers can vary significantly depending on the chosen input features. Additionally, information lost during feature extraction is often irretrievable, such as the neglect of phase information when using power spectrum features. Therefore, E2E architectures have gained more attention for speech Deepfake detection model development \cite{chintha2020recurrent, khan2023battling}. In E2E models, the entire process is encapsulated within a single network, eliminating the need for separate front-end feature processing and back-end classifier engineering stages. However, many proposed architectures claimed as E2E models actually utilize a learnable SincNet filter with fixed cutoff frequencies and Mel scales to extract a 2D spectral-temporal feature map. The use of learnable cutoff frequencies in the speech Deepfake detection task may lead to overfitting, and the pre-determined settings make them not truly representative of E2E models.

Ma et al. \cite{Ma2021RWResnet} present a genuinely E2E architecture by directly obtaining embedding from the raw waveform using a 1D convolutional layer with a kernel size of 3 followed by ResNet blocks. They also enhance the model by incorporating a 1D convolutional layer and BN layer into the skip connection of ResNet blocks, thus extending the perceptual information range. Hua et al. \cite{hua2021towards} modify the kernel size of the first 1D convolutional layer to 7 to capture longer-range dependencies. Wu et al. \cite{Wu2020FeatureGenuinization} construct a stack of convolutional layers and convolutional transpose layers acting as a genuinization transformer, which is similar to an autoencoder to learn the characteristics of bona fide speech and amplify differences between fake and real speech. Fang et al. \cite{fang2021voice} introduces an E2E model based on Res2Net blocks, wherein information from previous subsegments undergoes another convolutional layer with a kernel size of 1 before being added to the current subsegment. Conversely, the E2E ConvNet architecture proposed by \cite{ma2022convnext} alternates between 1D convolutional layers and original Res2Net blocks, while appending a channel attention layer to the end of each Res2Net block, assigning channel-dependent weights.

\subsection{Training Optimization Techniques}
We evaluate a range of training optimization techniques within the domains of data augmentation, loss function, and activation function. We aim to provide insights into their application and impact on enhancing the performance and generalization capabilities of speech Deepfake detection systems. TABLE \ref{TABLE:ABLATION} illustrates how integrating specific optimization techniques into detection algorithms leads to detection performance improvement.

\begin{table*}
  \caption{Ablation Study of Training Optimization Techniques on the ASVspoof2019-LA, ASVspoof2021-LA, ASVspoof2021-DF Datasets}
  \label{TABLE:ABLATION}
  \scriptsize
  \setlength{\tabcolsep}{3pt} 
  \renewcommand{\arraystretch}{1.3}
  \begin{tabularx}{\textwidth} {
   >{\centering\arraybackslash}m{0.165\textwidth}
  |>{\centering\arraybackslash}m{0.08\textwidth}
  |>{\centering\arraybackslash}m{0.08\textwidth}
  |>{\centering\arraybackslash}m{0.09\textwidth}
  |>{\centering\arraybackslash}m{0.09\textwidth}
  |>{\centering\arraybackslash}m{0.09\textwidth}
  |>{\centering\arraybackslash}m{0.09\textwidth}
  |>{\centering\arraybackslash}m{0.09\textwidth}
  |>{\centering\arraybackslash}m{0.09\textwidth}}  \clineB{1-9}{2.5}
    \multirow{3}{*}{Methods} & \multirow{3}{*}{\shortstack{First proposed \\year}} & \multirow{3}{*}{\shortstack{Used in \\literature}} & \multicolumn{6}{c}{EER (\%) $\downarrow$} \\ \cline{4-9}
    
    & & &\multicolumn{2}{c|}{ASVspoof2019-LA} & \multicolumn{2}{c|}{ASVspoof2021-LA} & \multicolumn{2}{c}{ASVspoof2021-DF} \\ \cline{4-9}

    & & & w/o & with & w/o & with & w/o & with\\ \clineB{1-9}{2.5}
    \multicolumn{8}{c}{Data augmentation} \\ \clineB{1-9}{2.5}

    Masking &2019& \cite{Gao2021} & 6.51 & 5.14 & - &-&-&-\\ \cline{1-9}
    Mix-up &2021&  \cite{dong2023multi} & - & - & 4.04 & 3.09  & - & - \\ \cline{1-9}
    \multirow{2}{*}{Noise addition} &\multirow{2}{*}{2021} & \cite{tak2022rawboost} & - & - & 9.50 & 5.31  & - & - \\ \cline{3-9}
     & & \cite{Tak2022autometic} & - & - & 4.48 & 0.82  & 4.57 & 2.85 \\ \cline{1-9}

     \multirow{2}{*}{Codec augmentation} & \multirow{2}{*}{2022} &\cite{wang2022multi} & 2.24 & 6.41 & 30.17 &  7.96 & - & - \\ \cline{3-9}
     & & \cite{cohen2022study} & - & - & 21.41 & 7.22  & 29.31 & 21.60 \\ \clineB{1-9}{2.5}
    
     \multicolumn{8}{c}{Loss function} \\ \clineB{1-9}{2.5}
     AM-Softmax &2020&  \cite{wang2022singleattention} & 1.41 & 1.29 & - & -  & - & - \\ \cline{1-9}
    LMCL &2020&  \cite{chen2020generalization} & 4.04 & 3.49 & - & -  & - & - \\ \cline{1-9}
    OC-Softmax & 2021& \cite{Zhang2021oneclass} & 4.69 & 2.19 & - & -  & - & - \\ \cline{1-9}
    MSE & 2021& \cite{Wang2021Comparative} & 3.04 & 1.92 & - &  - & - & - \\ \cline{1-9}
    Focal loss &2021& \cite{Caceres2021Vox} & - & - & 9.21 & 7.51  & - & - \\ \cline{1-9}
    Center loss* & 2023& \cite{huang2023frequency} & 0.68 & 0.52 & 3.85 & 3.38  & - & - \\ \clineB{1-9}{2.5}
\end{tabularx}
          \begin{minipage}{\textwidth}         
          \vspace{2pt}
        \linespread{0.6}\selectfont
        {\scriptsize  The evaluation metric is EER (\%). “$\downarrow$” indicates that a lower score corresponds to better detection performance. For data augmentation, "w/o" means that no data augmentation techniques are applied, whereas "with" denotes that only the specified data augmentation method is applied. For loss function, "w/o" means that the model utilizes the CE loss with Softmax as the loss function, while "with" indicates that the loss function switches to the specified alternative. “-” indicates that the authors do not report the performance with the corresponding dataset.
        
        $^{*}$ Under "with", Center loss is incorporated alongside the CE loss with Softmax.}
        \end{minipage}
\end{table*}

\subsubsection{Data Augmentation (DA) techniques} DA techniques serve as indispensable tools to enhance the robustness and diversity of datasets in training speech Deepfake detection models. This section assesses some commonly used DA techniques, including masking, mix-up, and codec variation, along with other DA methods.

\textbf{Masking} SpecAugment, initially introduced a DA method for speech recognition, involves randomly masking blocks of frequency bins and/or blocks of time steps on spectrograms. Over time, SpecAugment has been widely applied to improve the performance of Deepfake detectors \cite{Gao2021, alenin2022subnetwork, chen2020generalization, li2023voice}. Additionally, Zhou et al. \cite{zhou2022spoof} conduct experiments with random masking on the dimension of channels. Subsequently, SpecAverage \cite{cohen2022study} is proposed, to replace the random masking of the feature map with the average feature value rather than the value of zero. 

\textbf{Mix-up} To prevent information loss on critical speech segments, Kim et al. \cite{Kim2021specmix} introduce a cut-and-mix technique, known as SpecMix. This method combines two data samples, where up to three frequency bins or temporal bands are masked on one sample. Then, the masked regions are replaced with features from another sample to create a new training sample. Notably, the label for the new sample is determined through a weighted average of the labels from the two original samples, with the weights assigned by the extent of the masked areas. SpecMix is effective to prevent overfitting and generalize to the unseen attack \cite{Tomilov2021, hua2021towards}. However, determining the percentage of SpecMix over the entire training set is crucial. Experimental results indicate that either entirely omitting SpecMix operations or uniformly applying SpecMix to all data can lead to performance degradation \cite{dong2023multi}.

\textbf{Noise addition} The addition of noise and room impulse responses (RIRs) has been shown to improve the robustness of detection models \cite{tak2022rawboost, muller23_interspeech}. Notably, RawBoost \cite{tak2022rawboost} demonstrates strong performance under unknown encoding and transmission conditions. By perturbing a set of source utterances with a combination of linear and non-linear convolutive effects, along with both impulsive and stationary additive noise applied either serially or in parallel, RawBoost enhances the reliability of speech Deepfake detection especially for ASVspoof2021-LA set.

\textbf{Codec augmentation} This technique is utilized to replicate compression algorithms used in encoding speech signals, thereby enhancing the model's robustness to unseen coding and transmission artifacts, especially in datasets like ASVspoof2021-LA set \cite{ fathan2022mel, Caceres2021Vox}. Codec augmentation can be categorized into two main types: multimedia encoding and transmission encoding. In multimedia encoding, raw speech is transformed into various codecs with different sampling rates, before being resampled back to 16kHz. Transmission encoding involves employing multiple Voice over Internet Protocol (VoIP) or telephony transformation techniques, with varying bit rates. Cohen et al. \cite{cohen2022study} simulate the random packet loss as augmentation. Meanwhile, Tomilov et al. \cite{Tomilov2021} apply band-pass finite impulse response (FIR) filters to mimic speech codec, which may cause information loss at specific frequency bands.

Other commonly used DA techniques include speed perturbation \cite{wang2022multi}, time stretching \cite{chakravarty2023data}, pitch shifting \cite{chakravarty2023data}, and generation of new speech Deepfake samples using various vocoders \cite{wang2022multi}. It is common practice to combine multiple DA techniques, involving both online and offline implementations, to further enhance the model's robustness \cite{chen2020generalization, lee2023experimental}. 

\textit{\textbf{Discussion}} 
According to Table \ref{TABLE:ABLATION}, noise addition and codec augmentation emerge as the most promising DA techniques for SOTA methods due to their ability to simulate real-world conditions effectively. They excel in handling unknown encoding and transmission scenarios, making them particularly suitable for robust Deepfake detection models. However, experiment results indicate that the effectiveness of DA techniques may be feature-dependent or dataset-dependent. For instance, channel simulation, which includes applying codecs, adding additive noise and RIRs, is more effective for models leveraging wav2vec 2.0 features but less so for LFCC-based models. Future research should explore developing adaptive DA strategies, so that the choice and extent of augmentation can be dynamically adjusted based on model performance or dataset characteristics. Additionally, evaluating the effectiveness of DA techniques across multiple datasets and attack scenarios are essential to ensure robustness to real-world conditions.


\subsubsection{Loss function} The selection of loss functions is also crucial to detection performance. The cross-entropy (CE) loss with Softmax is the most widely used loss function for speech Deepfake detection tasks. Researchers also have proposed numerous variants tailored to this specific tasks. In this section, we first introduce CE-related loss functions, followed by an overview of other loss functions proposed in the literature with the formula provided.

\textbf{Cross-entropy (CE) loss with Softmax} This is the most commonly used objective function for speech Deepfake detection. Softmax is applied to generate a probability distribution over the classes. In most speech Deepfake detection tasks, the number of classes is typically limited to two: bona fide and Deepfake. The formula of binary version of CE loss with Softmax is given as:
\begin{equation}   \label{equation:CE} 
     \mathcal{L}_{BCE} = \frac{1}{N}\sum_{i=1}^{N}\log(1 + e^{(\boldsymbol{w}_{1-y_{i}} - \boldsymbol{w}_{y_{i}})^\top\boldsymbol{x}_{i}}),
\end{equation}
where $\boldsymbol{x}_{i} \in \mathbb{R}^{D}$ and $y_{i} \in \{0,1\}$ are the feature embedding and the corresponding label, respectively, and $\boldsymbol{w}_{0}, \boldsymbol{w}_{1} \in \mathbb{R}^{D}$ are the weight vectors for bona fide and Deepfake classes, $N$ is the batch size.

\textbf{Margin-based CE loss functions} To enhance discriminative power, margin-based loss functions have been developed as extensions of CE loss with Softmax, which encourage tighter intra-class clustering and larger inter-class separations.

(1) \textit{Additive Margin (AM)-Softmax} It introduces a margin $m$ in angular space to make both classes' embedding distributions more compact and encourage larger angular distances between feature vectors of different classes:
\begin{equation}   \label{equation:AM} 
     \mathcal{L}_{AM} = -\frac{1}{N}\sum_{i=1}^{N}\log \frac{e^{\alpha(cos(\theta_{y_{i}}+m))}}{e^{\alpha(cos(\theta_{y_{i}}+m))} + e^{\alpha(cos(\theta_{1-y_{i}}))}},
\end{equation}
where $cos\theta_{y_{i}} = \boldsymbol{\hat{w}}_{y_i}^\top\hat{\boldsymbol{x}_{i}}$ is the cosine distance between length normalized vector, $\hat{\boldsymbol{w}}$, $\hat{\boldsymbol{x}}$ are the normalized $\boldsymbol{w}$ and $\boldsymbol{x}$, and $\alpha$ is a scale factor.

(2) \textit{Large Margin Cosine loss (LMCL)} The LMCL also aims to maximize the inter-class variance and minimize the intra-class variance. In LMCL, the margin is added in the cosine space rather than angular space, as follow:
\begin{equation}   \label{equation:LMCL} 
\mathcal{L}_{LMCL} = \frac{1}{N}\sum_{i=1}^{N}\log(1 + e^{\alpha(m-(\boldsymbol{\hat{w}}_{y_i}-\boldsymbol{\hat{w}}_{1-y_i})^\top\hat{\boldsymbol{x}_{i}})}).
\end{equation}
 The margin term in LMCL helps make the model more robust to noisy samples.

(3) \textit{One Class (OC)-Softmax} The theoretical decision boundary of genuine speech and Deepfake speech remains the same angle to the weight vectors across Softmax, AM-Softmax, and LMCL. Zhang et al. \cite{Zhang2021oneclass} point out that employing a uniform compact margin for both genuine and speech Deepfake can lead to overfitting to known attacks. Therefore, OC-Softmax is proposed, which uses two different margins, $m_0$ and $m_1$, to compact the genuine speech as well as simultaneously isolate the Deepfake speech. OC-Softmax is denoted as:
 \begin{equation}   \label{equation:OC} 
     \mathcal{L}_{OC} = \frac{1}{N}\sum_{i=1}^{N}\log(1+e^{\alpha(m_{y_i}-\boldsymbol{\hat{w}}_{0}\boldsymbol{\hat{x}}_{i}(-1)^{y_i})}).
\end{equation}

Ding et al. \cite{ding2023samo} propose speaker attractor multicenter one-class learning (SAMO), which aims to construct multiple clusters for bona fide utterances based on individual speakers, rather than clustering all bona fide speeches into one group. 
Ren et al. \cite{ren2023lightweight} add a dispersion loss specifically tailored for Deepfake samples to the OC-Softmax. This addition aims to make the real speech space more compact than the originals, thereby encouraging known Deepfake samples to cover the entire Deepfake space. Lin et al. \cite{lin2024one} propose threshold One-class Softmax loss, which  tackles class imbalance issues by reducing the optimization weight of Deepfake samples during training.

\textbf{Alternative loss functions} Beyond CE and margin-based approaches, several alternative loss functions have been proposed to address unique challenges in speech Deepfake detection.

(1) \textit{Mean Square Error (MSE)} Since the margin-based loss is sensitive to the hyper-parameter settings, Wang et al. \cite{Wang2021Comparative} suggest utilizing the MSE between the model detection output and the ground truth, as a hyperparameter-free loss function:

\begin{equation}   \label{equation:MSE} 
     \mathcal{L}_{MSE} = \frac{1}{N}\sum_{i=1}^{N}\sum_{k=0}^{C-1}(\boldsymbol{\hat{w}}_{k}^\top\hat{\boldsymbol{x}_{i}} - \mathds{1}(y_{i}=k))^2.
\end{equation}
where $C$ indicates the total number of classes.


(2) \textit{Focal loss} Focal loss is another common objective function, especially when dealing with data imbalance during the training stage. The focal loss can be formulated by Eq. \ref{equation:Focal}.
\begin{equation}   \label{equation:Focal} 
     \mathcal{L}_{Focal} = -(1-p_{t})^{\gamma}\log(p_{t}).
\end{equation}
It adds $(1-p_{t})^{\gamma}$ to the standard CE loss, where $p_{t}$ is the predicted probability of the genuine class and $\gamma$ controls the rate at which the loss for well-classified examples is down-weighted.

(3) \textit{Center loss} Huang et al. \cite{huang2023frequency} propose to integrate Center loss alongside the CE loss to minimize intra-class variability of deep embeddings. This addition is significant because the CE loss primarily focuses on capturing inter-class differences. The definition of Center loss is indicated as follows:
\begin{equation}   \label{equation:CENTER} 
     \mathcal{L}_{Center} = \frac{1}{2}\sum_{i=1}^{N}||\boldsymbol{w}^\top_{y_i}\boldsymbol{x}_{i} - c_{y_i}||_{2}^{2},
\end{equation}
where $c_{y_i}$ is the centroid of the class to which the $i$-th input data belongs.

\textbf{\textit{Discussion}} Among the loss functions explored, OC-Softmax is shown as a highly effectively approach for enhancing detection performance. By introducing distinct margins for bona fide and Deepfake classes, it effectively addresses overfitting to known attacks, while improving generalization to unknown attacks. Furthermore, hybrid loss functions that combine the strengths of multiple approaches offer significant potential for further improving detection capabilities. For example, integrating Focal Loss with OC-Softmax can address data imbalance issues during training, particularly when certain attack types are underrepresented in the dataset. Prioritizing the development of OC-Softmax and hybrid loss functions in future research could lead to more robust and generalizable speech Deepfake detection systems.

\subsubsection{Activation function} Research has explored modifying the activation functions as well. Kang et al. \cite{kang2022attentive} conduct experiments with multiple E2E models and find that the learnable activation functions, such as parametric-ReLU (PReLU) and Attention-ReLU (ARelu), improve the performance greater than non-learnable activation functions. 

\subsection{Non-Machine Learning-Based Detection Models}
While most speech Deepfake detection models are DL-based, several notable approaches leverage non-machine learning (non-ML) techniques. For instance, Blue et al. \cite{blue2022you} address the challenges machine learning methods face in replicating micro-acoustic features, such as the fluid dynamics within the vocal tract. They propose reconstructing anatomical features of the vocal tract to detect Deepfake artifacts by analyzing the amplitudes of specific frequencies during certain phoneme transitions along the speaker's airway. This approach has shown high accuracy in detecting Deepfakes generated by Tacotron2. Similarly, Shirvanian et al. \cite{shirvanian2020voicefox} propose Voicefox, a method based on the observation that Deepfake speech is often transcribed less accurately by speech-to-text systems than natural human speech.

These non-ML methods offer distinct advantages in interpretability compared to DL-based models, which often operate as black boxes. For example, the vocal tract anomalies identified by \cite{blue2022you} are grounded in anatomical knowledge, providing clear, interpretable reasons for detection decisions. However, the generalizability of these non-ML methods remains uncertain, as they have not been rigorously evaluated on diverse datasets. Furthermore, it has not yet been demonstrated that the cues these models rely on are robust enough to generalize across the wide variety of Deepfake generation techniques and scenarios.

\section{Training and robustness advancements in fully Deepfake detection}
Beyond detection accuracy, substantial research has focused on improving model robustness, efficiency, and interpretability. The following section highlights recent advancements in these areas.

\subsection{Training strategies}
Various training techniques have been developed to tackle challenges such as model complexity, data scarcity, and computational efficiency. For instance, Xie et al. \cite{Xie2021SENET} employs a Siamese network with state-of-the-art architectures like LCNN, ResNet-18, and SE-Net to learn more compact and meaningful representations of speech samples without increasing model parameters. The Siamese network is also effective in handling highly imbalanced datasets. In addition to Siamese networks, Low-Rank Adaptation (LoRA) and knowledge distillation (KD) are commonly used in speech Deepfake detection to transfer knowledge into more efficient model structures. 

\textbf{Low-Rank Adaption (LoRA)} LoRA is a transfer learning method that introduces two low-rank adaptive matrices, specifically trained for new tasks or domains, without altering the main model's parameters \cite{zhang2023adaptive}. This allows for efficient incremental learning and prevents catastrophic forgetting. Wang et al. \cite{wang2023low} incorporate LoRA into the multi-head attention module within the wav2vec 2.0 Transformer architecture to generate new deep embeddings, better adapted to data from various domains.

\textbf{Knowledge Distillation (KD)} Large detection models are difficult to deploy on edge devices with limited resources. Knowledge distillation (KD) addresses this by transferring knowledge to a smaller model, facilitating deployment with minimal performance loss \cite{Liao2022distillation, ren2023lightweight}. Xue et al. \cite{xue2023learning} introduce a self-KD approach where the student model learns from the teacher by minimizing both the Kullback-Leibler (KL) divergence of softmax outputs and the MSE loss between feature maps. Lu et al. \cite{lu2024one} propose an offline KD method that freezes the teacher model’s parameters and trains the student model solely on bona fide data to focus on genuine speech features. Ren et al. \cite{ren2023voice} suggest that online KD allows the student model to learn additional knowledge through mutual learning with the teacher model.

\subsection{Interpretability of results} 
Explainable artificial intelligence (XAI) tools have been leveraged to uncover the behaviour of DL algorithms in detecting speech Deepfake. The work in \cite{ge2022explaining} is the first effort to utilize SHapley Additive exPlanations (SHAP) scores to explain detection results on the ASVspoof2019-LA dataset, by testing on speech and non-speech intervals and different subbands. The observation reveals that the classifier has learned to focus on non-speech intervals and highlighted the attention at the low-frequency sub-band at 0.5-0.6kHz. Lim et al. \cite{lim2022detecting} apply both Deep Taylor and layer-wise relevance propagation (LRP) to learn the attribution score of speech formats in spectrograms. Furthermore, the Gradient-weighted Class Activation Mapping (Grad-CAM) is used in \cite{Tak2020CQCC} to identify the significant frequency ranges in the spectrogram. Recently, Li et al. \cite{li24oa_interspeech} propose a class activation map as a visual interpretation of detection results, which provides internal justification for the decision-making process.

\subsection{Defense to adversarial attacks}
The majority of speech Deepfake detection models are vulnerable to adversarial attacks \cite{liu2019adversarial, zhang2020black}, such as the Fast Gradient Sign Method (FGSM) and Projected Gradient Descent (PGD). In response, various defence strategies have been proposed, primarily based on adversarial training, where Deepfake detection models are retrained using the adversarial examples generated by the PGD method \cite{wu2020defense} or the FGSM method \cite{ge2024spoofing}. Nguyen et al. \cite{nguyen2023defense} claim that attackers may focus on the range beyond human perception to maximize the attack effectiveness, therefore, they suggest augmenting training data with frequency band-pass filtering and denoising to defend such attacks. Furthermore, Liao et al. \cite{Liao2022distillation} utilize knowledge distillation as another defence method. This is because the soft targets learned by the student models capture more nuanced information about the decision boundary, enhancing robustness against adversarial attacks.

\subsection{Attacker source tracing}
Tracing and identifying the tools or algorithms behind speech Deepfake attacks is another challenging task in detection technology development. Categorizing attributes related to these attacks not only provides deeper insights into Deepfake system architectures but also offers forensic evidence that strengthens the reliability of detection outcomes. Research on source attribution began with \cite{muller22b_interspeech}, which explored low-level signal features and neural embeddings to capture attacker signatures, attributing Deepfake speech to specific sources. The results show that neural embeddings, especially with RNNs, outperform low-level features such as fundamental frequency, jitter, and shimmer in clustering and classification tasks. Yan et al. \cite{yan2022initial} utilize LFCC and ResNet blocks to detect vocoder fingerprints, as different vocoders leave distinct residual artifacts in specific frequency bands. The ADD2023 challenge \cite{yi2023add} introduces a track on Deepfake algorithm recognition, encouraging research in source tracing, particularly in identifying out-of-distribution (OOD) attackers. For example, Lu et al. \cite{lu2023detecting} propose a KNN-based OOD detection method using distances between utterance embeddings extracted by pre-trained models, while Qin et al. \cite{qin2023speaker} apply a center-based similarity maximization method, comparing speech embeddings to intra-class centroids to identify the source of Deepfake algorithms.

\subsection{Robustness on cross-dataset}
There are two practical scenarios involving multi-dataset. One is multi-dataset co-training, where datasets from different domains are combined as training data. Co-training encourages the training models to learn information from all provided domains.  The other one is cross-dataset evaluation, where models trained on a specific dataset are tested on various OOD datasets. This evaluation method helps assess the robustness and generalizability of models across different data domains. Although these generalization methods are inspired by the challenges of speech Deepfake detection, they can be applied more broadly to multimedia Deepfake detection and other machine learning fields, such as image recognition.

\textbf{Multi-dataset co-training / Continual learning} The experiments indicate that simply combining data from different domains does not guarantee an increase in the generalization of detection models due to domain mismatch. To address this challenge, Shim et al. \cite{Shi2023sharpness} propose a gradient-based method that considers reducing the curvature of neighbourhoods in the loss surface while minimizing the loss function. It effectively reduces the gap between variances of multi-domain datasets. Regularized Adaptive Weight Modification (RAWM) \cite{zhang2023you}  and Radian Weight Modification (RWM) \cite{zhang2024remember} are introduced to address the issue of catastrophic forgetting that occurs when a trained model is fine-tuned with an out-of-domain dataset. RWM accounts for the differences in feature distribution between bona fide speech and various types of speech Deepfakes by applying distinct strategies to each. It incorporates a self-attention mechanism that enables the model to determine the optimal direction for gradient modification based on the current input batch. When the data features significant variations across tasks, such as different types of Deepfake, RWM guides the model to adopt a direction orthogonal to the previous data plane, preserving previously learned knowledge while adapting to new Deepfake attacks. On the other hand, when the data shares similar characteristics, as seen with bona fide speech, the algorithm directs the model to learn a gradient modification direction that aligns with the previous data plane, reducing the interference from gradient modification. Wang et al. \cite{wang2023investigating} use the negative energy-based certainty score \cite{liu2020energy} to evaluate the usefulness of each data in the pool consisting of datasets from various domains. They then employ the active learning technique to select the useful data to be the training data for fine-tuning.

\textbf{Cross-dataset evaluation} Muller et al. \cite{muller22_interspeech} evaluate several SOTA speech Deepfake detection models trained using the ASVspoof2019-LA dataset on the ITW dataset. The results reveal a significant performance decline, with some models showing random guessing behaviour. Zhang et al. \cite{Zhang2021channeleffect} suggest the performance degradation may be due to the channel effect mismatch among different datasets, as evidenced by variations in the average spectra magnitude for each dataset. To address this issue, an additional channel classifier with a Gradient Reversal Layer (GRL) is added in \cite{Zhang2021channeleffect} as a discriminator. 
Besides, Salvi et al. \cite{salvi2023reliability} introduce a sub-network consisting of three layers of FC, dropout, BN, and LeakyReLU, operating in parallel to the detection classifier structure as a reliability estimator. This estimator evaluates each segment of the input speech, ensuring that all input segments contributing to the detection decision are discriminative and robust enough. 
According to Table \ref{TABLE:CROSSDATASET}, the knowledge distillation technique significantly improves the generalizability of student models in cross-dataset evaluations.

\begin{table*}[t]
  \caption{The Cross-dataset Performance of Single-System State-of-the-art Models. All Models are trained or fine-tuned on ASVspoof2019-LA Training and Development Set, and evaluated on ASVspoof2019-LA Evaluation Set and In-The-Wild Dataset.} 
  \label{TABLE:CROSSDATASET}
  \scriptsize
  \setlength{\tabcolsep}{1.2pt} 
  \renewcommand{\arraystretch}{1.2}
  \begin{tabularx}{\textwidth} {
   >{\raggedleft\arraybackslash}m{0.05\textwidth}
   |>{\centering\arraybackslash}m{0.12\textwidth}
  |>{\centering\arraybackslash}m{0.12\textwidth}
  |>{\raggedright\arraybackslash}m{0.2\textwidth}
  |>{\raggedright\arraybackslash}m{0.2\textwidth}
  |>{\centering\arraybackslash}m{0.11\textwidth}
  |>{\centering\arraybackslash}m{0.10\textwidth}
  |>{\centering\arraybackslash}m{0.05\textwidth}
  }  \clineB{1-8}{2.5}
    
    \multicolumn{2}{c|}{\multirow{2}{*}{Publication}} & 
    \multicolumn{1}{c|}{\multirow{2}{*}{\shortstack{\\Data \\augmentation}}} &
    \multicolumn{1}{c|}{\multirow{2}{*}{Feature}} & 
    \multicolumn{1}{c|}{\multirow{2}{*}{Classifier}} & 
    \multicolumn{1}{c|}{\multirow{2}{*}{Loss funcion}} & 
    \multicolumn{2}{c}{EER (\%)$\downarrow$}\\ \cline{7-8}
    \multicolumn{2}{c|}{}&\multicolumn{1}{c|}{}&\multicolumn{1}{c|}{}&\multicolumn{1}{c|}{}&\multicolumn{1}{c|}{}&\multicolumn{1}{c|}{ASVspoof19-LA} &
    \multicolumn{1}{c}{ITW} \\ \cline{1-8}

    \cite{Xie2023generailized}&INTERSPEECH'23&w/o&wav2vec2.0-XLSR&LCNN \textrightarrow Transformer&CE, Triplet, Adversarial&0.63&24.50\\
    \cite{Zhang2024temporal}&SPL'24&SpecAugment&ECAPA-TDNN&CNN\textrightarrow GRU \textrightarrow MLP&AM-Softmax&1.79&29.66\\[2pt]
    \cite{lu2024one}*&ICASSP'24&w/o&CNN \textrightarrow wav2vec2.0&AASIST&CE&0.39&\textbf{7.68}\\[1pt]
    \cite{Wang2024vocoderSSL}*&ICASSP'24&Rawboost&wav2vec2.0-XLSR-Vox&MLP&CE&\textbf{0.13}&12.50\\ \clineB{1-8}{2.5}

\end{tabularx}

          \begin{minipage}{\textwidth}         
          \vspace{2pt}
        \linespread{0.6}\selectfont
        {\scriptsize  The evaluation metric is EER (\%). “$\downarrow$” indicates that a lower score corresponds to better detection performance. The bold values refer to the best performance on the same dataset. "+” indicates multiple techniques processed in parallel, while "\textrightarrow" denotes sequential order. "w/o" means that no data augmentation techniques are applied. 
        
        $^{*}$ \cite{lu2024one} and \cite{Wang2024vocoderSSL} utilize knowledge distillation. The reported evaluation results on both datasets are produced by the student model.}
        \end{minipage}
\end{table*}

\section{Integration of ASV to Deepfake countermeasures}
ASV systems are widely used for biometric authentication, requiring not only speech authentication but also speaker identity verification. However, the rise of advanced Deepfake speech poses a greater threat to ASV systems than traditional human impersonation \cite{9383558}. This underscores the need to integrate ASV functionality with Deepfake detection architectures to enhance security.

The spoofing-aware speaker verification (SASV) challenge \cite{jung22c_interspeech} has been launched to encourage the development of single models which can detect the speech spoken by different speakers and speech Deepfake. Based on the challenge protocols, the VoxCeleb2 and the ASVspoof2019-LA database are used for training. The primary metric for evaluation is the SASV-EER, which treats bona fide speech from the target speaker as positive cases and all others as negative. The secondary metrics are speaker verification (SV)-EER and spoofing (SPF)-EER, which measure the capability of countermeasure and speaker verification respectively.



\subsection{SOTA for SASV models}
The current SASV models can be categorized into four types. 
Each category is discussed in the following.

\textbf{Cascaded systems} The cascaded system is the concatenation of the ASV and CM classifiers, both of which are pre-trained.  Wang et al. \cite{Wang2022DKU} propose a cascaded system beginning with an ASV binary classifier followed by the CM detector. The ASV and CM modules are trained separately. During testing, only the test speech labelled positive by the ASV classifier will be passed to the second CM module. However, they also find that the order of the modules affects the output performance. 

\textbf{Score-level Fusion} Score-level fusion indicates combining the individual scores from the ASV and CM classifiers to generate an ensemble score for the SASV system. Shim et al. \cite{Shim2022SASV} use a score-sum fusion technique to integrate the ASV and CM sub-systems, serving as the baseline for the SASV challenge. Wu et al. \cite{Wu2022modelfusion} suggest that the baseline does not consider the disparity in scale ranges between ASV and CM scores, so they utilize a fusion technique that concatenates scores from multiple sub-CM models and sub-ASV models before passing them to the prediction layer. This approach is refined in \cite{Wu2022levelfusion}, by applying the average pooling to CM embeddings to obtain a single CM score for concatenation with the rest of the sub-ASV scores. Utilizing the score-level fusion with the pooling strategy allows for flexible adjustment of the model’s scale. Zhang et al. \cite{Zhang2022sasvscorefusion} address the inconsistent score distribution of the CM and ASV subsystems by utilizing the L2-normalized inner product for speaker embeddings. Alenin et al. \cite{alenin2022subnetwork} use quality measurements of speech files, such as speech length and mean value of CM system scores, to penalize the ASV and CM scores. 

\begin{table*}[t]
   \begin{threeparttable}[b]
  \caption{The Performance of State-of-the-art SASV Models on the ASVspoof2019-LA evaluation set}
  \label{TABLE:SASV}
  \scriptsize
  \setlength{\tabcolsep}{3pt} 
  \renewcommand{\arraystretch}{1.3}
  \begin{tabularx}{\textwidth} {
   >{\centering\arraybackslash}m{0.04\textwidth}
  |>{\centering\arraybackslash}m{0.13\textwidth}
  |m{0.11\textwidth}
  |>{\raggedright}m{0.25\textwidth}
  |>{\raggedright}m{0.15\textwidth}
  |>{\centering\arraybackslash}m{0.05\textwidth}
  |>{\centering\arraybackslash}m{0.045\textwidth}
  |>{\centering\arraybackslash}m{0.045\textwidth}
  |>{\centering\arraybackslash}m{0.045\textwidth}}   \clineB{1-9}{2.5}
    
    \multicolumn{2}{c|}{Publication} & \multicolumn{1}{c|}{Category} & Algorithms for ASV& Algorithms for CM& SV-EER $\downarrow$ & SPF-EER $\downarrow$ & SASV-EER $\downarrow$ & Acces-sibility\\ \cline{1-9}

    \cite{Wang2022DKU} & INTERSPEECH'22 & Cascaded & SE-ResNet-34, ECAPA-TDNN & AASIST & \textbf{0.90}& \textbf{0.26}&\textbf{0.29} & No\\ \cline{1-9}
    
    
     \cite{Zhang2022sasvproba}&ODYSSEY'22 & Score Fusion&ECAPA-TDNN & AASIST&1.94 &0.80 &1.53 & Yes\tnote{1}\\ 
     
     \cite{Zhang2022sasvscorefusion}& INTERSPEECH'22& Score Fusion& SE-Res2Net-50& AASIST& 0.48& 0.78& 0.63& Yes\tnote{2}\\ 
     \cite{alenin2022subnetwork}& INTERSPEECH'22& Score Fusion& ResNet-48& ResNet-48& \textbf{0.19}& \textbf{0.25}& \textbf{0.22}& No\\
     
     \cline{1-9}
     
     \cite{Zhang2022sasvbackend}& INTERSPEECH'22& Embedding Fusion& ECAPA-TDNN& AASIST& 2.02& 0.50& 0.99& No\\ 
     \cite{Choi2022HYU}& INTERSPEECH'22& Embedding Fusion& Res2Net& AASIST& \textbf{0.28}& \textbf{0.28}& \textbf{0.28}& No\\ 
     \cite{Ge2023optimized}& ICASSP'23& Embedding Fusion& ResNet-34& AASIST& 2.34& 0.80& 1.49& Yes\tnote{3}\\ \cline{1-9}

    \cite{Teng2022SASASV}&INTERSPEECH'22 & Integrated System& \multicolumn{2}{l|}{ECAPA-TDNN, AResNet} & 8.06& \textbf{0.50}& 4.86& No\\ 
    \cite{Mun2023sasvembedding}& INTERSPEECH'23& Integrated System& \multicolumn{2}{l|}{MFA-Conformer} &\textbf{1.83} &0.58 &\textbf{1.19} &Yes\tnote{4} \\  \clineB{1-9}{2.5}
     
\end{tabularx}
     \begin{tablenotes}
       \item [1] \url{https://github.com/yzyouzhang/SASV_PR}
       \item [2] \url{https://github.com/WebPrague/SASV2022_DoubleRoc}
       \item [3] \url{https://github.com/eurecom-asp/sasv-joint-optimisation}
       \item [4] \url{https://github.com/sasv-challenge/ASVSpoof5-SASVBaseline}
     \end{tablenotes}
  \end{threeparttable}
          \begin{minipage}{\textwidth}         
        \linespread{0.6}\selectfont
        {\scriptsize  All models are trained using the ASVspoof2019-LA training and development set, as well as the VoxCeleb2 dataset. “$\downarrow$” indicates that a lower score corresponds to better detection performance for all evaluation metrics. The bold values refer to the best performance of each subcategory}
        \end{minipage}

\end{table*}

\textbf{Feature Embedding-level Fusion} Feature-level fusion involves either the concatenation of the ASV and CM embeddings from their extractors or the extraction of an integrated embedding to represent both ASV and CM information. Another baseline of the SASV challenge mentioned in \cite{Shim2022SASV} is to fuse the embeddings from the ASV and CM module together by concatenation and pass them into a simple DNN-based classifier containing three FC layers. Zhang et al. \cite{Zhang2022sasvbackend} apply circulant matrix transformation to ASV and CM embeddings before stacking them together. Parallel attention and SE attention are added to the back-end classifier to learn the global relationship between these two embeddings. 
Feature-wise Linear Modulation (FiLM) technique is utilized to obtain the spoofing-aware speaker embedding (SASE), conditioning on both speaker and CM embeddings through affine transformation. 

In previously mentioned models, embeddings are separately optimized. However, Gomez-Alanis et al. \cite{gomez2020joint} introduce the concept of joint optimization by proposing DNN-based integration methods to train embeddings from ASV and CM systems jointly. They suggest that this integrated network can be trained as a multi-class classifier. Similarly, Ge et al. \cite{Ge2023optimized} suggest that through joint optimization, the strength of one subsystem may compensate for the weaknesses of the other. 

\textbf{Integrated/ E2E system} Mun et al. \cite{Mun2023sasvembedding} propose a multi-stage training scheme on a multi-scale feature aggregation Conformer (MFA-Conformer) to obtain an SASV embedding directly rather than using separate ASV and CM models. The embedding encoder captures the mutual characteristics of ASV and CM throughout the multiple training and fine-tuning. This design not only contributes to developing a single integrated system for the SASV task but also addresses the lack of Deepfake data, reducing the impact of data imbalance. However, its computational cost has yet to be discussed in the literature. In the E2E model proposed by \cite{Teng2022SASASV}, Deepfake speeches are labelled as TTS and VC separately, based on their generating methods, while bona fide speeches receive different labels based on the speaker's identity. The boundaries between different labels are also distinct. 

\subsection{Performance discussion on SASV models}
Based on the same training and evaluation datasets, the performance of the SOTA methods is presented in Table \ref{TABLE:SASV}.  Currently, the SOTAs still highly rely on the capability of independent ASV and CM subsystems. Surprisingly, according to the EER metrics, simple ensemble mechanisms, such as score-fusion and cascaded systems, achieve a better performance than a single integrated SASV system. However, directly concatenating and stacking the ASV and CM embeddings reduces the ability of speaker verification, as reflected by SV-EER, even though feature-level fusion algorithms have less false alarm rate than cascaded systems. These results suggest that a single system may require a new latent space to effectively represent both speakers and Deepfake information. Therefore, the future direction should focus on joint optimization and the development of an integrated SASV system to improve overall performance.

\section{Partially Deepfake detection}
Partially Deepfake utterances can be created by inserting one or more clips of Deepfake speech generated by speech synthesis and voice conversion technologies into the original bona fide utterance, such as by altering specific words within an expression while the rest of the speech remains authentic. 
The series of ASVspoof Challenges have not yet addressed this type of Deepfake speech. The concept of partially Deepfake speech was initially investigated by \cite{zhang2021initialpartial}. Subsequently, the PF attack was introduced in the ADD 2022 Challenge \cite{yi2022add}, and ADD 2023 \cite{yi2023add}. The challenges are then extended to not only detect manipulated intervals but also to localize the boundaries within the utterance. 
The latest literature, evaluated using the partially Deepfake datasets, are summarized and presented in TABLE \ref{TABLE:PARTIAL}.

While this section focuses on partially Deepfakes and their detection techniques, it's important to distinguish between partially Deepfakes and half-truths. Partially Deepfakes involve the manipulation or alteration of speech using advanced Deepfake generation technologies, producing speech that may sound convincing but is ultimately deceptive. In contrast, half-truths consist of factually correct information that is incomplete or presented in a misleading context, created through selective presentation or omission of details, without altering the actual speech content or using synthetic speech technologies. Although half-truths are a significant element of misinformation, they fall outside the scope of this study, which focuses on the technical detection of speech Deepfakes.


\begin{table*}[t]
  \caption{The Performance of State-of-the-art Partially Deepfake Detection Models on the evaluation set of ADD2022-PF, ADD2023-PF, and PartialSpoof datasets}
  \label{TABLE:PARTIAL}
  \scriptsize
  \setlength{\tabcolsep}{3pt} 
  \renewcommand{\arraystretch}{1.2}
  \begin{tabularx}{\textwidth} {
   >{\centering\arraybackslash}m{0.04\textwidth}
   |>{\centering\arraybackslash}m{0.13\textwidth}
  |>{\centering\arraybackslash}m{0.15\textwidth}
  |>{\raggedright}m{0.15\textwidth}
  |>{\raggedright}m{0.14\textwidth}
  |>{\centering\arraybackslash}m{0.07\textwidth}
  |>{\centering\arraybackslash}m{0.07\textwidth}
  |>{\centering\arraybackslash}m{0.055\textwidth}
  |>{\centering\arraybackslash}m{0.05\textwidth}}  \clineB{1-9}{2.5}
    
    \multicolumn{2}{c|}{\multirow{2}{*}{Publication}} & 
    \multicolumn{1}{c|}{\multirow{2}{*}{Category}} &
    \multicolumn{1}{c|}{\multirow{2}{*}{Feature}} & 
    \multicolumn{1}{c|}{\multirow{2}{*}{Classifier}} & 
    \multicolumn{2}{c|}{PartialSpoof (EER \%)$\downarrow$} & \multirow{2}{*}{\shortstack{\\ADD \\2022-PF \\(EER \%)$\downarrow$}} &\multirow{2}{*}{\shortstack{\\ADD \\2023-PF \\(Acc+F1)$\uparrow$}}\\ \cline{6-7}

    \multicolumn{2}{c|}{}& \multicolumn{1}{c|}{}& \multicolumn{1}{c|}{} & \multicolumn{1}{c|}{}&Utterance-level&Segment-level&&\\ \cline{1-9}
    
    \cite{zhang2021initialpartial}&INTERSPEECH'21&Frame-level&LFCC&LCNN-LSTM&6.19&16.21&-&-\\ 
    \cite{donas2023partial}&DADA'23&Frame-level&wav2vec2.0&LSTM&-&-&-&59.62\\ 
    \cite{liu2023transsionadd}&DADA'23&Frame-level&ResNet&Bi-LSTM, CNN&-&-&-&62.49\\ \cline{1-9}
    
    \cite{zhang2021multitask}&INTERSPEECH'21&Multi-task&LFCC&SE-LCNN, LSTM&5.90&17.55&-&-\\
    \cite{zhang2022partialspoof}&TASLP'22&Multi-task&wav2vec2.0-large&Gated-MLP&\textbf{0.49}&\textbf{9.24}&-&-\\ 
    \cite{li2023multitask}&DADA'23&Multi-task&Mel-Spec&CNN, RNN&-&-&-&62.02\\ 
    \cite{li2023partiallocation}&DADA'23&Multi-task&\multicolumn{2}{c|}{E2E: wav2vec2.0, AASIST}&-&-&-&58.65\\ \cline{1-9}
    \cite{wu2022partiallyspan}&ICASSP'22&Boundary detection&LFCC&SE-Net, Transformer &-&-&11.1&-\\ 
    \cite{lv2022fake}&ICASSP'22&Boundary detection&wav2vec2.0-XLSR &MLP, Transformer &-&-&\textbf{4.80}&-\\ 
    \cite{cai2023location}&DADA'23&Boundary detection&WavLM, ResNet&Transformer, Bi-LST&-&-&-&\textbf{67.13}\\  \clineB{1-9}{2.5}

\end{tabularx}

          \begin{minipage}{\textwidth}         
          \vspace{2pt}
        \linespread{0.6}\selectfont
        {\scriptsize  The evaluation metric for PartialSpoof and ADD2022-PF is EER (\%). The evaluation metric for ADD2023-PF dataset is a weighted sum of utterance-level accuracy and frame-level F1 score. “$\downarrow$” indicates that a lower score corresponds to better detection performance, while "$\uparrow$" means the opposite. "-” indicates that the authors do not report the performance with the corresponding dataset. The bold values refer to the best performance on the same dataset.}
        \end{minipage}

\end{table*}

\subsection{SOTA models for partially Deepfake detection}
The SOTA for detecting partially Deepfake speeches can be categorized into three main categories: frame-level detection, multi-task learning strategies, and boundary detection. 

\textbf{Frame-level} The speech can be divided into small segments and the frame-level detection algorithms assign a genuine or Deepfake label to each segment. Kumar et al. \cite{kumar2021speech} utilize the GMM to calculate the log-likelihood score for each frame and assign frame-level labels. Originally designed to detect full Deepfake speech, this system has shown potential for detecting partially Deepfake speech. Zhang et al. \cite{zhang2021initialpartial} also attempt to calculate the frame-level score by utilizing the PLDA classifier, LCNN backbone with Bi-LSTM for smoothing. The frame-level algorithms often face the challenge of identifying small frames as Deepfake within genuine segments. However, it's reasonable to expect that the manipulated part of the attacked speech should be longer than just a few frames. For instance, Deepfake segmentS should ideally be no shorter than the duration of a phoneme. A swap algorithm is proposed \cite{zhang2022localizing} to switch labels when only one fake frame occurs between real labels in the sequence, and vice versa. This approach obtains longer Deepfake segments. 

\textbf{Multi-task} In addition to the frame-level labels, utterance-level labels are also considered during the training process, as multi-task learning. Zhang et al. \cite{zhang2021multitask} first implement multi-task learning frameworks to construct a binary-branch structure, where frame-level and utterance-level tasks have individual classifiers and loss functions after the jointly embedding encoder. 
Li et al. \cite{li2023partiallocation} integrate two back-end models, ASSIST and wav2vec 2.0. This decision is based on the authors’ observation that ASSIST tends to misidentify Deepfake segments, while wav2vec 2.0 exhibits a bias towards the real class. Therefore, the proposed model combines the detection results from AASIST at the utterance level and wav2vec 2.0 at the frame level. 

\textbf{Boundary detection} The method of boundary detection aims to identify the transition boundaries between genuine and Deepfake segments. Wu et al. \cite{wu2022partiallyspan} add one FC layer as the question-answer (QA) layer to assess each frame's potential as the start or end position of fake clips. Softmax is applied as a QA loss function, and the QA loss and CE loss are summed up for training. This model achieves second place in the PF track of the ADD 2022 Challenge. Utilizing boundary detection can also address post-processing needs in frame-level score-based algorithms. Cai et al. \cite{cai2023location} train two systems with identical architecture for the task of boundary detection and frame-level detection respectively. The scores from the boundary detection system are utilized as a reference to determine the outline frame labels. The proposed model achieves the first rank in the PF track of ADD 2023. 

\subsection{Performance discussion on partially Deepfake detection }
Recent research agrees that manipulations in PF speech primarily occur in the time domain, resulting in more noticeable artifacts. Consequently, incorporating RNN architectures into detectors is preferred for leveraging temporal information. RNNs can also provide global context to CNNs, which are limited by their fixed receptive fields. Additionally, SSL features with fine-tuning are widely used in PF detection due to their effectiveness in capturing temporal characteristics. 

Manipulated segments in PF utterances can be either Deepfake segments or some arbitrary real segments from other speakers, which raises challenges for speaker verification. Some approaches, such as the clustering module in \cite{donas2023partial} and the Variational Autoencoder module in \cite{cai2023location}, serve as supplementary speaker verification systems to identify outliers. However, current testing datasets do not yet include scenarios with real segments from different speakers, making it difficult to fully assess the effectiveness of these methods.

\section{Current Challenge and Future Work}
While the presented works have made significant efforts in improving detection performance, generalization, interpretability, and robustness, there are still certain limitations in current research that need to be addressed. For instance, some model-agnostic transformations on speech signals, such as removing inter-word redundant silence or enhancing the spectral center, can enable speech Deepfake to bypass existing detection models \cite{kassis2023breaking}. This section summarizes the challenges identified in the review and outlines future directions for advancing speech Deepfake detection.

\textbf{Reproducibility of detection models} Only about 10\% of  published papers in the field of speech Deepfake detection provide source code for replication. The lack of details, such as loss functions and hyperparameter configurations, in some publications is especially particularly concerning. Improving reproducibility is critical for advancing research and ensuring the reliability of findings. Transparent detection models allow researchers to build on previous work, helping to identify and correct errors, inconsistencies, or biases, ultimately enhancing the quality of future models.


\textbf{Cross-dataset generalization} Along with the development of dataset diversity, speech Deepfake detection models must improve their ability to generalize across different datasets and domains. Utilizing transfer learning and domain adaptation offers a promising direction for enhancing generalization to unseen Deepfake attacks, speech conditions, and diverse languages, while maintaining strong performance on known conditions.

\textbf{Interpretability of detection results} Current efforts to explore the interpretability of detection results mainly involve applying various XAI tools to trained models. However, balancing detection performance with interpretability during the design of detection architectures remains a challenge. Techniques like attention mechanisms and feature visualization offer insights into the decision-making process of Deepfake detection. Enhancing interpretability ultimately builds trust in speech Deepfake detection systems.

\textbf{Robustness to multilingual with multiple accents} Most current research on speech Deepfake detection focuses on English, while multilingual detection is still in its early stages and faces two key challenges. The first is the lack of diverse speech datasets in languages other than English. To date, only one dataset includes more than two languages. This highlights the need for large-scale multilingual datasets, both for fully and partially Deepfake scenarios, to develop and evaluate robust detection models. Additionally, accent variation, which is currently addressed only in English-based datasets, should be considered for other languages as well. The second challenge is the limited research on evaluating detection models in multilingual environments. Most studies train and test models on individual languages rather than using multilingual data or assessing cross-lingual robustness. A potential research direction is to explore the unique phonetic and acoustic characteristics of different languages and accents, applying transfer learning or domain adaptation techniques to build models that are robust across languages and sensitive to accent and dialect variations.

\textbf{Robustness to adversarial attacks} Most existing defenses against adversarial attacks rely on adversarial training, which involves generating adversarial samples from known attacks to retrain the model, demanding significant computational resources. Future directions could include deploying generator-and-discriminator mechanisms to learn from domain-invariant attacks. Additionally, exploring adversarial attacks across different modalities may reveal potential vulnerabilities and strengthen the resilience of current models. Another approach to enhancing the robustness of detection models is to reduce their reliance on specific parts of the signal, such as silence, low-frequency regions, or continuous background noise, in order to mitigate the risk of adversarial attacks that manipulate the speech signal.

\textbf{Robustness to neural codec-based Deepfake speech} As neural codec methods evolve and produce increasingly convincing synthesized speech, detection models, which are often trained on vocoder-based techniques, may struggle to generalize effectively. While new datasets incorporating neural codec deepfakes are gradually emerging, there remains a significant need for more diverse training datasets that encompass a wide range of acoustic conditions. Additionally, since the generation of codec-based speech is influenced by various preset parameters, there is potential to leverage advanced feature extraction techniques to capture the subtle artifacts associated with these parameter settings more effectively.

\textbf{Streaming / Real-time detection} Limited research has focused on real-time detection, including scenarios like fake phone calls, IoT edge devices, or other low-latency conditions, which require detection models to be computationally efficient. Techniques such as model pruning, distributed computing, and real-time incremental learning can be incorporated into speech Deepfake detection systems.

\textbf{Privacy preservation} Developing and deploying real-time speech Deepfake detection systems on smart devices could raise privacy concerns about access to users' biometric data and model parameters. To mitigate these concerns, privacy-preserving techniques such as secure multiparty computation can be implemented to restrict access from both client and server sides.

\textbf{Integration of multi-modal Deepfake detection} As Deepfake technology continues to evolve, integrating speech Deepfake detection within the broader framework of Deepfake research is emerging as a critical future direction. For instance, advanced techniques used in image and video Deepfake detection could be adapted to identify anomalies in the spectrograms of speech signals. Conversely, techniques developed for speech Deepfake detection could enhance multi-modal Deepfake detection by identifying synchronization issues between speech and video, or uncovering mismatches in emotional tone between speech and visual content.

\section{Conclusion}
In conclusion, this survey reviews advanced speech Deepfake detection algorithms, covering key aspects such as model architectures, training techniques, application scenarios, and available datasets. While many surveys on speech Deepfake detection exist, they often focus solely on model architectures without considering the broader detection pipeline. This survey is the first to offer a comprehensive review of all stages in speech Deepfake detection. We present a detailed comparison and discussion of current advancements in feature engineering and classifier design across diverse applications. Additionally, we assess the effectiveness of various optimization techniques in model training, including data augmentation, activation functions, and loss functions. We also present performance evaluations and open-source information on state-of-the-art models to help establish strong baselines for future experiments. Lastly, we highlight current challenges and suggest promising research directions. We hope this survey serves as a valuable guide and offers insights for future development of in-depth solutions to combat malicious speech Deepfakes.  Furthermore, we encourage positioning speech Deepfake detection within the larger context of Deepfake research, aiming to integrate these techniques into comprehensive, multimodal detection systems capable of addressing the full range of Deepfake threats.



\begin{thebibliography}{224}


\ifx \showCODEN    \undefined \def \showCODEN     #1{\unskip}     \fi
\ifx \showDOI      \undefined \def \showDOI       #1{#1}\fi
\ifx \showISBNx    \undefined \def \showISBNx     #1{\unskip}     \fi
\ifx \showISBNxiii \undefined \def \showISBNxiii  #1{\unskip}     \fi
\ifx \showISSN     \undefined \def \showISSN      #1{\unskip}     \fi
\ifx \showLCCN     \undefined \def \showLCCN      #1{\unskip}     \fi
\ifx \shownote     \undefined \def \shownote      #1{#1}          \fi
\ifx \showarticletitle \undefined \def \showarticletitle #1{#1}   \fi
\ifx \showURL      \undefined \def \showURL       {\relax}        \fi
\providecommand\bibfield[2]{#2}
\providecommand\bibinfo[2]{#2}
\providecommand\natexlab[1]{#1}
\providecommand\showeprint[2][]{arXiv:#2}

\bibitem[Abdzadeh and Veisi(2023)]%
        {abdzadeh2023comparison}
\bibfield{author}{\bibinfo{person}{P Abdzadeh} {and} \bibinfo{person}{Hadi Veisi}.} \bibinfo{year}{2023}\natexlab{}.
\newblock \showarticletitle{A Comparison of CQT Spectrogram with STFT-based Acoustic Features in Deep Learning-based Synthetic Speech Detection}.
\newblock \bibinfo{journal}{\emph{Journal of AI and Data Mining}} \bibinfo{volume}{11}, \bibinfo{number}{1} (\bibinfo{year}{2023}), \bibinfo{pages}{119--129}.
\newblock


\bibitem[AlBadawy et~al\mbox{.}(2019)]%
        {albadawy2019detecting}
\bibfield{author}{\bibinfo{person}{Ehab~A AlBadawy}, \bibinfo{person}{Siwei Lyu}, {and} \bibinfo{person}{Hany Farid}.} \bibinfo{year}{2019}\natexlab{}.
\newblock \showarticletitle{Detecting AI-Synthesized Speech Using Bispectral Analysis.}. In \bibinfo{booktitle}{\emph{CVPR workshops}}. \bibinfo{pages}{104--109}.
\newblock


\bibitem[Alegre et~al\mbox{.}(2013)]%
        {alegre2013one}
\bibfield{author}{\bibinfo{person}{Federico Alegre}, \bibinfo{person}{Asmaa Amehraye}, {and} \bibinfo{person}{Nicholas Evans}.} \bibinfo{year}{2013}\natexlab{}.
\newblock \showarticletitle{A one-class classification approach to generalised speaker verification spoofing countermeasures using local binary patterns}. In \bibinfo{booktitle}{\emph{2013 IEEE Sixth International Conference on Biometrics: Theory, Applications and Systems (BTAS)}}. IEEE, \bibinfo{pages}{1--8}.
\newblock


\bibitem[Alenin et~al\mbox{.}(2022)]%
        {alenin2022subnetwork}
\bibfield{author}{\bibinfo{person}{Alexander Alenin}, \bibinfo{person}{Nikita Torgashov}, \bibinfo{person}{Anton Okhotnikov}, \bibinfo{person}{Rostislav Makarov}, {and} \bibinfo{person}{Ivan Yakovlev}.} \bibinfo{year}{2022}\natexlab{}.
\newblock \showarticletitle{A Subnetwork Approach for Spoofing Aware Speaker Verification.}. In \bibinfo{booktitle}{\emph{INTERSPEECH}}. \bibinfo{pages}{2888--2892}.
\newblock


\bibitem[Almutairi and Elgibreen(2022)]%
        {almutairi2022review}
\bibfield{author}{\bibinfo{person}{Zaynab Almutairi} {and} \bibinfo{person}{Hebah Elgibreen}.} \bibinfo{year}{2022}\natexlab{}.
\newblock \showarticletitle{A review of modern audio deepfake detection methods: Challenges and future directions}.
\newblock \bibinfo{journal}{\emph{Algorithms}} \bibinfo{volume}{15}, \bibinfo{number}{5} (\bibinfo{year}{2022}), \bibinfo{pages}{155}.
\newblock


\bibitem[Alzantot et~al\mbox{.}(2019)]%
        {alzantot19_interspeech}
\bibfield{author}{\bibinfo{person}{Moustafa Alzantot}, \bibinfo{person}{Ziqi Wang}, {and} \bibinfo{person}{Mani~B. Srivastava}.} \bibinfo{year}{2019}\natexlab{}.
\newblock \showarticletitle{{Deep Residual Neural Networks for Audio Spoofing Detection}}. In \bibinfo{booktitle}{\emph{Proc. Interspeech 2019}}. \bibinfo{pages}{1078--1082}.
\newblock


\bibitem[Attorresi et~al\mbox{.}(2022)]%
        {attorresi2022combining}
\bibfield{author}{\bibinfo{person}{Luigi Attorresi}, \bibinfo{person}{Davide Salvi}, \bibinfo{person}{Clara Borrelli}, \bibinfo{person}{Paolo Bestagini}, {and} \bibinfo{person}{Stefano Tubaro}.} \bibinfo{year}{2022}\natexlab{}.
\newblock \showarticletitle{Combining automatic speaker verification and prosody analysis for synthetic speech detection}. In \bibinfo{booktitle}{\emph{International Conference on Pattern Recognition}}. Springer, \bibinfo{pages}{247--263}.
\newblock


\bibitem[Babu et~al\mbox{.}(2022)]%
        {babu22_interspeech}
\bibfield{author}{\bibinfo{person}{Arun Babu}, \bibinfo{person}{Changhan Wang}, \bibinfo{person}{Andros Tjandra}, \bibinfo{person}{Kushal Lakhotia}, \bibinfo{person}{Qiantong Xu}, \bibinfo{person}{Naman Goyal}, \bibinfo{person}{Kritika Singh}, \bibinfo{person}{Patrick {von Platen}}, \bibinfo{person}{Yatharth Saraf}, \bibinfo{person}{Juan Pino}, \bibinfo{person}{Alexei Baevski}, \bibinfo{person}{Alexis Conneau}, {and} \bibinfo{person}{Michael Auli}.} \bibinfo{year}{2022}\natexlab{}.
\newblock \showarticletitle{XLS-R: Self-supervised Cross-lingual Speech Representation Learning at Scale}. In \bibinfo{booktitle}{\emph{Interspeech 2022}}. \bibinfo{pages}{2278--2282}.
\newblock
\showISSN{2958-1796}


\bibitem[Balamurali et~al\mbox{.}(2019)]%
        {balamurali2019toward}
\bibfield{author}{\bibinfo{person}{BT Balamurali}, \bibinfo{person}{Kinwah~Edward Lin}, \bibinfo{person}{Simon Lui}, \bibinfo{person}{Jer-Ming Chen}, {and} \bibinfo{person}{Dorien Herremans}.} \bibinfo{year}{2019}\natexlab{}.
\newblock \showarticletitle{Toward robust audio spoofing detection: A detailed comparison of traditional and learned features}.
\newblock \bibinfo{journal}{\emph{IEEE Access}}  \bibinfo{volume}{7} (\bibinfo{year}{2019}), \bibinfo{pages}{84229--84241}.
\newblock


\bibitem[Bartusiak and Delp(2021)]%
        {bartusiak2021synthesized}
\bibfield{author}{\bibinfo{person}{Emily~R Bartusiak} {and} \bibinfo{person}{Edward~J Delp}.} \bibinfo{year}{2021}\natexlab{}.
\newblock \showarticletitle{Synthesized speech detection using convolutional transformer-based spectrogram analysis}. In \bibinfo{booktitle}{\emph{2021 55th Asilomar Conference on Signals, Systems, and Computers}}. IEEE, \bibinfo{pages}{1426--1430}.
\newblock


\bibitem[Bird and Lotfi(2023)]%
        {bird2023real}
\bibfield{author}{\bibinfo{person}{Jordan~J Bird} {and} \bibinfo{person}{Ahmad Lotfi}.} \bibinfo{year}{2023}\natexlab{}.
\newblock \showarticletitle{Real-time Detection of AI-Generated Speech for DeepFake Voice Conversion}.
\newblock \bibinfo{journal}{\emph{arXiv preprint arXiv:2308.12734}} (\bibinfo{year}{2023}).
\newblock


\bibitem[Blue et~al\mbox{.}(2022)]%
        {blue2022you}
\bibfield{author}{\bibinfo{person}{Logan Blue}, \bibinfo{person}{Kevin Warren}, \bibinfo{person}{Hadi Abdullah}, \bibinfo{person}{Cassidy Gibson}, \bibinfo{person}{Luis Vargas}, \bibinfo{person}{Jessica O'Dell}, \bibinfo{person}{Kevin Butler}, {and} \bibinfo{person}{Patrick Traynor}.} \bibinfo{year}{2022}\natexlab{}.
\newblock \showarticletitle{Who are you (i really wanna know)? detecting audio DeepFakes through vocal tract reconstruction}. In \bibinfo{booktitle}{\emph{31st USENIX Security Symposium (USENIX Security 22)}}. \bibinfo{pages}{2691--2708}.
\newblock


\bibitem[Cai et~al\mbox{.}(2023)]%
        {cai2023location}
\bibfield{author}{\bibinfo{person}{Zexin Cai}, \bibinfo{person}{Weiqing Wang}, \bibinfo{person}{Yikang Wang}, {and} \bibinfo{person}{Ming Li}.} \bibinfo{year}{2023}\natexlab{}.
\newblock \showarticletitle{The DKU-DUKEECE System for the Manipulation Region Location Task of ADD 2023}.
\newblock \bibinfo{journal}{\emph{Proceedings of IJCAI 2023 Workshop on Deepfake Audio Detection and Analysis (DADA 2023)}} (\bibinfo{year}{2023}).
\newblock


\bibitem[Castan et~al\mbox{.}(2022)]%
        {Castan2022speaker}
\bibfield{author}{\bibinfo{person}{Diego Castan}, \bibinfo{person}{Md~Hafizur Rahman}, \bibinfo{person}{Sarah Bakst}, \bibinfo{person}{Chris Cobo-Kroenke}, \bibinfo{person}{Mitchell McLaren}, \bibinfo{person}{Martin Graciarena}, {and} \bibinfo{person}{Aaron Lawson}.} \bibinfo{year}{2022}\natexlab{}.
\newblock \showarticletitle{Speaker-targeted synthetic speech detection}.
\newblock \bibinfo{journal}{\emph{The Speaker and Language Recognition Workshop (Odyssey 2022)}} (\bibinfo{date}{Jun} \bibinfo{year}{2022}).
\newblock


\bibitem[Chakravarty and Dua(2023)]%
        {chakravarty2023data}
\bibfield{author}{\bibinfo{person}{Nidhi Chakravarty} {and} \bibinfo{person}{Mohit Dua}.} \bibinfo{year}{2023}\natexlab{}.
\newblock \showarticletitle{Data augmentation and hybrid feature amalgamation to detect audio deep fake attacks}.
\newblock \bibinfo{journal}{\emph{Physica Scripta}} \bibinfo{volume}{98}, \bibinfo{number}{9} (\bibinfo{year}{2023}), \bibinfo{pages}{096001}.
\newblock


\bibitem[Chang et~al\mbox{.}(2019)]%
        {Chang2019TransferRepresentation}
\bibfield{author}{\bibinfo{person}{Su-Yu Chang}, \bibinfo{person}{Kai-Cheng Wu}, {and} \bibinfo{person}{Chia-Ping Chen}.} \bibinfo{year}{2019}\natexlab{}.
\newblock \showarticletitle{Transfer-representation learning for detecting spoofing attacks with converted and synthesized speech in automatic speaker verification system}.
\newblock \bibinfo{journal}{\emph{Interspeech 2019}} (\bibinfo{date}{Sep} \bibinfo{year}{2019}).
\newblock


\bibitem[Chen et~al\mbox{.}(2023b)]%
        {chen2023twice}
\bibfield{author}{\bibinfo{person}{Chen Chen}, \bibinfo{person}{Yaozu Song}, \bibinfo{person}{Bohan Dai}, {and} \bibinfo{person}{Deyun Chen}.} \bibinfo{year}{2023}\natexlab{b}.
\newblock \showarticletitle{Twice attention networks for synthetic speech detection}.
\newblock \bibinfo{journal}{\emph{Neurocomputing}}  \bibinfo{volume}{559} (\bibinfo{year}{2023}), \bibinfo{pages}{126799}.
\newblock


\bibitem[Chen et~al\mbox{.}(2023a)]%
        {chen2023graph}
\bibfield{author}{\bibinfo{person}{Feng Chen}, \bibinfo{person}{Shiwen Deng}, \bibinfo{person}{Tieran Zheng}, \bibinfo{person}{Yongjun He}, {and} \bibinfo{person}{Jiqing Han}.} \bibinfo{year}{2023}\natexlab{a}.
\newblock \showarticletitle{Graph-based spectro-temporal dependency modeling for anti-spoofing}. In \bibinfo{booktitle}{\emph{ICASSP 2023-2023 IEEE International Conference on Acoustics, Speech and Signal Processing (ICASSP)}}. IEEE, \bibinfo{pages}{1--5}.
\newblock


\bibitem[Chen et~al\mbox{.}(2020)]%
        {chen2020generalization}
\bibfield{author}{\bibinfo{person}{Tianxiang Chen}, \bibinfo{person}{Avrosh Kumar}, \bibinfo{person}{Parav Nagarsheth}, \bibinfo{person}{Ganesh Sivaraman}, {and} \bibinfo{person}{Elie Khoury}.} \bibinfo{year}{2020}\natexlab{}.
\newblock \showarticletitle{Generalization of Audio Deepfake Detection.}. In \bibinfo{booktitle}{\emph{Odyssey}}. \bibinfo{pages}{132--137}.
\newblock


\bibitem[Chen et~al\mbox{.}(2021)]%
        {Chen2021urchannel}
\bibfield{author}{\bibinfo{person}{Xinhui Chen}, \bibinfo{person}{You Zhang}, \bibinfo{person}{Ge Zhu}, {and} \bibinfo{person}{Zhiyao Duan}.} \bibinfo{year}{2021}\natexlab{}.
\newblock \showarticletitle{Ur channel-robust synthetic speech detection system for ASVSPOOF 2021}.
\newblock \bibinfo{journal}{\emph{2021 Edition of the Automatic Speaker Verification and Spoofing Countermeasures Challenge}} (\bibinfo{date}{Sep} \bibinfo{year}{2021}).
\newblock


\bibitem[Chettri et~al\mbox{.}(2020)]%
        {chettri2020dataset}
\bibfield{author}{\bibinfo{person}{Bhusan Chettri}, \bibinfo{person}{Emmanouil Benetos}, {and} \bibinfo{person}{Bob~LT Sturm}.} \bibinfo{year}{2020}\natexlab{}.
\newblock \showarticletitle{Dataset artefacts in anti-spoofing systems: a case study on the ASVspoof 2017 benchmark}.
\newblock \bibinfo{journal}{\emph{IEEE/ACM Transactions on Audio, Speech, and Language Processing}}  \bibinfo{volume}{28} (\bibinfo{year}{2020}), \bibinfo{pages}{3018--3028}.
\newblock


\bibitem[{Cheuk} et~al\mbox{.}(2020)]%
        {cheuk2020nnaudio}
\bibfield{author}{\bibinfo{person}{K.~W. {Cheuk}}, \bibinfo{person}{H. {Anderson}}, \bibinfo{person}{K. {Agres}}, {and} \bibinfo{person}{D. {Herremans}}.} \bibinfo{year}{2020}\natexlab{}.
\newblock \showarticletitle{nnAudio: An on-the-Fly GPU Audio to Spectrogram Conversion Toolbox Using 1D Convolutional Neural Networks}.
\newblock \bibinfo{journal}{\emph{IEEE Access}}  \bibinfo{volume}{8} (\bibinfo{year}{2020}), \bibinfo{pages}{161981--162003}.
\newblock


\bibitem[Chintha et~al\mbox{.}(2020)]%
        {chintha2020recurrent}
\bibfield{author}{\bibinfo{person}{Akash Chintha}, \bibinfo{person}{Bao Thai}, \bibinfo{person}{Saniat~Javid Sohrawardi}, \bibinfo{person}{Kartavya Bhatt}, \bibinfo{person}{Andrea Hickerson}, \bibinfo{person}{Matthew Wright}, {and} \bibinfo{person}{Raymond Ptucha}.} \bibinfo{year}{2020}\natexlab{}.
\newblock \showarticletitle{Recurrent convolutional structures for audio spoof and video deepfake detection}.
\newblock \bibinfo{journal}{\emph{IEEE Journal of Selected Topics in Signal Processing}} \bibinfo{volume}{14}, \bibinfo{number}{5} (\bibinfo{year}{2020}), \bibinfo{pages}{1024--1037}.
\newblock


\bibitem[Choi et~al\mbox{.}(2022b)]%
        {Choi2022HYU}
\bibfield{author}{\bibinfo{person}{Jeong-Hwan Choi}, \bibinfo{person}{Joon-Young Yang}, \bibinfo{person}{Ye-Rin Jeoung}, {and} \bibinfo{person}{Joon-Hyuk Chang}.} \bibinfo{year}{2022}\natexlab{b}.
\newblock \showarticletitle{Hyu submission for the SASV challenge 2022: Reforming speaker embeddings with spoofing-aware conditioning}.
\newblock \bibinfo{journal}{\emph{Interspeech 2022}} (\bibinfo{date}{Sep} \bibinfo{year}{2022}).
\newblock


\bibitem[Choi et~al\mbox{.}(2022a)]%
        {Choi2022overlapped}
\bibfield{author}{\bibinfo{person}{Sunmook Choi}, \bibinfo{person}{Il-Youp Kwak}, {and} \bibinfo{person}{Seungsang Oh}.} \bibinfo{year}{2022}\natexlab{a}.
\newblock \showarticletitle{Overlapped Frequency-distributed network: Frequency-Aware Voice spoofing countermeasure}.
\newblock \bibinfo{journal}{\emph{Interspeech 2022}} (\bibinfo{date}{Sep} \bibinfo{year}{2022}).
\newblock


\bibitem[Cohen et~al\mbox{.}(2022)]%
        {cohen2022study}
\bibfield{author}{\bibinfo{person}{Ariel Cohen}, \bibinfo{person}{Inbal Rimon}, \bibinfo{person}{Eran Aflalo}, {and} \bibinfo{person}{Haim~H Permuter}.} \bibinfo{year}{2022}\natexlab{}.
\newblock \showarticletitle{A study on data augmentation in voice anti-spoofing}.
\newblock \bibinfo{journal}{\emph{Speech Communication}}  \bibinfo{volume}{141} (\bibinfo{year}{2022}), \bibinfo{pages}{56--67}.
\newblock


\bibitem[Conti et~al\mbox{.}(2022)]%
        {conti2022deepfake}
\bibfield{author}{\bibinfo{person}{Emanuele Conti}, \bibinfo{person}{Davide Salvi}, \bibinfo{person}{Clara Borrelli}, \bibinfo{person}{Brian Hosler}, \bibinfo{person}{Paolo Bestagini}, \bibinfo{person}{Fabio Antonacci}, \bibinfo{person}{Augusto Sarti}, \bibinfo{person}{Matthew~C Stamm}, {and} \bibinfo{person}{Stefano Tubaro}.} \bibinfo{year}{2022}\natexlab{}.
\newblock \showarticletitle{Deepfake speech detection through emotion recognition: a semantic approach}. In \bibinfo{booktitle}{\emph{ICASSP 2022-2022 IEEE International Conference on Acoustics, Speech and Signal Processing (ICASSP)}}. IEEE, \bibinfo{pages}{8962--8966}.
\newblock


\bibitem[Cuccovillo et~al\mbox{.}(2022)]%
        {cuccovillo2022open}
\bibfield{author}{\bibinfo{person}{Luca Cuccovillo}, \bibinfo{person}{Christoforos Papastergiopoulos}, \bibinfo{person}{Anastasios Vafeiadis}, \bibinfo{person}{Artem Yaroshchuk}, \bibinfo{person}{Patrick Aichroth}, \bibinfo{person}{Konstantinos Votis}, {and} \bibinfo{person}{Dimitrios Tzovaras}.} \bibinfo{year}{2022}\natexlab{}.
\newblock \showarticletitle{Open challenges in synthetic speech detection}. In \bibinfo{booktitle}{\emph{2022 IEEE International Workshop on Information Forensics and Security (WIFS)}}. IEEE, \bibinfo{pages}{1--6}.
\newblock


\bibitem[Cáceres et~al\mbox{.}(2021)]%
        {Caceres2021Vox}
\bibfield{author}{\bibinfo{person}{Joaquín Cáceres}, \bibinfo{person}{Roberto Font}, \bibinfo{person}{Teresa Grau}, {and} \bibinfo{person}{Javier Molina}.} \bibinfo{year}{2021}\natexlab{}.
\newblock \showarticletitle{The biometric Vox system for the ASVSPOOF 2021 challenge}.
\newblock \bibinfo{journal}{\emph{2021 Edition of the Automatic Speaker Verification and Spoofing Countermeasures Challenge}} (\bibinfo{date}{Sep} \bibinfo{year}{2021}).
\newblock


\bibitem[Davis and Mermelstein(1980)]%
        {davis1980comparison}
\bibfield{author}{\bibinfo{person}{Steven Davis} {and} \bibinfo{person}{Paul Mermelstein}.} \bibinfo{year}{1980}\natexlab{}.
\newblock \showarticletitle{Comparison of parametric representations for monosyllabic word recognition in continuously spoken sentences}.
\newblock \bibinfo{journal}{\emph{IEEE transactions on acoustics, speech, and signal processing}} \bibinfo{volume}{28}, \bibinfo{number}{4} (\bibinfo{year}{1980}), \bibinfo{pages}{357--366}.
\newblock


\bibitem[Deng et~al\mbox{.}(2022)]%
        {deng2022detection}
\bibfield{author}{\bibinfo{person}{Jiacheng Deng}, \bibinfo{person}{Terui Mao}, \bibinfo{person}{Diqun Yan}, \bibinfo{person}{Li Dong}, {and} \bibinfo{person}{Mingyu Dong}.} \bibinfo{year}{2022}\natexlab{}.
\newblock \showarticletitle{Detection of synthetic speech based on spectrum defects}. In \bibinfo{booktitle}{\emph{Proceedings of the 1st International Workshop on Deepfake Detection for Audio Multimedia}}. \bibinfo{pages}{3--8}.
\newblock


\bibitem[Dhamyal et~al\mbox{.}(2021)]%
        {dhamyal2021fake}
\bibfield{author}{\bibinfo{person}{Hira Dhamyal}, \bibinfo{person}{Ayesha Ali}, \bibinfo{person}{Ihsan~Ayyub Qazi}, {and} \bibinfo{person}{Agha~Ali Raza}.} \bibinfo{year}{2021}\natexlab{}.
\newblock \showarticletitle{Fake Audio Detection in Resource-Constrained Settings Using Microfeatures.}. In \bibinfo{booktitle}{\emph{Interspeech}}. \bibinfo{pages}{4149--4153}.
\newblock


\bibitem[Ding et~al\mbox{.}(2023)]%
        {ding2023samo}
\bibfield{author}{\bibinfo{person}{Siwen Ding}, \bibinfo{person}{You Zhang}, {and} \bibinfo{person}{Zhiyao Duan}.} \bibinfo{year}{2023}\natexlab{}.
\newblock \showarticletitle{Samo: Speaker attractor multi-center one-class learning for voice anti-spoofing}. In \bibinfo{booktitle}{\emph{ICASSP 2023-2023 IEEE International Conference on Acoustics, Speech and Signal Processing (ICASSP)}}. IEEE, \bibinfo{pages}{1--5}.
\newblock


\bibitem[Dixit et~al\mbox{.}(2023)]%
        {dixit2023review}
\bibfield{author}{\bibinfo{person}{Abhishek Dixit}, \bibinfo{person}{Nirmal Kaur}, {and} \bibinfo{person}{Staffy Kingra}.} \bibinfo{year}{2023}\natexlab{}.
\newblock \showarticletitle{Review of audio deepfake detection techniques: Issues and prospects}.
\newblock \bibinfo{journal}{\emph{Expert Systems}} \bibinfo{volume}{40}, \bibinfo{number}{8} (\bibinfo{year}{2023}), \bibinfo{pages}{e13322}.
\newblock


\bibitem[Doan et~al\mbox{.}(2023)]%
        {doan2023bts}
\bibfield{author}{\bibinfo{person}{Thien-Phuc Doan}, \bibinfo{person}{Long Nguyen-Vu}, \bibinfo{person}{Souhwan Jung}, {and} \bibinfo{person}{Kihun Hong}.} \bibinfo{year}{2023}\natexlab{}.
\newblock \showarticletitle{Bts-e: Audio deepfake detection using breathing-talking-silence encoder}. In \bibinfo{booktitle}{\emph{ICASSP 2023-2023 IEEE International Conference on Acoustics, Speech and Signal Processing (ICASSP)}}. IEEE, \bibinfo{pages}{1--5}.
\newblock


\bibitem[Dong et~al\mbox{.}(2023)]%
        {dong2023multi}
\bibfield{author}{\bibinfo{person}{Shunbo Dong}, \bibinfo{person}{Jun Xue}, \bibinfo{person}{Cunhang Fan}, \bibinfo{person}{Kang Zhu}, \bibinfo{person}{Yujie Chen}, {and} \bibinfo{person}{Zhao Lv}.} \bibinfo{year}{2023}\natexlab{}.
\newblock \showarticletitle{Multi-perspective Information Fusion Res2Net with RandomSpecmix for Fake Speech Detection}.
\newblock \bibinfo{journal}{\emph{Proceedings of IJCAI 2023 Workshop on Deepfake Audio Detection and Analysis (DADA 2023)}} (\bibinfo{year}{2023}).
\newblock


\bibitem[Fan et~al\mbox{.}(2023)]%
        {Fan2023F0}
\bibfield{author}{\bibinfo{person}{Cunhang Fan}, \bibinfo{person}{Jun Xue}, \bibinfo{person}{Shunbo Dong}, \bibinfo{person}{Mingming Ding}, \bibinfo{person}{Jiangyan Yi}, \bibinfo{person}{Jinpeng Li}, {and} \bibinfo{person}{Zhao Lv}.} \bibinfo{year}{2023}\natexlab{}.
\newblock \showarticletitle{Subband fusion of complex spectrogram for fake speech detection}.
\newblock \bibinfo{journal}{\emph{Speech Communication}}  \bibinfo{volume}{155} (\bibinfo{date}{Nov} \bibinfo{year}{2023}), \bibinfo{pages}{102988}.
\newblock


\bibitem[Fang et~al\mbox{.}(2021)]%
        {fang2021voice}
\bibfield{author}{\bibinfo{person}{Xin Fang}, \bibinfo{person}{Haijia Du}, \bibinfo{person}{Tian Gao}, \bibinfo{person}{Liang Zou}, {and} \bibinfo{person}{Zhenhua Ling}.} \bibinfo{year}{2021}\natexlab{}.
\newblock \showarticletitle{Voice spoofing detection with raw waveform based on Dual Path Res2net}. In \bibinfo{booktitle}{\emph{5th International Conference on Crowd Science and Engineering}}. \bibinfo{pages}{160--165}.
\newblock


\bibitem[Fathan et~al\mbox{.}(2022)]%
        {fathan2022mel}
\bibfield{author}{\bibinfo{person}{Abderrahim Fathan}, \bibinfo{person}{Jahangir Alam}, {and} \bibinfo{person}{Woo~Hyun Kang}.} \bibinfo{year}{2022}\natexlab{}.
\newblock \showarticletitle{Mel-spectrogram image-based end-to-end audio deepfake detection under channel-mismatched conditions}. In \bibinfo{booktitle}{\emph{2022 IEEE International Conference on Multimedia and Expo (ICME)}}. IEEE, \bibinfo{pages}{1--6}.
\newblock


\bibitem[Frank and Sch{\"o}nherr(2021)]%
        {frank2021wavefake}
\bibfield{author}{\bibinfo{person}{Joel Frank} {and} \bibinfo{person}{Lea Sch{\"o}nherr}.} \bibinfo{year}{2021}\natexlab{}.
\newblock \showarticletitle{WaveFake: A Data Set to Facilitate Audio Deepfake Detection}. In \bibinfo{booktitle}{\emph{Thirty-fifth Conference on Neural Information Processing Systems Datasets and Benchmarks Track}}.
\newblock


\bibitem[Fu et~al\mbox{.}(2022)]%
        {fu2022fastaudio}
\bibfield{author}{\bibinfo{person}{Quchen Fu}, \bibinfo{person}{Zhongwei Teng}, \bibinfo{person}{Jules White}, \bibinfo{person}{Maria~E Powell}, {and} \bibinfo{person}{Douglas~C Schmidt}.} \bibinfo{year}{2022}\natexlab{}.
\newblock \showarticletitle{Fastaudio: A learnable audio front-end for spoof speech detection}. In \bibinfo{booktitle}{\emph{ICASSP 2022-2022 IEEE International Conference on Acoustics, Speech and Signal Processing (ICASSP)}}. IEEE, \bibinfo{pages}{3693--3697}.
\newblock


\bibitem[Gao et~al\mbox{.}(2021a)]%
        {9383558}
\bibfield{author}{\bibinfo{person}{Yang Gao}, \bibinfo{person}{Jiachen Lian}, \bibinfo{person}{Bhiksha Raj}, {and} \bibinfo{person}{Rita Singh}.} \bibinfo{year}{2021}\natexlab{a}.
\newblock \showarticletitle{Detection and Evaluation of Human and Machine Generated Speech in Spoofing Attacks on Automatic Speaker Verification Systems}. In \bibinfo{booktitle}{\emph{2021 IEEE Spoken Language Technology Workshop (SLT)}}. \bibinfo{pages}{544--551}.
\newblock


\bibitem[Gao et~al\mbox{.}(2021b)]%
        {Gao2021}
\bibfield{author}{\bibinfo{person}{Yang Gao}, \bibinfo{person}{Tyler Vuong}, \bibinfo{person}{Mahsa Elyasi}, \bibinfo{person}{Gaurav Bharaj}, {and} \bibinfo{person}{Rita Singh}.} \bibinfo{year}{2021}\natexlab{b}.
\newblock \showarticletitle{Generalized spoofing detection inspired from audio generation artifacts}.
\newblock \bibinfo{journal}{\emph{Interspeech 2021}} (\bibinfo{date}{Aug} \bibinfo{year}{2021}).
\newblock


\bibitem[Gasenzer and Wolter(2023)]%
        {gasenzer2023towards}
\bibfield{author}{\bibinfo{person}{Konstantin Gasenzer} {and} \bibinfo{person}{Moritz Wolter}.} \bibinfo{year}{2023}\natexlab{}.
\newblock \showarticletitle{Towards generalizing deep-audio fake detection networks}.
\newblock \bibinfo{journal}{\emph{arXiv preprint arXiv:2305.13033}} (\bibinfo{year}{2023}).
\newblock


\bibitem[Ge et~al\mbox{.}(2021)]%
        {Ge2021pcdart}
\bibfield{author}{\bibinfo{person}{Wanying Ge}, \bibinfo{person}{Michele Panariello}, \bibinfo{person}{Jose Patino}, \bibinfo{person}{Massimiliano Todisco}, {and} \bibinfo{person}{Nicholas Evans}.} \bibinfo{year}{2021}\natexlab{}.
\newblock \showarticletitle{Partially-connected differentiable architecture search for Deepfake and spoofing detection}.
\newblock \bibinfo{journal}{\emph{Interspeech 2021}} (\bibinfo{date}{Aug} \bibinfo{year}{2021}).
\newblock


\bibitem[Ge et~al\mbox{.}(2022)]%
        {ge2022explaining}
\bibfield{author}{\bibinfo{person}{Wanying Ge}, \bibinfo{person}{Jose Patino}, \bibinfo{person}{Massimiliano Todisco}, {and} \bibinfo{person}{Nicholas Evans}.} \bibinfo{year}{2022}\natexlab{}.
\newblock \showarticletitle{Explaining deep learning models for spoofing and deepfake detection with SHapley Additive exPlanations}. In \bibinfo{booktitle}{\emph{ICASSP 2022-2022 IEEE International Conference on Acoustics, Speech and Signal Processing (ICASSP)}}. IEEE, \bibinfo{pages}{6387--6391}.
\newblock


\bibitem[Ge et~al\mbox{.}(2023)]%
        {Ge2023optimized}
\bibfield{author}{\bibinfo{person}{Wanying Ge}, \bibinfo{person}{Hemlata Tak}, \bibinfo{person}{Massimiliano Todisco}, {and} \bibinfo{person}{Nicholas Evans}.} \bibinfo{year}{2023}\natexlab{}.
\newblock \showarticletitle{Can spoofing countermeasure and speaker verification systems be jointly optimised?}
\newblock \bibinfo{journal}{\emph{ICASSP 2023 - 2023 IEEE International Conference on Acoustics, Speech and Signal Processing (ICASSP)}} (\bibinfo{date}{Jun} \bibinfo{year}{2023}).
\newblock


\bibitem[Ge et~al\mbox{.}(2024)]%
        {ge2024spoofing}
\bibfield{author}{\bibinfo{person}{Wanying Ge}, \bibinfo{person}{Xin Wang}, \bibinfo{person}{Junichi Yamagishi}, \bibinfo{person}{Massimiliano Todisco}, {and} \bibinfo{person}{Nicholas Evans}.} \bibinfo{year}{2024}\natexlab{}.
\newblock \showarticletitle{Spoofing attack augmentation: can differently-trained attack models improve generalisation?}. In \bibinfo{booktitle}{\emph{ICASSP 2024-2024 IEEE International Conference on Acoustics, Speech and Signal Processing (ICASSP)}}. IEEE, \bibinfo{pages}{12531--12535}.
\newblock


\bibitem[Gomez-Alanis et~al\mbox{.}(2020)]%
        {gomez2020joint}
\bibfield{author}{\bibinfo{person}{Alejandro Gomez-Alanis}, \bibinfo{person}{Jose~A Gonzalez-Lopez}, \bibinfo{person}{S~Pavankumar Dubagunta}, \bibinfo{person}{Antonio~M Peinado}, {and} \bibinfo{person}{Mathew~Magimai Doss}.} \bibinfo{year}{2020}\natexlab{}.
\newblock \showarticletitle{On joint optimization of automatic speaker verification and anti-spoofing in the embedding space}.
\newblock \bibinfo{journal}{\emph{IEEE Transactions on Information Forensics and Security}}  \bibinfo{volume}{16} (\bibinfo{year}{2020}), \bibinfo{pages}{1579--1593}.
\newblock


\bibitem[Gonzalez-Soler et~al\mbox{.}(2020)]%
        {gonzalez2020texture}
\bibfield{author}{\bibinfo{person}{Lazaro~J Gonzalez-Soler}, \bibinfo{person}{Jose Patino}, \bibinfo{person}{Marta Gomez-Barrero}, \bibinfo{person}{Massimiliano Todisco}, \bibinfo{person}{Christoph Busch}, {and} \bibinfo{person}{Nicholas Evans}.} \bibinfo{year}{2020}\natexlab{}.
\newblock \showarticletitle{Texture-based presentation attack detection for automatic speaker verification}. In \bibinfo{booktitle}{\emph{2020 IEEE International Workshop on Information Forensics and Security (WIFS)}}. IEEE, \bibinfo{pages}{1--6}.
\newblock


\bibitem[Gupta et~al\mbox{.}(2022)]%
        {gupta2022significance}
\bibfield{author}{\bibinfo{person}{Priyanka Gupta}, \bibinfo{person}{Piyushkumar~K Chodingala}, {and} \bibinfo{person}{Hemant~A Patil}.} \bibinfo{year}{2022}\natexlab{}.
\newblock \showarticletitle{Significance of Quadrature and In-Phase Components for Synthetic Spoofed Speech Detection}. In \bibinfo{booktitle}{\emph{2022 Asia-Pacific Signal and Information Processing Association Annual Summit and Conference (APSIPA ASC)}}. IEEE, \bibinfo{pages}{1252--1258}.
\newblock


\bibitem[Hansen and Wang(2022)]%
        {wang2022singleattention}
\bibfield{author}{\bibinfo{person}{John~H.L. Hansen} {and} \bibinfo{person}{Zhenyu Wang}.} \bibinfo{year}{2022}\natexlab{}.
\newblock \showarticletitle{Audio anti-spoofing using simple attention module and joint optimization based on additive angular margin loss and meta-learning}.
\newblock \bibinfo{journal}{\emph{Interspeech 2022}} (\bibinfo{date}{Sep} \bibinfo{year}{2022}).
\newblock


\bibitem[Hasan et~al\mbox{.}(2013)]%
        {hasan2013crss}
\bibfield{author}{\bibinfo{person}{Taufiq Hasan}, \bibinfo{person}{Seyed~Omid Sadjadi}, \bibinfo{person}{Gang Liu}, \bibinfo{person}{Navid Shokouhi}, \bibinfo{person}{Hynek Bo{\v{r}}il}, {and} \bibinfo{person}{John~HL Hansen}.} \bibinfo{year}{2013}\natexlab{}.
\newblock \showarticletitle{CRSS systems for 2012 NIST speaker recognition evaluation}. In \bibinfo{booktitle}{\emph{2013 IEEE International Conference on Acoustics, Speech and Signal Processing}}. IEEE, \bibinfo{pages}{6783--6787}.
\newblock


\bibitem[Hua et~al\mbox{.}(2021)]%
        {hua2021towards}
\bibfield{author}{\bibinfo{person}{Guang Hua}, \bibinfo{person}{Andrew Beng~Jin Teoh}, {and} \bibinfo{person}{Haijian Zhang}.} \bibinfo{year}{2021}\natexlab{}.
\newblock \showarticletitle{Towards end-to-end synthetic speech detection}.
\newblock \bibinfo{journal}{\emph{IEEE Signal Processing Letters}}  \bibinfo{volume}{28} (\bibinfo{year}{2021}), \bibinfo{pages}{1265--1269}.
\newblock


\bibitem[Huang et~al\mbox{.}(2023)]%
        {huang2023frequency}
\bibfield{author}{\bibinfo{person}{Bingyuan Huang}, \bibinfo{person}{Sanshuai Cui}, \bibinfo{person}{Jiwu Huang}, {and} \bibinfo{person}{Xiangui Kang}.} \bibinfo{year}{2023}\natexlab{}.
\newblock \showarticletitle{Discriminative frequency information learning for end-to-end speech anti-spoofing}.
\newblock \bibinfo{journal}{\emph{IEEE Signal Processing Letters}}  \bibinfo{volume}{30} (\bibinfo{year}{2023}), \bibinfo{pages}{185--189}.
\newblock


\bibitem[Ibrar et~al\mbox{.}(2023)]%
        {ibrar2023voice}
\bibfield{author}{\bibinfo{person}{Sundas Ibrar}, \bibinfo{person}{Ali Javed}, {and} \bibinfo{person}{Hafsa Ilyas}.} \bibinfo{year}{2023}\natexlab{}.
\newblock \showarticletitle{Voice presentation attacks detection using acoustic MLTP Features and BiLSTM}. In \bibinfo{booktitle}{\emph{2023 International Conference on Communication, Computing and Digital Systems (C-CODE)}}. IEEE, \bibinfo{pages}{1--5}.
\newblock


\bibitem[Ji et~al\mbox{.}(2017)]%
        {ji17_interspeech}
\bibfield{author}{\bibinfo{person}{Zhe Ji}, \bibinfo{person}{Zhi-Yi Li}, \bibinfo{person}{Peng Li}, \bibinfo{person}{Maobo An}, \bibinfo{person}{Shengxiang Gao}, \bibinfo{person}{Dan Wu}, {and} \bibinfo{person}{Faru Zhao}.} \bibinfo{year}{2017}\natexlab{}.
\newblock \showarticletitle{{Ensemble Learning for Countermeasure of Audio Replay Spoofing Attack in ASVspoof2017}}. In \bibinfo{booktitle}{\emph{Proc. Interspeech 2017}}. \bibinfo{pages}{87--91}.
\newblock


\bibitem[Jung et~al\mbox{.}(2022a)]%
        {jung2022aasist}
\bibfield{author}{\bibinfo{person}{Jee-weon Jung}, \bibinfo{person}{Hee-Soo Heo}, \bibinfo{person}{Hemlata Tak}, \bibinfo{person}{Hye-jin Shim}, \bibinfo{person}{Joon~Son Chung}, \bibinfo{person}{Bong-Jin Lee}, \bibinfo{person}{Ha-Jin Yu}, {and} \bibinfo{person}{Nicholas Evans}.} \bibinfo{year}{2022}\natexlab{a}.
\newblock \showarticletitle{Aasist: Audio anti-spoofing using integrated spectro-temporal graph attention networks}. In \bibinfo{booktitle}{\emph{ICASSP 2022-2022 IEEE international conference on acoustics, speech and signal processing (ICASSP)}}. IEEE, \bibinfo{pages}{6367--6371}.
\newblock


\bibitem[Jung et~al\mbox{.}(2019)]%
        {Jung_Shim_Heo_Yu_2019}
\bibfield{author}{\bibinfo{person}{Jee-weon Jung}, \bibinfo{person}{Hye-jin Shim}, \bibinfo{person}{Hee-Soo Heo}, {and} \bibinfo{person}{Ha-Jin Yu}.} \bibinfo{year}{2019}\natexlab{}.
\newblock \showarticletitle{Replay attack detection with complementary high-resolution information using end-to-end DNN for the ASVSPOOF 2019 challenge}.
\newblock \bibinfo{journal}{\emph{Interspeech 2019}} (\bibinfo{date}{Sep} \bibinfo{year}{2019}).
\newblock


\bibitem[Jung et~al\mbox{.}(2022b)]%
        {jung22c_interspeech}
\bibfield{author}{\bibinfo{person}{Jee-weon Jung}, \bibinfo{person}{Hemlata Tak}, \bibinfo{person}{Hye-jin Shim}, \bibinfo{person}{Hee-Soo Heo}, \bibinfo{person}{Bong-Jin Lee}, \bibinfo{person}{Soo-Whan Chung}, \bibinfo{person}{Ha-Jin Yu}, \bibinfo{person}{Nicholas Evans}, {and} \bibinfo{person}{Tomi Kinnunen}.} \bibinfo{year}{2022}\natexlab{b}.
\newblock \showarticletitle{SASV 2022: The First Spoofing-Aware Speaker Verification Challenge}. In \bibinfo{booktitle}{\emph{Interspeech 2022}}. \bibinfo{pages}{2893--2897}.
\newblock
\showISSN{2958-1796}


\bibitem[Kandari et~al\mbox{.}(2023)]%
        {kandari2023comprehensive}
\bibfield{author}{\bibinfo{person}{Megha Kandari}, \bibinfo{person}{Vikas Tripathi}, {and} \bibinfo{person}{Bhaskar Pant}.} \bibinfo{year}{2023}\natexlab{}.
\newblock \showarticletitle{A Comprehensive Review of Media Forensics and Deepfake Detection Technique}. In \bibinfo{booktitle}{\emph{2023 10th International Conference on Computing for Sustainable Global Development (INDIACom)}}. IEEE, \bibinfo{pages}{392--395}.
\newblock


\bibitem[Kang et~al\mbox{.}(2022)]%
        {kang2022attentive}
\bibfield{author}{\bibinfo{person}{Woo~Hyun Kang}, \bibinfo{person}{Jahangir Alam}, {and} \bibinfo{person}{Abderrahim Fathan}.} \bibinfo{year}{2022}\natexlab{}.
\newblock \showarticletitle{Attentive activation function for improving end-to-end spoofing countermeasure systems}.
\newblock \bibinfo{journal}{\emph{arXiv preprint arXiv:2205.01528}} (\bibinfo{year}{2022}).
\newblock


\bibitem[Kassis and Hengartner(2023)]%
        {kassis2023breaking}
\bibfield{author}{\bibinfo{person}{Andre Kassis} {and} \bibinfo{person}{Urs Hengartner}.} \bibinfo{year}{2023}\natexlab{}.
\newblock \showarticletitle{Breaking Security-Critical Voice Authentication}. In \bibinfo{booktitle}{\emph{2023 IEEE Symposium on Security and Privacy (SP)}}. IEEE, \bibinfo{pages}{951--968}.
\newblock


\bibitem[Khan and Malik(2023)]%
        {khan2023spotnet}
\bibfield{author}{\bibinfo{person}{Awais Khan} {and} \bibinfo{person}{Khalid~Mahmood Malik}.} \bibinfo{year}{2023}\natexlab{}.
\newblock \showarticletitle{SpoTNet: A spoofing-aware Transformer Network for Effective Synthetic Speech Detection}. In \bibinfo{booktitle}{\emph{Proceedings of the 2nd ACM International Workshop on Multimedia AI against Disinformation}}. \bibinfo{pages}{10--18}.
\newblock


\bibitem[Khan et~al\mbox{.}(2024)]%
        {khan2024frame}
\bibfield{author}{\bibinfo{person}{Awais Khan}, \bibinfo{person}{Khalid~Mahmood Malik}, {and} \bibinfo{person}{Shah Nawaz}.} \bibinfo{year}{2024}\natexlab{}.
\newblock \showarticletitle{Frame-to-Utterance Convergence: A Spectra-Temporal Approach for Unified Spoofing Detection}. In \bibinfo{booktitle}{\emph{ICASSP 2024-2024 IEEE International Conference on Acoustics, Speech and Signal Processing (ICASSP)}}. IEEE, \bibinfo{pages}{10761--10765}.
\newblock


\bibitem[Khan et~al\mbox{.}(2023)]%
        {khan2023battling}
\bibfield{author}{\bibinfo{person}{Awais Khan}, \bibinfo{person}{Khalid~Mahmood Malik}, \bibinfo{person}{James Ryan}, {and} \bibinfo{person}{Mikul Saravanan}.} \bibinfo{year}{2023}\natexlab{}.
\newblock \showarticletitle{Battling voice spoofing: a review, comparative analysis, and generalizability evaluation of state-of-the-art voice spoofing counter measures}.
\newblock \bibinfo{journal}{\emph{Artificial Intelligence Review}} \bibinfo{volume}{56}, \bibinfo{number}{Suppl 1} (\bibinfo{year}{2023}), \bibinfo{pages}{513--566}.
\newblock


\bibitem[Khanjani et~al\mbox{.}(2021)]%
        {khanjani2021deep}
\bibfield{author}{\bibinfo{person}{Zahra Khanjani}, \bibinfo{person}{Gabrielle Watson}, {and} \bibinfo{person}{Vandana~P Janeja}.} \bibinfo{year}{2021}\natexlab{}.
\newblock \showarticletitle{How deep are the fakes? focusing on audio deepfake: A survey}.
\newblock \bibinfo{journal}{\emph{arXiv preprint arXiv:2111.14203}} (\bibinfo{year}{2021}).
\newblock


\bibitem[Kim et~al\mbox{.}(2021)]%
        {Kim2021specmix}
\bibfield{author}{\bibinfo{person}{Gwantae Kim}, \bibinfo{person}{David~K. Han}, {and} \bibinfo{person}{Hanseok Ko}.} \bibinfo{year}{2021}\natexlab{}.
\newblock \showarticletitle{SpecMix : A mixed sample data augmentation method for training with time-frequency domain features}.
\newblock \bibinfo{journal}{\emph{Interspeech 2021}} (\bibinfo{date}{Aug} \bibinfo{year}{2021}).
\newblock


\bibitem[Kinnunen et~al\mbox{.}(2020)]%
        {kinnunen2020tandem}
\bibfield{author}{\bibinfo{person}{Tomi Kinnunen}, \bibinfo{person}{H{\'e}ctor Delgado}, \bibinfo{person}{Nicholas Evans}, \bibinfo{person}{Kong~Aik Lee}, \bibinfo{person}{Ville Vestman}, \bibinfo{person}{Andreas Nautsch}, \bibinfo{person}{Massimiliano Todisco}, \bibinfo{person}{Xin Wang}, \bibinfo{person}{Md Sahidullah}, \bibinfo{person}{Junichi Yamagishi}, {et~al\mbox{.}}} \bibinfo{year}{2020}\natexlab{}.
\newblock \showarticletitle{Tandem assessment of spoofing countermeasures and automatic speaker verification: Fundamentals}.
\newblock \bibinfo{journal}{\emph{IEEE/ACM Transactions on Audio, Speech, and Language Processing}}  \bibinfo{volume}{28} (\bibinfo{year}{2020}), \bibinfo{pages}{2195--2210}.
\newblock


\bibitem[Kumar et~al\mbox{.}(2021)]%
        {kumar2021speech}
\bibfield{author}{\bibinfo{person}{A~Kishore Kumar}, \bibinfo{person}{Dipjyoti Paul}, \bibinfo{person}{Monisankha Pal}, \bibinfo{person}{Md Sahidullah}, {and} \bibinfo{person}{Goutam Saha}.} \bibinfo{year}{2021}\natexlab{}.
\newblock \showarticletitle{Speech frame selection for spoofing detection with an application to partially spoofed audio-data}.
\newblock \bibinfo{journal}{\emph{International Journal of Speech Technology}}  \bibinfo{volume}{24} (\bibinfo{year}{2021}), \bibinfo{pages}{193--203}.
\newblock


\bibitem[Kwak et~al\mbox{.}(2021)]%
        {kwak2021resmax}
\bibfield{author}{\bibinfo{person}{Il-Youp Kwak}, \bibinfo{person}{Sungsu Kwag}, \bibinfo{person}{Junhee Lee}, \bibinfo{person}{Jun~Ho Huh}, \bibinfo{person}{Choong-Hoon Lee}, \bibinfo{person}{Youngbae Jeon}, \bibinfo{person}{Jeonghwan Hwang}, {and} \bibinfo{person}{Ji~Won Yoon}.} \bibinfo{year}{2021}\natexlab{}.
\newblock \showarticletitle{ResMax: Detecting voice spoofing attacks with residual network and max feature map}. In \bibinfo{booktitle}{\emph{2020 25th International Conference on Pattern Recognition (ICPR)}}. IEEE, \bibinfo{pages}{4837--4844}.
\newblock


\bibitem[Lai et~al\mbox{.}(2019)]%
        {Lai2019ASSERT}
\bibfield{author}{\bibinfo{person}{Cheng-I Lai}, \bibinfo{person}{Nanxin Chen}, \bibinfo{person}{Jesús Villalba}, {and} \bibinfo{person}{Najim Dehak}.} \bibinfo{year}{2019}\natexlab{}.
\newblock \showarticletitle{Assert: Anti-spoofing with squeeze-excitation and residual networks}.
\newblock \bibinfo{journal}{\emph{Interspeech 2019}} (\bibinfo{date}{Sep} \bibinfo{year}{2019}).
\newblock


\bibitem[Lavrentyeva et~al\mbox{.}(2019)]%
        {Lavrentyeva2019}
\bibfield{author}{\bibinfo{person}{Galina Lavrentyeva}, \bibinfo{person}{Sergey Novoselov}, \bibinfo{person}{Andzhukaev Tseren}, \bibinfo{person}{Marina Volkova}, \bibinfo{person}{Artem Gorlanov}, {and} \bibinfo{person}{Alexandr Kozlov}.} \bibinfo{year}{2019}\natexlab{}.
\newblock \showarticletitle{STC antispoofing systems for the ASVSPOOF2019 challenge}.
\newblock \bibinfo{journal}{\emph{Interspeech 2019}} (\bibinfo{date}{Sep} \bibinfo{year}{2019}).
\newblock


\bibitem[Layton et~al\mbox{.}(2024)]%
        {layton2024every}
\bibfield{author}{\bibinfo{person}{Seth Layton}, \bibinfo{person}{Thiago De~Andrade}, \bibinfo{person}{Daniel Olszewski}, \bibinfo{person}{Kevin Warren}, \bibinfo{person}{Kevin Butler}, {and} \bibinfo{person}{Patrick Traynor}.} \bibinfo{year}{2024}\natexlab{}.
\newblock \showarticletitle{Every Breath You Don't Take: Deepfake Speech Detection Using Breath}.
\newblock \bibinfo{journal}{\emph{arXiv preprint arXiv:2404.15143}} (\bibinfo{year}{2024}).
\newblock


\bibitem[Lee et~al\mbox{.}(2022)]%
        {Lee2022sasv}
\bibfield{author}{\bibinfo{person}{Jin~Woo Lee}, \bibinfo{person}{Eungbeom Kim}, \bibinfo{person}{Junghyun Koo}, {and} \bibinfo{person}{Kyogu Lee}.} \bibinfo{year}{2022}\natexlab{}.
\newblock \showarticletitle{Representation selective self-distillation and WAV2VEC 2.0 feature exploration for spoof-aware speaker verification}.
\newblock \bibinfo{journal}{\emph{Interspeech 2022}} (\bibinfo{date}{Sep} \bibinfo{year}{2022}).
\newblock


\bibitem[Lee et~al\mbox{.}(2023)]%
        {lee2023experimental}
\bibfield{author}{\bibinfo{person}{Yerin Lee}, \bibinfo{person}{Narin Kim}, \bibinfo{person}{Jaehong Jeong}, {and} \bibinfo{person}{Il-Youp Kwak}.} \bibinfo{year}{2023}\natexlab{}.
\newblock \showarticletitle{Experimental Case Study of Self-Supervised Learning for Voice Spoofing Detection}.
\newblock \bibinfo{journal}{\emph{IEEE Access}}  \bibinfo{volume}{11} (\bibinfo{year}{2023}), \bibinfo{pages}{24216--24226}.
\newblock


\bibitem[Li et~al\mbox{.}(2022d)]%
        {li2022role}
\bibfield{author}{\bibinfo{person}{Changtao Li}, \bibinfo{person}{Feiran Yang}, {and} \bibinfo{person}{Jun Yang}.} \bibinfo{year}{2022}\natexlab{d}.
\newblock \showarticletitle{The role of long-term dependency in synthetic speech detection}.
\newblock \bibinfo{journal}{\emph{IEEE Signal Processing Letters}}  \bibinfo{volume}{29} (\bibinfo{year}{2022}), \bibinfo{pages}{1142--1146}.
\newblock


\bibitem[Li et~al\mbox{.}(2023b)]%
        {li2023partiallocation}
\bibfield{author}{\bibinfo{person}{Jun Li}, \bibinfo{person}{Lin Li}, \bibinfo{person}{Mengjie Luo}, \bibinfo{person}{Xiaoqin Wang}, \bibinfo{person}{Shushan Qiao}, {and} \bibinfo{person}{Yumei Zhou}.} \bibinfo{year}{2023}\natexlab{b}.
\newblock \showarticletitle{Multi-grained Backend Fusion for Manipulation Region Location of Partially Fake Audio}. In \bibinfo{booktitle}{\emph{Proceedings of IJCAI 2023 Workshop on Deepfake Audio Detection and Analysis}}, Vol.~\bibinfo{volume}{755}.
\newblock


\bibitem[Li et~al\mbox{.}(2022b)]%
        {li2022long}
\bibfield{author}{\bibinfo{person}{Jialong Li}, \bibinfo{person}{Hongxia Wang}, \bibinfo{person}{Peisong He}, \bibinfo{person}{Sani~M Abdullahi}, {and} \bibinfo{person}{Bin Li}.} \bibinfo{year}{2022}\natexlab{b}.
\newblock \showarticletitle{Long-term variable Q transform: A novel time-frequency transform algorithm for synthetic speech detection}.
\newblock \bibinfo{journal}{\emph{Digital Signal Processing}}  \bibinfo{volume}{120} (\bibinfo{year}{2022}), \bibinfo{pages}{103256}.
\newblock


\bibitem[Li et~al\mbox{.}(2023c)]%
        {li2023contributions}
\bibfield{author}{\bibinfo{person}{Kai Li}, \bibinfo{person}{Xugang Lu}, \bibinfo{person}{Masato Akagi}, {and} \bibinfo{person}{Masashi Unoki}.} \bibinfo{year}{2023}\natexlab{c}.
\newblock \showarticletitle{Contributions of Jitter and Shimmer in the Voice for Fake Audio Detection}.
\newblock \bibinfo{journal}{\emph{IEEE Access}} (\bibinfo{year}{2023}).
\newblock


\bibitem[Li et~al\mbox{.}(2022c)]%
        {li2022analysis}
\bibfield{author}{\bibinfo{person}{Kai Li}, \bibinfo{person}{Yao Wang}, \bibinfo{person}{Minh Le~Nguyen}, \bibinfo{person}{Masato Akagi}, {and} \bibinfo{person}{Masashi Unoki}.} \bibinfo{year}{2022}\natexlab{c}.
\newblock \showarticletitle{Analysis of amplitude and frequency perturbation in the voice for fake audio detection}. In \bibinfo{booktitle}{\emph{2022 Asia-Pacific Signal and Information Processing Association Annual Summit and Conference (APSIPA ASC)}}. IEEE, \bibinfo{pages}{929--936}.
\newblock


\bibitem[Li et~al\mbox{.}(2023f)]%
        {li2023multitask}
\bibfield{author}{\bibinfo{person}{Kang Li}, \bibinfo{person}{Xiao-Min Zeng}, \bibinfo{person}{Jian-Tao Zhang}, {and} \bibinfo{person}{Yan Song}.} \bibinfo{year}{2023}\natexlab{f}.
\newblock \showarticletitle{Convolutional Recurrent Neural Network and Multitask Learning for Manipulation Region Location}. In \bibinfo{booktitle}{\emph{Proceedings of IJCAI 2023 Workshop on Deepfake Audio Detection and Analysis}}, Vol.~\bibinfo{volume}{750}.
\newblock


\bibitem[Li et~al\mbox{.}(2023d)]%
        {li2023voice}
\bibfield{author}{\bibinfo{person}{Lanting Li}, \bibinfo{person}{Tianliang Lu}, \bibinfo{person}{Xingbang Ma}, \bibinfo{person}{Mengjiao Yuan}, {and} \bibinfo{person}{Da Wan}.} \bibinfo{year}{2023}\natexlab{d}.
\newblock \showarticletitle{Voice Deepfake Detection Using the Self-Supervised Pre-Training Model HuBERT}.
\newblock \bibinfo{journal}{\emph{Applied Sciences}} \bibinfo{volume}{13}, \bibinfo{number}{14} (\bibinfo{year}{2023}), \bibinfo{pages}{8488}.
\newblock


\bibitem[Li et~al\mbox{.}(2022a)]%
        {li2022comparative}
\bibfield{author}{\bibinfo{person}{Menglu Li}, \bibinfo{person}{Yasaman Ahmadiadli}, {and} \bibinfo{person}{Xiao-Ping Zhang}.} \bibinfo{year}{2022}\natexlab{a}.
\newblock \showarticletitle{A comparative study on physical and perceptual features for deepfake audio detection}. In \bibinfo{booktitle}{\emph{Proceedings of the 1st International Workshop on Deepfake Detection for Audio Multimedia}}. \bibinfo{pages}{35--41}.
\newblock


\bibitem[Li et~al\mbox{.}(2023a)]%
        {li2023bilevel}
\bibfield{author}{\bibinfo{person}{Menglu Li}, \bibinfo{person}{Yasaman Ahmadiadli}, {and} \bibinfo{person}{Xiao-Ping Zhang}.} \bibinfo{year}{2023}\natexlab{a}.
\newblock \showarticletitle{Robust Deepfake Audio Detection via Bi-Level Optimization}. In \bibinfo{booktitle}{\emph{2023 IEEE 25th International Workshop on Multimedia Signal Processing (MMSP)}}. IEEE, \bibinfo{pages}{1--6}.
\newblock


\bibitem[Li and Zhang(2023)]%
        {li2023robust}
\bibfield{author}{\bibinfo{person}{Menglu Li} {and} \bibinfo{person}{Xiao-Ping Zhang}.} \bibinfo{year}{2023}\natexlab{}.
\newblock \showarticletitle{Robust Audio Anti-Spoofing System Based on Low-Frequency Sub-Band Information}. In \bibinfo{booktitle}{\emph{2023 IEEE Workshop on Applications of Signal Processing to Audio and Acoustics (WASPAA)}}. IEEE, \bibinfo{pages}{1--5}.
\newblock


\bibitem[Li and Zhang(2024)]%
        {li24oa_interspeech}
\bibfield{author}{\bibinfo{person}{Menglu Li} {and} \bibinfo{person}{Xiao-Ping Zhang}.} \bibinfo{year}{2024}\natexlab{}.
\newblock \showarticletitle{Interpretable Temporal Class Activation Representation for Audio Spoofing Detection}. In \bibinfo{booktitle}{\emph{Interspeech 2024}}. \bibinfo{pages}{1120--1124}.
\newblock
\showISSN{2958-1796}


\bibitem[Li et~al\mbox{.}(2023e)]%
        {li2023investigation}
\bibfield{author}{\bibinfo{person}{Wei Li}, \bibinfo{person}{Jichen Yang}, {and} \bibinfo{person}{Pei Lin}.} \bibinfo{year}{2023}\natexlab{e}.
\newblock \showarticletitle{Investigation of the influence of blocks on the linear spectrum for synthetic speech detection}.
\newblock \bibinfo{journal}{\emph{Electronics Letters}} \bibinfo{volume}{59}, \bibinfo{number}{9} (\bibinfo{year}{2023}), \bibinfo{pages}{e12797}.
\newblock


\bibitem[Li et~al\mbox{.}(2021)]%
        {li2021replay}
\bibfield{author}{\bibinfo{person}{Xu Li}, \bibinfo{person}{Na Li}, \bibinfo{person}{Chao Weng}, \bibinfo{person}{Xunying Liu}, \bibinfo{person}{Dan Su}, \bibinfo{person}{Dong Yu}, {and} \bibinfo{person}{Helen Meng}.} \bibinfo{year}{2021}\natexlab{}.
\newblock \showarticletitle{Replay and synthetic speech detection with res2net architecture}. In \bibinfo{booktitle}{\emph{ICASSP 2021-2021 IEEE international conference on acoustics, speech and signal processing (ICASSP)}}. IEEE, \bibinfo{pages}{6354--6358}.
\newblock


\bibitem[Liao et~al\mbox{.}(2022)]%
        {Liao2022distillation}
\bibfield{author}{\bibinfo{person}{Yen-Lun Liao}, \bibinfo{person}{Xuanjun Chen}, \bibinfo{person}{Chung-Che Wang}, {and} \bibinfo{person}{Jyh-Shing~Roger Jang}.} \bibinfo{year}{2022}\natexlab{}.
\newblock \showarticletitle{Adversarial speaker distillation for countermeasure model on Automatic speaker verification}.
\newblock \bibinfo{journal}{\emph{2nd Symposium on Security and Privacy in Speech Communication}} (\bibinfo{date}{Sep} \bibinfo{year}{2022}).
\newblock


\bibitem[Lim et~al\mbox{.}(2022)]%
        {lim2022detecting}
\bibfield{author}{\bibinfo{person}{Suk-Young Lim}, \bibinfo{person}{Dong-Kyu Chae}, {and} \bibinfo{person}{Sang-Chul Lee}.} \bibinfo{year}{2022}\natexlab{}.
\newblock \showarticletitle{Detecting deepfake voice using explainable deep learning techniques}.
\newblock \bibinfo{journal}{\emph{Applied Sciences}} \bibinfo{volume}{12}, \bibinfo{number}{8} (\bibinfo{year}{2022}), \bibinfo{pages}{3926}.
\newblock


\bibitem[Lin et~al\mbox{.}(2024)]%
        {lin2024one}
\bibfield{author}{\bibinfo{person}{Guoyuan Lin}, \bibinfo{person}{Weiqi Luo}, \bibinfo{person}{Da Luo}, {and} \bibinfo{person}{Jiwu Huang}.} \bibinfo{year}{2024}\natexlab{}.
\newblock \showarticletitle{One-class neural network with directed statistics pooling for spoofing speech detection}.
\newblock \bibinfo{journal}{\emph{IEEE Transactions on Information Forensics and Security}} (\bibinfo{year}{2024}).
\newblock


\bibitem[Liu et~al\mbox{.}(2023b)]%
        {liu2023transsionadd}
\bibfield{author}{\bibinfo{person}{Jie Liu}, \bibinfo{person}{Zhiba Su}, \bibinfo{person}{Hui Huang}, \bibinfo{person}{Caiyan Wan}, \bibinfo{person}{Quanxiu Wang}, \bibinfo{person}{Jiangli Hong}, \bibinfo{person}{Benlai Tang}, {and} \bibinfo{person}{Fengjie Zhu}.} \bibinfo{year}{2023}\natexlab{b}.
\newblock \showarticletitle{TranssionADD: A multi-frame reinforcement based sequence tagging model for audio deepfake detection}.
\newblock \bibinfo{journal}{\emph{Proceedings of IJCAI 2023 Workshop on Deepfake Audio Detection and Analysis (DADA 2023)}} (\bibinfo{year}{2023}).
\newblock


\bibitem[Liu et~al\mbox{.}(2023d)]%
        {Liu2023betray}
\bibfield{author}{\bibinfo{person}{Rui Liu}, \bibinfo{person}{Jinhua Zhang}, \bibinfo{person}{Guanglai Gao}, {and} \bibinfo{person}{Haizhou Li}.} \bibinfo{year}{2023}\natexlab{d}.
\newblock \showarticletitle{Betray oneself: A novel audio deepfake detection model via mono-to-stereo conversion}.
\newblock \bibinfo{journal}{\emph{INTERSPEECH 2023}} (\bibinfo{date}{Aug} \bibinfo{year}{2023}).
\newblock


\bibitem[Liu et~al\mbox{.}(2019)]%
        {liu2019adversarial}
\bibfield{author}{\bibinfo{person}{Songxiang Liu}, \bibinfo{person}{Haibin Wu}, \bibinfo{person}{Hung-yi Lee}, {and} \bibinfo{person}{Helen Meng}.} \bibinfo{year}{2019}\natexlab{}.
\newblock \showarticletitle{Adversarial attacks on spoofing countermeasures of automatic speaker verification}. In \bibinfo{booktitle}{\emph{2019 IEEE Automatic Speech Recognition and Understanding Workshop (ASRU)}}. IEEE, \bibinfo{pages}{312--319}.
\newblock


\bibitem[Liu et~al\mbox{.}(2020)]%
        {liu2020energy}
\bibfield{author}{\bibinfo{person}{Weitang Liu}, \bibinfo{person}{Xiaoyun Wang}, \bibinfo{person}{John Owens}, {and} \bibinfo{person}{Yixuan Li}.} \bibinfo{year}{2020}\natexlab{}.
\newblock \showarticletitle{Energy-based out-of-distribution detection}.
\newblock \bibinfo{journal}{\emph{Advances in neural information processing systems}}  \bibinfo{volume}{33} (\bibinfo{year}{2020}), \bibinfo{pages}{21464--21475}.
\newblock


\bibitem[Liu et~al\mbox{.}(2023a)]%
        {liu2023leveraging}
\bibfield{author}{\bibinfo{person}{Xiaohui Liu}, \bibinfo{person}{Meng Liu}, \bibinfo{person}{Longbiao Wang}, \bibinfo{person}{Kong~Aik Lee}, \bibinfo{person}{Hanyi Zhang}, {and} \bibinfo{person}{Jianwu Dang}.} \bibinfo{year}{2023}\natexlab{a}.
\newblock \showarticletitle{Leveraging positional-related local-global dependency for synthetic speech detection}. In \bibinfo{booktitle}{\emph{ICASSP 2023-2023 IEEE International Conference on Acoustics, Speech and Signal Processing (ICASSP)}}. IEEE, \bibinfo{pages}{1--5}.
\newblock


\bibitem[Liu et~al\mbox{.}(2023c)]%
        {liu2023asvspoof}
\bibfield{author}{\bibinfo{person}{Xuechen Liu}, \bibinfo{person}{Xin Wang}, \bibinfo{person}{Md Sahidullah}, \bibinfo{person}{Jose Patino}, \bibinfo{person}{H{\'e}ctor Delgado}, \bibinfo{person}{Tomi Kinnunen}, \bibinfo{person}{Massimiliano Todisco}, \bibinfo{person}{Junichi Yamagishi}, \bibinfo{person}{Nicholas Evans}, \bibinfo{person}{Andreas Nautsch}, {et~al\mbox{.}}} \bibinfo{year}{2023}\natexlab{c}.
\newblock \showarticletitle{Asvspoof 2021: Towards spoofed and deepfake speech detection in the wild}.
\newblock \bibinfo{journal}{\emph{IEEE/ACM Transactions on Audio, Speech, and Language Processing}} (\bibinfo{year}{2023}).
\newblock


\bibitem[Lorenzo-Trueba et~al\mbox{.}(2018)]%
        {lorenzo2018voice}
\bibfield{author}{\bibinfo{person}{Jaime Lorenzo-Trueba}, \bibinfo{person}{Junichi Yamagishi}, \bibinfo{person}{Tomoki Toda}, \bibinfo{person}{Daisuke Saito}, \bibinfo{person}{Fernando Villavicencio}, \bibinfo{person}{Tomi Kinnunen}, {and} \bibinfo{person}{Zhenhua Ling}.} \bibinfo{year}{2018}\natexlab{}.
\newblock \showarticletitle{The Voice Conversion Challenge 2018: Promoting Development of Parallel and Nonparallel Methods}. In \bibinfo{booktitle}{\emph{The Speaker and Language Recognition Workshop}}. ISCA, \bibinfo{pages}{195--202}.
\newblock


\bibitem[Lu et~al\mbox{.}(2022)]%
        {lu2022acoustic}
\bibfield{author}{\bibinfo{person}{Jingze Lu}, \bibinfo{person}{Zhuo Li}, \bibinfo{person}{Yuxiang Zhang}, \bibinfo{person}{Wenchao Wang}, {and} \bibinfo{person}{Pengyuan Zhang}.} \bibinfo{year}{2022}\natexlab{}.
\newblock \showarticletitle{Acoustic or pattern? speech spoofing countermeasure based on image pre-training models}. In \bibinfo{booktitle}{\emph{Proceedings of the 1st International Workshop on Deepfake Detection for Audio Multimedia}}. \bibinfo{pages}{77--84}.
\newblock


\bibitem[Lu et~al\mbox{.}(2023)]%
        {lu2023detecting}
\bibfield{author}{\bibinfo{person}{Jingze Lu}, \bibinfo{person}{Yuxiang Zhang}, \bibinfo{person}{Zhuo Li}, \bibinfo{person}{Zengqiang Shang}, \bibinfo{person}{Wenchao Wang}, {and} \bibinfo{person}{Pengyuan Zhang}.} \bibinfo{year}{2023}\natexlab{}.
\newblock \showarticletitle{Detecting Unknown Speech Spoofing Algorithms with Nearest Neighbors.}. In \bibinfo{booktitle}{\emph{Proceedings of IJCAI 2023 Workshop on Deepfake Audio Detection and Analysis (DADA 2023)}}. \bibinfo{pages}{89--94}.
\newblock


\bibitem[Lu et~al\mbox{.}(2024)]%
        {lu2024one}
\bibfield{author}{\bibinfo{person}{Jingze Lu}, \bibinfo{person}{Yuxiang Zhang}, \bibinfo{person}{Wenchao Wang}, \bibinfo{person}{Zengqiang Shang}, {and} \bibinfo{person}{Pengyuan Zhang}.} \bibinfo{year}{2024}\natexlab{}.
\newblock \showarticletitle{One-Class Knowledge Distillation for Spoofing Speech Detection}. In \bibinfo{booktitle}{\emph{ICASSP 2024-2024 IEEE International Conference on Acoustics, Speech and Signal Processing (ICASSP)}}. IEEE, \bibinfo{pages}{11251--11255}.
\newblock


\bibitem[Luo et~al\mbox{.}(2021)]%
        {Luo2021capsule}
\bibfield{author}{\bibinfo{person}{Anwei Luo}, \bibinfo{person}{Enlei Li}, \bibinfo{person}{Yongliang Liu}, \bibinfo{person}{Xiangui Kang}, {and} \bibinfo{person}{Z.~Jane Wang}.} \bibinfo{year}{2021}\natexlab{}.
\newblock \showarticletitle{A capsule network based approach for detection of audio spoofing attacks}.
\newblock \bibinfo{journal}{\emph{ICASSP 2021 - 2021 IEEE International Conference on Acoustics, Speech and Signal Processing (ICASSP)}} (\bibinfo{date}{Jun} \bibinfo{year}{2021}).
\newblock


\bibitem[Lv et~al\mbox{.}(2022)]%
        {lv2022fake}
\bibfield{author}{\bibinfo{person}{Zhiqiang Lv}, \bibinfo{person}{Shanshan Zhang}, \bibinfo{person}{Kai Tang}, {and} \bibinfo{person}{Pengfei Hu}.} \bibinfo{year}{2022}\natexlab{}.
\newblock \showarticletitle{Fake audio detection based on unsupervised pretraining models}. In \bibinfo{booktitle}{\emph{ICASSP 2022-2022 IEEE International Conference on Acoustics, Speech and Signal Processing (ICASSP)}}. IEEE, \bibinfo{pages}{9231--9235}.
\newblock


\bibitem[Ma et~al\mbox{.}(2024)]%
        {ma2024cfad}
\bibfield{author}{\bibinfo{person}{Haoxin Ma}, \bibinfo{person}{Jiangyan Yi}, \bibinfo{person}{Chenglong Wang}, \bibinfo{person}{Xinrui Yan}, \bibinfo{person}{Jianhua Tao}, \bibinfo{person}{Tao Wang}, \bibinfo{person}{Shiming Wang}, {and} \bibinfo{person}{Ruibo Fu}.} \bibinfo{year}{2024}\natexlab{}.
\newblock \showarticletitle{CFAD: A Chinese dataset for fake audio detection}.
\newblock \bibinfo{journal}{\emph{Speech Communication}}  \bibinfo{volume}{164} (\bibinfo{year}{2024}), \bibinfo{pages}{103122}.
\newblock


\bibitem[Ma et~al\mbox{.}(2023a)]%
        {ma2023dualbranch}
\bibfield{author}{\bibinfo{person}{Kaijie Ma}, \bibinfo{person}{Yifan Feng}, \bibinfo{person}{Beijing Chen}, {and} \bibinfo{person}{Guoying Zhao}.} \bibinfo{year}{2023}\natexlab{a}.
\newblock \showarticletitle{End-to-end dual-branch network towards synthetic speech detection}.
\newblock \bibinfo{journal}{\emph{IEEE Signal Processing Letters}}  \bibinfo{volume}{30} (\bibinfo{year}{2023}), \bibinfo{pages}{359--363}.
\newblock


\bibitem[Ma et~al\mbox{.}(2022)]%
        {ma2022convnext}
\bibfield{author}{\bibinfo{person}{Qiaowei Ma}, \bibinfo{person}{Jinghui Zhong}, \bibinfo{person}{Yitao Yang}, \bibinfo{person}{Weiheng Liu}, \bibinfo{person}{Ying Gao}, {and} \bibinfo{person}{Wing~WY Ng}.} \bibinfo{year}{2022}\natexlab{}.
\newblock \showarticletitle{ConvNeXt Based Neural Network for Audio Anti-Spoofing}.
\newblock \bibinfo{journal}{\emph{arXiv preprint arXiv:2209.06434}} (\bibinfo{year}{2022}).
\newblock


\bibitem[Ma et~al\mbox{.}(2023b)]%
        {Ma2023xvector}
\bibfield{author}{\bibinfo{person}{Xinyue Ma}, \bibinfo{person}{Shanshan Zhang}, \bibinfo{person}{Shen Huang}, \bibinfo{person}{Ji Gao}, \bibinfo{person}{Ying Hu}, {and} \bibinfo{person}{Liang He}.} \bibinfo{year}{2023}\natexlab{b}.
\newblock \showarticletitle{How to boost anti-spoofing with X-vectors}.
\newblock \bibinfo{journal}{\emph{2022 IEEE Spoken Language Technology Workshop (SLT)}} (\bibinfo{date}{Jan} \bibinfo{year}{2023}).
\newblock


\bibitem[Ma et~al\mbox{.}(2021)]%
        {Ma2021RWResnet}
\bibfield{author}{\bibinfo{person}{Youxuan Ma}, \bibinfo{person}{Zongze Ren}, {and} \bibinfo{person}{Shugong Xu}.} \bibinfo{year}{2021}\natexlab{}.
\newblock \showarticletitle{RW-resnet: A novel speech anti-spoofing model using Raw Waveform}.
\newblock  (\bibinfo{date}{Aug} \bibinfo{year}{2021}).
\newblock


\bibitem[Martin-Donas and Alvarez(2022)]%
        {donas2022vicomtech}
\bibfield{author}{\bibinfo{person}{Juan~M. Martin-Donas} {and} \bibinfo{person}{Aitor Alvarez}.} \bibinfo{year}{2022}\natexlab{}.
\newblock \showarticletitle{The Vicomtech audio deepfake detection system based on WAV2VEC2 for the 2022 add challenge}.
\newblock \bibinfo{journal}{\emph{ICASSP 2022 - 2022 IEEE International Conference on Acoustics, Speech and Signal Processing (ICASSP)}} (\bibinfo{date}{May} \bibinfo{year}{2022}).
\newblock


\bibitem[Mart{\'\i}n-Do{\~n}as and {\'A}lvarez(2023)]%
        {donas2023partial}
\bibfield{author}{\bibinfo{person}{Juan~Manuel Mart{\'\i}n-Do{\~n}as} {and} \bibinfo{person}{Aitor {\'A}lvarez}.} \bibinfo{year}{2023}\natexlab{}.
\newblock \showarticletitle{The Vicomtech partial deepfake detection and location system for the 2023 ADD Challenge}. In \bibinfo{booktitle}{\emph{Proceedings of IJCAI 2023 Workshop on Deepfake Audio Detection and Analysis}}.
\newblock


\bibitem[Masood et~al\mbox{.}(2023)]%
        {masood2023deepfakes}
\bibfield{author}{\bibinfo{person}{Momina Masood}, \bibinfo{person}{Mariam Nawaz}, \bibinfo{person}{Khalid~Mahmood Malik}, \bibinfo{person}{Ali Javed}, \bibinfo{person}{Aun Irtaza}, {and} \bibinfo{person}{Hafiz Malik}.} \bibinfo{year}{2023}\natexlab{}.
\newblock \showarticletitle{Deepfakes generation and detection: State-of-the-art, open challenges, countermeasures, and way forward}.
\newblock \bibinfo{journal}{\emph{Applied intelligence}} \bibinfo{volume}{53}, \bibinfo{number}{4} (\bibinfo{year}{2023}), \bibinfo{pages}{3974--4026}.
\newblock


\bibitem[M{\"u}ller et~al\mbox{.}(2022)]%
        {muller22_interspeech}
\bibfield{author}{\bibinfo{person}{Nicolas~M M{\"u}ller}, \bibinfo{person}{Pavel Czempin}, \bibinfo{person}{Franziska Dieckmann}, \bibinfo{person}{Adam Froghyar}, {and} \bibinfo{person}{Konstantin B{\"o}ttinger}.} \bibinfo{year}{2022}\natexlab{}.
\newblock \showarticletitle{Does Audio Deepfake Detection Generalize?}. In \bibinfo{booktitle}{\emph{Interspeech 2022}}. \bibinfo{pages}{2783--2787}.
\newblock
\showISSN{2958-1796}


\bibitem[M{\"u}ller et~al\mbox{.}(2024)]%
        {muller2024mlaad}
\bibfield{author}{\bibinfo{person}{Nicolas~M M{\"u}ller}, \bibinfo{person}{Piotr Kawa}, \bibinfo{person}{Wei~Herng Choong}, \bibinfo{person}{Edresson Casanova}, \bibinfo{person}{Eren G{\"o}lge}, \bibinfo{person}{Thorsten M{\"u}ller}, \bibinfo{person}{Piotr Syga}, \bibinfo{person}{Philip Sperl}, {and} \bibinfo{person}{Konstantin B{\"o}ttinger}.} \bibinfo{year}{2024}\natexlab{}.
\newblock \showarticletitle{MLAAD: The Multi-Language Audio Anti-Spoofing Dataset}.
\newblock \bibinfo{journal}{\emph{arXiv preprint arXiv:2401.09512}} (\bibinfo{year}{2024}).
\newblock


\bibitem[Mun et~al\mbox{.}(2023)]%
        {Mun2023sasvembedding}
\bibfield{author}{\bibinfo{person}{Sung~Hwan Mun}, \bibinfo{person}{Hye-jin Shim}, \bibinfo{person}{Hemlata Tak}, \bibinfo{person}{Xin Wang}, \bibinfo{person}{Xuechen Liu}, \bibinfo{person}{Md Sahidullah}, \bibinfo{person}{Myeonghun Jeong}, \bibinfo{person}{Min~Hyun Han}, \bibinfo{person}{Massimiliano Todisco}, \bibinfo{person}{Kong~Aik Lee}, {and} \bibinfo{person}{et al.}} \bibinfo{year}{2023}\natexlab{}.
\newblock \showarticletitle{Towards single integrated spoofing-aware speaker Verification Embeddings}.
\newblock \bibinfo{journal}{\emph{INTERSPEECH 2023}} (\bibinfo{date}{Aug} \bibinfo{year}{2023}).
\newblock


\bibitem[Muttathu Sivasankara~Pillai et~al\mbox{.}(2022)]%
        {Sankar2022voicedforVC}
\bibfield{author}{\bibinfo{person}{Arun~Sankar Muttathu Sivasankara~Pillai}, \bibinfo{person}{Phillip L.~De~Leon}, {and} \bibinfo{person}{Utz Roedig}.} \bibinfo{year}{2022}\natexlab{}.
\newblock \showarticletitle{Detection of voice conversion spoofing attacks using voiced speech}.
\newblock \bibinfo{journal}{\emph{Secure IT Systems}} (\bibinfo{year}{2022}), \bibinfo{pages}{159–175}.
\newblock


\bibitem[Müller et~al\mbox{.}(2021)]%
        {Muller2021silence}
\bibfield{author}{\bibinfo{person}{Nicolas Müller}, \bibinfo{person}{Franziska Dieckmann}, \bibinfo{person}{Pavel Czempin}, \bibinfo{person}{Roman Canals}, \bibinfo{person}{Konstantin Böttinger}, {and} \bibinfo{person}{Jennifer Williams}.} \bibinfo{year}{2021}\natexlab{}.
\newblock \showarticletitle{Speech is silver, silence is golden: What do ASVSPOOF-trained models really learn?}
\newblock \bibinfo{journal}{\emph{2021 Edition of the Automatic Speaker Verification and Spoofing Countermeasures Challenge}} (\bibinfo{date}{Sep} \bibinfo{year}{2021}).
\newblock


\bibitem[Müller et~al\mbox{.}(2022)]%
        {muller22b_interspeech}
\bibfield{author}{\bibinfo{person}{Nicolas Müller}, \bibinfo{person}{Franziska Diekmann}, {and} \bibinfo{person}{Jennifer Williams}.} \bibinfo{year}{2022}\natexlab{}.
\newblock \showarticletitle{Attacker Attribution of Audio Deepfakes}. In \bibinfo{booktitle}{\emph{Interspeech 2022}}. \bibinfo{pages}{2788--2792}.
\newblock
\showISSN{2958-1796}


\bibitem[Müller et~al\mbox{.}(2023)]%
        {muller23_interspeech}
\bibfield{author}{\bibinfo{person}{Nicolas~M. Müller}, \bibinfo{person}{Philip Sperl}, {and} \bibinfo{person}{Konstantin Böttinger}.} \bibinfo{year}{2023}\natexlab{}.
\newblock \showarticletitle{{Complex-valued neural networks for voice anti-spoofing}}. In \bibinfo{booktitle}{\emph{Proc. INTERSPEECH 2023}}. \bibinfo{pages}{3814--3818}.
\newblock


\bibitem[Nguyen-Vu et~al\mbox{.}(2023)]%
        {nguyen2023defense}
\bibfield{author}{\bibinfo{person}{Long Nguyen-Vu}, \bibinfo{person}{Thien-Phuc Doan}, \bibinfo{person}{Mai Bui}, \bibinfo{person}{Kihun Hong}, {and} \bibinfo{person}{Souhwan Jung}.} \bibinfo{year}{2023}\natexlab{}.
\newblock \showarticletitle{On the defense of spoofing countermeasures against adversarial attacks}.
\newblock \bibinfo{journal}{\emph{IEEE Access}} (\bibinfo{year}{2023}).
\newblock


\bibitem[Panayotov et~al\mbox{.}(2015)]%
        {panayotov2015librispeech}
\bibfield{author}{\bibinfo{person}{Vassil Panayotov}, \bibinfo{person}{Guoguo Chen}, \bibinfo{person}{Daniel Povey}, {and} \bibinfo{person}{Sanjeev Khudanpur}.} \bibinfo{year}{2015}\natexlab{}.
\newblock \showarticletitle{Librispeech: an asr corpus based on public domain audio books}. In \bibinfo{booktitle}{\emph{2015 IEEE international conference on acoustics, speech and signal processing (ICASSP)}}. IEEE, \bibinfo{pages}{5206--5210}.
\newblock


\bibitem[Pasini(2019)]%
        {pasini2019melgan}
\bibfield{author}{\bibinfo{person}{Marco Pasini}.} \bibinfo{year}{2019}\natexlab{}.
\newblock \showarticletitle{MelGAN-VC: Voice conversion and audio style transfer on arbitrarily long samples using spectrograms}.
\newblock \bibinfo{journal}{\emph{arXiv preprint arXiv:1910.03713}} (\bibinfo{year}{2019}).
\newblock


\bibitem[Patil et~al\mbox{.}(2022)]%
        {patil2022effectiveness}
\bibfield{author}{\bibinfo{person}{Ankur~T Patil}, \bibinfo{person}{Hemant~A Patil}, {and} \bibinfo{person}{Kuldeep Khoria}.} \bibinfo{year}{2022}\natexlab{}.
\newblock \showarticletitle{Effectiveness of energy separation-based instantaneous frequency estimation for cochlear cepstral features for synthetic and voice-converted spoofed speech detection}.
\newblock \bibinfo{journal}{\emph{Computer Speech \& Language}}  \bibinfo{volume}{72} (\bibinfo{year}{2022}), \bibinfo{pages}{101301}.
\newblock


\bibitem[Phukan et~al\mbox{.}(2024)]%
        {phukan2024heterogeneity}
\bibfield{author}{\bibinfo{person}{Orchid~Chetia Phukan}, \bibinfo{person}{Gautam~Siddharth Kashyap}, \bibinfo{person}{Arun~Balaji Buduru}, {and} \bibinfo{person}{Rajesh Sharma}.} \bibinfo{year}{2024}\natexlab{}.
\newblock \showarticletitle{Heterogeneity over Homogeneity: Investigating Multilingual Speech Pre-Trained Models for Detecting Audio Deepfake}. In \bibinfo{booktitle}{\emph{Findings of the Association for Computational Linguistics: NAACL 2024}}. \bibinfo{pages}{2496--2506}.
\newblock


\bibitem[Pilia et~al\mbox{.}(2021)]%
        {pilia2021time}
\bibfield{author}{\bibinfo{person}{Michele Pilia}, \bibinfo{person}{Sara Mandelli}, \bibinfo{person}{Paolo Bestagini}, {and} \bibinfo{person}{Stefano Tubaro}.} \bibinfo{year}{2021}\natexlab{}.
\newblock \showarticletitle{Time scaling detection and estimation in audio recordings}. In \bibinfo{booktitle}{\emph{2021 IEEE International Workshop on Information Forensics and Security (WIFS)}}. IEEE, \bibinfo{pages}{1--6}.
\newblock


\bibitem[Pratap et~al\mbox{.}(2024)]%
        {pratap2024scaling}
\bibfield{author}{\bibinfo{person}{Vineel Pratap}, \bibinfo{person}{Andros Tjandra}, \bibinfo{person}{Bowen Shi}, \bibinfo{person}{Paden Tomasello}, \bibinfo{person}{Arun Babu}, \bibinfo{person}{Sayani Kundu}, \bibinfo{person}{Ali Elkahky}, \bibinfo{person}{Zhaoheng Ni}, \bibinfo{person}{Apoorv Vyas}, \bibinfo{person}{Maryam Fazel-Zarandi}, {et~al\mbox{.}}} \bibinfo{year}{2024}\natexlab{}.
\newblock \showarticletitle{Scaling speech technology to 1,000+ languages}.
\newblock \bibinfo{journal}{\emph{Journal of Machine Learning Research}} \bibinfo{volume}{25}, \bibinfo{number}{97} (\bibinfo{year}{2024}), \bibinfo{pages}{1--52}.
\newblock


\bibitem[Qin et~al\mbox{.}(2023)]%
        {qin2023speaker}
\bibfield{author}{\bibinfo{person}{Xiaoyi Qin}, \bibinfo{person}{Xingming Wang}, \bibinfo{person}{Yanli Chen}, \bibinfo{person}{Qinglin Meng}, {and} \bibinfo{person}{Ming Li}.} \bibinfo{year}{2023}\natexlab{}.
\newblock \showarticletitle{From Speaker Verification to Deepfake Algorithm Recognition: Our Learned Lessons from ADD2023 Track 3.}. In \bibinfo{booktitle}{\emph{Proceedings of IJCAI 2023 Workshop on Deepfake Audio Detection and Analysis (DADA 2023)}}. \bibinfo{pages}{107--112}.
\newblock


\bibitem[Radford et~al\mbox{.}(2023)]%
        {radford2023robust}
\bibfield{author}{\bibinfo{person}{Alec Radford}, \bibinfo{person}{Jong~Wook Kim}, \bibinfo{person}{Tao Xu}, \bibinfo{person}{Greg Brockman}, \bibinfo{person}{Christine McLeavey}, {and} \bibinfo{person}{Ilya Sutskever}.} \bibinfo{year}{2023}\natexlab{}.
\newblock \showarticletitle{Robust speech recognition via large-scale weak supervision}. In \bibinfo{booktitle}{\emph{International conference on machine learning}}. PMLR, \bibinfo{pages}{28492--28518}.
\newblock


\bibitem[Ranjan et~al\mbox{.}(2023a)]%
        {ranjan2023sv}
\bibfield{author}{\bibinfo{person}{Rishabh Ranjan}, \bibinfo{person}{Mayank Vatsa}, {and} \bibinfo{person}{Richa Singh}.} \bibinfo{year}{2023}\natexlab{a}.
\newblock \showarticletitle{SV-DeiT: Speaker Verification with DeiTCap Spoofing Detection}. In \bibinfo{booktitle}{\emph{2023 IEEE International Joint Conference on Biometrics (IJCB)}}. IEEE, \bibinfo{pages}{1--10}.
\newblock


\bibitem[Ranjan et~al\mbox{.}(2023b)]%
        {ijcai2023p756}
\bibfield{author}{\bibinfo{person}{Rishabh Ranjan}, \bibinfo{person}{Mayank Vatsa}, {and} \bibinfo{person}{Richa Singh}.} \bibinfo{year}{2023}\natexlab{b}.
\newblock \showarticletitle{Uncovering the Deceptions: An Analysis on Audio Spoofing Detection and Future Prospects}. In \bibinfo{booktitle}{\emph{Proceedings of the Thirty-Second International Joint Conference on Artificial Intelligence, {IJCAI-23}}}, \bibfield{editor}{\bibinfo{person}{Edith Elkind}} (Ed.). \bibinfo{publisher}{International Joint Conferences on Artificial Intelligence Organization}, \bibinfo{pages}{6750--6758}.
\newblock
\newblock
\shownote{Survey Track}.


\bibitem[Ray et~al\mbox{.}(2021)]%
        {ray2021feature}
\bibfield{author}{\bibinfo{person}{Ruchira Ray}, \bibinfo{person}{Sanka Karthik}, \bibinfo{person}{Vinayak Mathur}, \bibinfo{person}{Prashant Kumar}, \bibinfo{person}{G Maragatham}, \bibinfo{person}{Sourabh Tiwari}, {and} \bibinfo{person}{Rashmi~T Shankarappa}.} \bibinfo{year}{2021}\natexlab{}.
\newblock \showarticletitle{Feature genuinization based residual squeeze-and-excitation for audio anti-spoofing in sound AI}. In \bibinfo{booktitle}{\emph{2021 12th International Conference on Computing Communication and Networking Technologies (ICCCNT)}}. IEEE, \bibinfo{pages}{1--5}.
\newblock


\bibitem[Reimao and Tzerpos(2019)]%
        {reimao2019dataset}
\bibfield{author}{\bibinfo{person}{Ricardo Reimao} {and} \bibinfo{person}{Vassilios Tzerpos}.} \bibinfo{year}{2019}\natexlab{}.
\newblock \showarticletitle{For: A dataset for synthetic speech detection}. In \bibinfo{booktitle}{\emph{2019 International Conference on Speech Technology and Human-Computer Dialogue (SpeD)}}. IEEE, \bibinfo{pages}{1--10}.
\newblock


\bibitem[Ren et~al\mbox{.}(2020)]%
        {renfastspeech}
\bibfield{author}{\bibinfo{person}{Yi Ren}, \bibinfo{person}{Chenxu Hu}, \bibinfo{person}{Xu Tan}, \bibinfo{person}{Tao Qin}, \bibinfo{person}{Sheng Zhao}, \bibinfo{person}{Zhou Zhao}, {and} \bibinfo{person}{Tie-Yan Liu}.} \bibinfo{year}{2020}\natexlab{}.
\newblock \showarticletitle{FastSpeech 2: Fast and High-Quality End-to-End Text to Speech}. In \bibinfo{booktitle}{\emph{International Conference on Learning Representations}}.
\newblock


\bibitem[Ren et~al\mbox{.}(2023b)]%
        {ren2023voice}
\bibfield{author}{\bibinfo{person}{Yeqing Ren}, \bibinfo{person}{Haipeng Peng}, \bibinfo{person}{Lixiang Li}, \bibinfo{person}{Xiaopeng Xue}, \bibinfo{person}{Yang Lan}, {and} \bibinfo{person}{Yixian Yang}.} \bibinfo{year}{2023}\natexlab{b}.
\newblock \showarticletitle{A voice spoofing detection framework for IoT systems with feature pyramid and online knowledge distillation}.
\newblock \bibinfo{journal}{\emph{Journal of Systems Architecture}}  \bibinfo{volume}{143} (\bibinfo{year}{2023}), \bibinfo{pages}{102981}.
\newblock


\bibitem[Ren et~al\mbox{.}(2023a)]%
        {ren2023lightweight}
\bibfield{author}{\bibinfo{person}{Yeqing Ren}, \bibinfo{person}{Haipeng Peng}, \bibinfo{person}{Lixiang Li}, {and} \bibinfo{person}{Yixian Yang}.} \bibinfo{year}{2023}\natexlab{a}.
\newblock \showarticletitle{Lightweight Voice Spoofing Detection using Improved One-Class Learning and Knowledge Distillation}.
\newblock \bibinfo{journal}{\emph{IEEE Transactions on Multimedia}} (\bibinfo{year}{2023}).
\newblock


\bibitem[Sahidullah et~al\mbox{.}(2015)]%
        {sahidullah15_interspeech}
\bibfield{author}{\bibinfo{person}{Md. Sahidullah}, \bibinfo{person}{Tomi Kinnunen}, {and} \bibinfo{person}{Cemal Hanilçi}.} \bibinfo{year}{2015}\natexlab{}.
\newblock \showarticletitle{{A comparison of features for synthetic speech detection}}. In \bibinfo{booktitle}{\emph{Proc. Interspeech 2015}}. \bibinfo{pages}{2087--2091}.
\newblock


\bibitem[Salvi et~al\mbox{.}(2023a)]%
        {salvi2023reliability}
\bibfield{author}{\bibinfo{person}{Davide Salvi}, \bibinfo{person}{Paolo Bestagini}, {and} \bibinfo{person}{Stefano Tubaro}.} \bibinfo{year}{2023}\natexlab{a}.
\newblock \showarticletitle{Reliability Estimation for Synthetic Speech Detection}. In \bibinfo{booktitle}{\emph{ICASSP 2023-2023 IEEE International Conference on Acoustics, Speech and Signal Processing (ICASSP)}}. IEEE, \bibinfo{pages}{1--5}.
\newblock


\bibitem[Salvi et~al\mbox{.}(2023b)]%
        {salvi2023timit}
\bibfield{author}{\bibinfo{person}{Davide Salvi}, \bibinfo{person}{Brian Hosler}, \bibinfo{person}{Paolo Bestagini}, \bibinfo{person}{Matthew~C Stamm}, {and} \bibinfo{person}{Stefano Tubaro}.} \bibinfo{year}{2023}\natexlab{b}.
\newblock \showarticletitle{TIMIT-TTS: a Text-to-Speech Dataset for Multimodal Synthetic Media Detection}.
\newblock \bibinfo{journal}{\emph{IEEE Access}} (\bibinfo{year}{2023}).
\newblock


\bibitem[Service(2023)]%
        {misinfo}
\bibfield{author}{\bibinfo{person}{Canadian Security~Intelligence Service}.} \bibinfo{year}{2023}\natexlab{}.
\newblock \showarticletitle{Deepfakes: A Real Threat to a Canadian Future}.
\newblock In \bibinfo{booktitle}{\emph{The evolution of disinformation: A deepfake future}}. \bibinfo{publisher}{Canadian Security Intelligence Service}, \bibinfo{pages}{11--18}.
\newblock


\bibitem[Shim et~al\mbox{.}(2022a)]%
        {shim2022graph}
\bibfield{author}{\bibinfo{person}{Hye-jin Shim}, \bibinfo{person}{Jungwoo Heo}, \bibinfo{person}{Jae-Han Park}, \bibinfo{person}{Ga-Hui Lee}, {and} \bibinfo{person}{Ha-Jin Yu}.} \bibinfo{year}{2022}\natexlab{a}.
\newblock \showarticletitle{Graph attentive feature aggregation for text-independent speaker verification}. In \bibinfo{booktitle}{\emph{ICASSP 2022-2022 IEEE International Conference on Acoustics, Speech and Signal Processing (ICASSP)}}. IEEE, \bibinfo{pages}{7972--7976}.
\newblock


\bibitem[Shim et~al\mbox{.}(2023)]%
        {Shi2023sharpness}
\bibfield{author}{\bibinfo{person}{Hye-jin Shim}, \bibinfo{person}{Jee-weon Jung}, {and} \bibinfo{person}{Tomi Kinnunen}.} \bibinfo{year}{2023}\natexlab{}.
\newblock \showarticletitle{Multi-dataset co-training with sharpness-aware optimization for audio anti-spoofing}.
\newblock \bibinfo{journal}{\emph{INTERSPEECH 2023}} (\bibinfo{date}{Aug} \bibinfo{year}{2023}).
\newblock


\bibitem[Shim et~al\mbox{.}(2024)]%
        {shim2024dcf}
\bibfield{author}{\bibinfo{person}{Hye-jin Shim}, \bibinfo{person}{Jee-weon Jung}, \bibinfo{person}{Tomi Kinnunen}, \bibinfo{person}{Nicholas Evans}, \bibinfo{person}{Jean-Francois Bonastre}, {and} \bibinfo{person}{Itshak Lapidot}.} \bibinfo{year}{2024}\natexlab{}.
\newblock \showarticletitle{a-DCF: an architecture agnostic metric with application to spoofing-robust speaker verification}.
\newblock \bibinfo{journal}{\emph{arXiv preprint arXiv:2403.01355}} (\bibinfo{year}{2024}).
\newblock


\bibitem[Shim et~al\mbox{.}(2022b)]%
        {Shim2022SASV}
\bibfield{author}{\bibinfo{person}{Hye-jin Shim}, \bibinfo{person}{Hemlata Tak}, \bibinfo{person}{Xuechen Liu}, \bibinfo{person}{Hee-Soo Heo}, \bibinfo{person}{Jee-weon Jung}, \bibinfo{person}{Joon~Son Chung}, \bibinfo{person}{Soo-Whan Chung}, \bibinfo{person}{Ha-Jin Yu}, \bibinfo{person}{Bong-Jin Lee}, \bibinfo{person}{Massimiliano Todisco}, {and} \bibinfo{person}{et al.}} \bibinfo{year}{2022}\natexlab{b}.
\newblock \showarticletitle{Baseline Systems for the first spoofing-aware speaker Verification Challenge: Score and Embedding Fusion}.
\newblock \bibinfo{journal}{\emph{The Speaker and Language Recognition Workshop (Odyssey 2022)}} (\bibinfo{date}{Jun} \bibinfo{year}{2022}).
\newblock


\bibitem[Shirvanian et~al\mbox{.}(2020)]%
        {shirvanian2020voicefox}
\bibfield{author}{\bibinfo{person}{Maliheh Shirvanian}, \bibinfo{person}{Manar Mohammed}, \bibinfo{person}{Nitesh Saxena}, {and} \bibinfo{person}{S~Abhishek Anand}.} \bibinfo{year}{2020}\natexlab{}.
\newblock \showarticletitle{Voicefox: Leveraging inbuilt transcription to enhance the security of machine-human speaker verification against voice synthesis attacks}. In \bibinfo{booktitle}{\emph{Proceedings of the 36th Annual Computer Security Applications Conference}}. \bibinfo{pages}{870--883}.
\newblock


\bibitem[Snyder et~al\mbox{.}(2017)]%
        {snyder17_interspeech}
\bibfield{author}{\bibinfo{person}{David Snyder}, \bibinfo{person}{Daniel Garcia-Romero}, \bibinfo{person}{Daniel Povey}, {and} \bibinfo{person}{Sanjeev Khudanpur}.} \bibinfo{year}{2017}\natexlab{}.
\newblock \showarticletitle{{Deep Neural Network Embeddings for Text-Independent Speaker Verification}}. In \bibinfo{booktitle}{\emph{Proc. Interspeech 2017}}. \bibinfo{pages}{999--1003}.
\newblock


\bibitem[Sriskandaraja et~al\mbox{.}(2016)]%
        {Sriskandaraja2016}
\bibfield{author}{\bibinfo{person}{Kaavya Sriskandaraja}, \bibinfo{person}{Vidhyasaharan Sethu}, \bibinfo{person}{Phu~Ngoc Le}, {and} \bibinfo{person}{Eliathamby Ambikairajah}.} \bibinfo{year}{2016}\natexlab{}.
\newblock \showarticletitle{Investigation of sub-band discriminative information between spoofed and genuine speech}.
\newblock \bibinfo{journal}{\emph{Interspeech 2016}} (\bibinfo{date}{Sep} \bibinfo{year}{2016}).
\newblock


\bibitem[Tak et~al\mbox{.}(2021a)]%
        {Tak2021Graph}
\bibfield{author}{\bibinfo{person}{Hemlata Tak}, \bibinfo{person}{Jee-weon Jung}, \bibinfo{person}{Jose Patino}, \bibinfo{person}{Massimiliano Todisco}, {and} \bibinfo{person}{Nicholas Evans}.} \bibinfo{year}{2021}\natexlab{a}.
\newblock \showarticletitle{Graph attention networks for anti-spoofing}.
\newblock \bibinfo{journal}{\emph{Interspeech 2021}} (\bibinfo{date}{Aug} \bibinfo{year}{2021}).
\newblock


\bibitem[Tak et~al\mbox{.}(2022a)]%
        {tak2022rawboost}
\bibfield{author}{\bibinfo{person}{Hemlata Tak}, \bibinfo{person}{Madhu Kamble}, \bibinfo{person}{Jose Patino}, \bibinfo{person}{Massimiliano Todisco}, {and} \bibinfo{person}{Nicholas Evans}.} \bibinfo{year}{2022}\natexlab{a}.
\newblock \showarticletitle{Rawboost: A raw data boosting and augmentation method applied to automatic speaker verification anti-spoofing}. In \bibinfo{booktitle}{\emph{ICASSP 2022-2022 IEEE International Conference on Acoustics, Speech and Signal Processing (ICASSP)}}. IEEE, \bibinfo{pages}{6382--6386}.
\newblock


\bibitem[Tak et~al\mbox{.}(2020)]%
        {Tak2020CQCC}
\bibfield{author}{\bibinfo{person}{Hemlata Tak}, \bibinfo{person}{Jose Patino}, \bibinfo{person}{Andreas Nautsch}, \bibinfo{person}{Nicholas W.~D. Evans}, {and} \bibinfo{person}{Massimiliano Todisco}.} \bibinfo{year}{2020}\natexlab{}.
\newblock \showarticletitle{An Explainability Study of the Constant Q Cepstral Coefficient Spoofing Countermeasure for Automatic Speaker Verification}. In \bibinfo{booktitle}{\emph{Odyssey 2020: The Speaker and Language Recognition Workshop, 1-5 November 2020, Tokyo, Japan}}, \bibfield{editor}{\bibinfo{person}{Kong-Aik Lee}, \bibinfo{person}{Takafumi Koshinaka}, {and} \bibinfo{person}{Koichi Shinoda}} (Eds.). \bibinfo{publisher}{ISCA}, \bibinfo{pages}{333--340}.
\newblock


\bibitem[Tak et~al\mbox{.}(2021b)]%
        {tak2021end}
\bibfield{author}{\bibinfo{person}{Hemlata Tak}, \bibinfo{person}{Jose Patino}, \bibinfo{person}{Massimiliano Todisco}, \bibinfo{person}{Andreas Nautsch}, \bibinfo{person}{Nicholas Evans}, {and} \bibinfo{person}{Anthony Larcher}.} \bibinfo{year}{2021}\natexlab{b}.
\newblock \showarticletitle{End-to-end anti-spoofing with rawnet2}. In \bibinfo{booktitle}{\emph{ICASSP 2021-2021 IEEE International Conference on Acoustics, Speech and Signal Processing (ICASSP)}}. IEEE, \bibinfo{pages}{6369--6373}.
\newblock


\bibitem[Tak et~al\mbox{.}(2022b)]%
        {Tak2022autometic}
\bibfield{author}{\bibinfo{person}{Hemlata Tak}, \bibinfo{person}{Massimiliano Todisco}, \bibinfo{person}{Xin Wang}, \bibinfo{person}{Jee-weon Jung}, \bibinfo{person}{Junichi Yamagishi}, {and} \bibinfo{person}{Nicholas Evans}.} \bibinfo{year}{2022}\natexlab{b}.
\newblock \showarticletitle{Automatic speaker verification spoofing and Deepfake detection using WAV2VEC 2.0 and data augmentation}.
\newblock \bibinfo{journal}{\emph{The Speaker and Language Recognition Workshop (Odyssey 2022)}} (\bibinfo{date}{Jun} \bibinfo{year}{2022}).
\newblock


\bibitem[Tak et~al\mbox{.}(2021c)]%
        {tak21_asvspoof}
\bibfield{author}{\bibinfo{person}{Hemlata Tak}, \bibinfo{person}{Jee weon Jung}, \bibinfo{person}{Jose Patino}, \bibinfo{person}{Madhu Kamble}, \bibinfo{person}{Massimiliano Todisco}, {and} \bibinfo{person}{Nicholas Evans}.} \bibinfo{year}{2021}\natexlab{c}.
\newblock \showarticletitle{End-to-end spectro-temporal graph attention networks for speaker verification anti-spoofing and speech deepfake detection}. In \bibinfo{booktitle}{\emph{2021 Edition of the Automatic Speaker Verification and Spoofing Countermeasures Challenge}}. \bibinfo{pages}{1--8}.
\newblock


\bibitem[Tamayo~Fl{\'o}rez et~al\mbox{.}(2022)]%
        {tamayo2022voice}
\bibfield{author}{\bibinfo{person}{Pablo~Andr{\'e}s Tamayo~Fl{\'o}rez} {et~al\mbox{.}}} \bibinfo{year}{2022}\natexlab{}.
\newblock \showarticletitle{Voice anti-spoofing data-set built from Latin American Spanish accents implementing voice conversion and text-to-speech techniques}.
\newblock  (\bibinfo{year}{2022}).
\newblock


\bibitem[Teng et~al\mbox{.}(2022a)]%
        {Teng2022SASASV}
\bibfield{author}{\bibinfo{person}{Zhongwei Teng}, \bibinfo{person}{Quchen Fu}, \bibinfo{person}{Jules White}, \bibinfo{person}{Maria Powell}, {and} \bibinfo{person}{Douglas Schmidt}.} \bibinfo{year}{2022}\natexlab{a}.
\newblock \showarticletitle{Sa-SASV: An end-to-end spoof-aggregated spoofing-aware speaker verification system}.
\newblock \bibinfo{journal}{\emph{Interspeech 2022}} (\bibinfo{date}{Sep} \bibinfo{year}{2022}).
\newblock


\bibitem[Teng et~al\mbox{.}(2022b)]%
        {teng2022arawnet}
\bibfield{author}{\bibinfo{person}{Zhongwei Teng}, \bibinfo{person}{Quchen Fu}, \bibinfo{person}{Jules White}, \bibinfo{person}{Maria~E Powell}, {and} \bibinfo{person}{Douglas~C Schmidt}.} \bibinfo{year}{2022}\natexlab{b}.
\newblock \showarticletitle{ARawNet: A lightweight solution for leveraging raw waveforms in spoof speech detection}. In \bibinfo{booktitle}{\emph{2022 26th International Conference on Pattern Recognition (ICPR)}}. IEEE, \bibinfo{pages}{692--698}.
\newblock


\bibitem[Tomilov et~al\mbox{.}(2021)]%
        {Tomilov2021}
\bibfield{author}{\bibinfo{person}{Anton Tomilov}, \bibinfo{person}{Aleksei Svishchev}, \bibinfo{person}{Marina Volkova}, \bibinfo{person}{Artem Chirkovskiy}, \bibinfo{person}{Alexander Kondratev}, {and} \bibinfo{person}{Galina Lavrentyeva}.} \bibinfo{year}{2021}\natexlab{}.
\newblock \showarticletitle{STC antispoofing systems for the ASVSPOOF2021 challenge}.
\newblock \bibinfo{journal}{\emph{2021 Edition of the Automatic Speaker Verification and Spoofing Countermeasures Challenge}} (\bibinfo{date}{Sep} \bibinfo{year}{2021}).
\newblock


\bibitem[Wang et~al\mbox{.}(2022d)]%
        {wang2022fully}
\bibfield{author}{\bibinfo{person}{Chenglong Wang}, \bibinfo{person}{Jiangyan Yi}, \bibinfo{person}{Jianhua Tao}, \bibinfo{person}{Haiyang Sun}, \bibinfo{person}{Xun Chen}, \bibinfo{person}{Zhengkun Tian}, \bibinfo{person}{Haoxin Ma}, \bibinfo{person}{Cunhang Fan}, {and} \bibinfo{person}{Ruibo Fu}.} \bibinfo{year}{2022}\natexlab{d}.
\newblock \showarticletitle{Fully automated end-to-end fake audio detection}. In \bibinfo{booktitle}{\emph{Proceedings of the 1st International Workshop on Deepfake Detection for Audio Multimedia}}. \bibinfo{pages}{27--33}.
\newblock


\bibitem[Wang et~al\mbox{.}(2023b)]%
        {Wang2023prodoscic}
\bibfield{author}{\bibinfo{person}{Chenglong Wang}, \bibinfo{person}{Jiangyan Yi}, \bibinfo{person}{Jianhua Tao}, \bibinfo{person}{Chu~Yuan Zhang}, \bibinfo{person}{Shuai Zhang}, {and} \bibinfo{person}{Xun Chen}.} \bibinfo{year}{2023}\natexlab{b}.
\newblock \showarticletitle{Detection of cross-dataset fake audio based on prosodic and pronunciation features}.
\newblock \bibinfo{journal}{\emph{INTERSPEECH 2023}} (\bibinfo{date}{Aug} \bibinfo{year}{2023}).
\newblock


\bibitem[Wang et~al\mbox{.}(2023c)]%
        {wang2023low}
\bibfield{author}{\bibinfo{person}{Chenglong Wang}, \bibinfo{person}{Jiangyan Yi}, \bibinfo{person}{Xiaohui Zhang}, \bibinfo{person}{Jianhua Tao}, \bibinfo{person}{Le Xu}, {and} \bibinfo{person}{Ruibo Fu}.} \bibinfo{year}{2023}\natexlab{c}.
\newblock \showarticletitle{Low-rank adaptation method for wav2vec2-based fake audio detection}.
\newblock \bibinfo{journal}{\emph{Proceedings of IJCAI 2023 Workshop on Deepfake Audio Detection and Analysis (DADA 2023)}} (\bibinfo{year}{2023}).
\newblock


\bibitem[Wang et~al\mbox{.}(2015)]%
        {wang2015relative}
\bibfield{author}{\bibinfo{person}{Longbiao Wang}, \bibinfo{person}{Yohei Yoshida}, \bibinfo{person}{Yuta Kawakami}, {and} \bibinfo{person}{Seiichi Nakagawa}.} \bibinfo{year}{2015}\natexlab{}.
\newblock \showarticletitle{Relative phase information for detecting human speech and spoofed speech.}. In \bibinfo{booktitle}{\emph{INTERSPEECH}}. \bibinfo{pages}{2092--2096}.
\newblock


\bibitem[Wang et~al\mbox{.}(2023a)]%
        {wang2023quantum}
\bibfield{author}{\bibinfo{person}{Ruoyu Wang}, \bibinfo{person}{Jun Du}, {and} \bibinfo{person}{Tian Gao}.} \bibinfo{year}{2023}\natexlab{a}.
\newblock \showarticletitle{Quantum transfer learning using the large-scale unsupervised pre-trained model wavlm-large for synthetic speech detection}. In \bibinfo{booktitle}{\emph{ICASSP 2023-2023 IEEE International Conference on Acoustics, Speech and Signal Processing (ICASSP)}}. IEEE, \bibinfo{pages}{1--5}.
\newblock


\bibitem[Wang et~al\mbox{.}(2022a)]%
        {wang2022multi}
\bibfield{author}{\bibinfo{person}{Ruoyu Wang}, \bibinfo{person}{Jun Du}, {and} \bibinfo{person}{Chang Wang}.} \bibinfo{year}{2022}\natexlab{a}.
\newblock \showarticletitle{Multi-branch Network with Circle Loss Using Voice Conversion and Channel Robust Data Augmentation for Synthetic Speech Detection}. In \bibinfo{booktitle}{\emph{Chinese Conference on Biometric Recognition}}. Springer, \bibinfo{pages}{613--620}.
\newblock


\bibitem[Wang et~al\mbox{.}(2024a)]%
        {wang2024asvspoof}
\bibfield{author}{\bibinfo{person}{Xin Wang}, \bibinfo{person}{Hector Delgado}, \bibinfo{person}{Hemlata Tak}, \bibinfo{person}{Jee-weon Jung}, \bibinfo{person}{Hye-jin Shim}, \bibinfo{person}{Massimiliano Todisco}, \bibinfo{person}{Ivan Kukanov}, \bibinfo{person}{Xuechen Liu}, \bibinfo{person}{Md Sahidullah}, \bibinfo{person}{Tomi Kinnunen}, {et~al\mbox{.}}} \bibinfo{year}{2024}\natexlab{a}.
\newblock \showarticletitle{ASVspoof 5: Crowdsourced speech data, deepfakes, and adversarial attacks at scale}.
\newblock \bibinfo{journal}{\emph{arXiv preprint arXiv:2408.08739}} (\bibinfo{year}{2024}).
\newblock


\bibitem[Wang et~al\mbox{.}(2022b)]%
        {Wang2022DKU}
\bibfield{author}{\bibinfo{person}{Xingming Wang}, \bibinfo{person}{Xiaoyi Qin}, \bibinfo{person}{Yikang Wang}, \bibinfo{person}{Yunfei Xu}, {and} \bibinfo{person}{Ming Li}.} \bibinfo{year}{2022}\natexlab{b}.
\newblock \showarticletitle{The DKU-oppo system for the 2022 spoofing-aware speaker Verification Challenge}.
\newblock \bibinfo{journal}{\emph{Interspeech 2022}} (\bibinfo{date}{Sep} \bibinfo{year}{2022}).
\newblock


\bibitem[Wang et~al\mbox{.}(2024b)]%
        {wang2024speechx}
\bibfield{author}{\bibinfo{person}{Xiaofei Wang}, \bibinfo{person}{Manthan Thakker}, \bibinfo{person}{Zhuo Chen}, \bibinfo{person}{Naoyuki Kanda}, \bibinfo{person}{Sefik~Emre Eskimez}, \bibinfo{person}{Sanyuan Chen}, \bibinfo{person}{Min Tang}, \bibinfo{person}{Shujie Liu}, \bibinfo{person}{Jinyu Li}, {and} \bibinfo{person}{Takuya Yoshioka}.} \bibinfo{year}{2024}\natexlab{b}.
\newblock \showarticletitle{Speechx: Neural codec language model as a versatile speech transformer}.
\newblock \bibinfo{journal}{\emph{IEEE/ACM Transactions on Audio, Speech, and Language Processing}} (\bibinfo{year}{2024}).
\newblock


\bibitem[Wang and Yamagishi(2021)]%
        {Wang2021Comparative}
\bibfield{author}{\bibinfo{person}{Xin Wang} {and} \bibinfo{person}{Junichi Yamagishi}.} \bibinfo{year}{2021}\natexlab{}.
\newblock \showarticletitle{A comparative study on recent neural spoofing countermeasures for synthetic speech detection}.
\newblock \bibinfo{journal}{\emph{Interspeech 2021}} (\bibinfo{date}{Aug} \bibinfo{year}{2021}).
\newblock


\bibitem[Wang and Yamagishi(2022a)]%
        {Wang2022SSL}
\bibfield{author}{\bibinfo{person}{Xin Wang} {and} \bibinfo{person}{Junichi Yamagishi}.} \bibinfo{year}{2022}\natexlab{a}.
\newblock \showarticletitle{Investigating self-supervised front ends for speech spoofing countermeasures}.
\newblock \bibinfo{journal}{\emph{The Speaker and Language Recognition Workshop (Odyssey 2022)}} (\bibinfo{date}{Jun} \bibinfo{year}{2022}).
\newblock


\bibitem[Wang and Yamagishi(2022b)]%
        {wang2022practical}
\bibfield{author}{\bibinfo{person}{Xin Wang} {and} \bibinfo{person}{Junichi Yamagishi}.} \bibinfo{year}{2022}\natexlab{b}.
\newblock \showarticletitle{A practical guide to logical access voice presentation attack detection}.
\newblock In \bibinfo{booktitle}{\emph{Frontiers in Fake Media Generation and Detection}}. \bibinfo{publisher}{Springer}, \bibinfo{pages}{169--214}.
\newblock


\bibitem[Wang and Yamagishi(2023)]%
        {wang2023investigating}
\bibfield{author}{\bibinfo{person}{Xin Wang} {and} \bibinfo{person}{Junichi Yamagishi}.} \bibinfo{year}{2023}\natexlab{}.
\newblock \showarticletitle{Investigating active-learning-based training data selection for speech spoofing countermeasure}. In \bibinfo{booktitle}{\emph{2022 IEEE Spoken Language Technology Workshop (SLT)}}. IEEE, \bibinfo{pages}{585--592}.
\newblock


\bibitem[Wang and Yamagishi(2024)]%
        {Wang2024vocoderSSL}
\bibfield{author}{\bibinfo{person}{Xin Wang} {and} \bibinfo{person}{Junichi Yamagishi}.} \bibinfo{year}{2024}\natexlab{}.
\newblock \showarticletitle{Can large-scale vocoded spoofed data improve speech spoofing countermeasure with a self-supervised front end?}
\newblock \bibinfo{journal}{\emph{ICASSP 2024 - 2024 IEEE International Conference on Acoustics, Speech and Signal Processing (ICASSP)}} (\bibinfo{date}{Apr} \bibinfo{year}{2024}).
\newblock


\bibitem[Wang et~al\mbox{.}(2020)]%
        {wang2020densely}
\bibfield{author}{\bibinfo{person}{Zheng Wang}, \bibinfo{person}{Sanshuai Cui}, \bibinfo{person}{Xiangui Kang}, \bibinfo{person}{Wei Sun}, {and} \bibinfo{person}{Zhonghua Li}.} \bibinfo{year}{2020}\natexlab{}.
\newblock \showarticletitle{Densely connected convolutional network for audio spoofing detection}. In \bibinfo{booktitle}{\emph{2020 Asia-Pacific Signal and Information Processing Association Annual Summit and Conference (APSIPA ASC)}}. IEEE, \bibinfo{pages}{1352--1360}.
\newblock


\bibitem[Wang et~al\mbox{.}(2022c)]%
        {wang2022npu}
\bibfield{author}{\bibinfo{person}{Ziqian Wang}, \bibinfo{person}{Qing Wang}, \bibinfo{person}{Jixun Yao}, {and} \bibinfo{person}{Lei Xie}.} \bibinfo{year}{2022}\natexlab{c}.
\newblock \showarticletitle{The NPU-ASLP System for Deepfake Algorithm Recognition in ADD 2023 Challenge}.
\newblock \bibinfo{journal}{\emph{Proceedings of IJCAI 2023 Workshop on Deepfake Audio Detection and Analysis (DADA 2023)}} (\bibinfo{year}{2022}).
\newblock


\bibitem[Wu et~al\mbox{.}(2022a)]%
        {Wu2022modelfusion}
\bibfield{author}{\bibinfo{person}{Haibin Wu}, \bibinfo{person}{Jiawen Kang}, \bibinfo{person}{Lingwei Meng}, \bibinfo{person}{Yang Zhang}, \bibinfo{person}{Xixin Wu}, \bibinfo{person}{Zhiyong Wu}, \bibinfo{person}{Hung-yi Lee}, {and} \bibinfo{person}{Helen Meng}.} \bibinfo{year}{2022}\natexlab{a}.
\newblock \showarticletitle{Tackling spoofing-aware speaker verification with multi-model fusion}.
\newblock \bibinfo{journal}{\emph{The Speaker and Language Recognition Workshop (Odyssey 2022)}} (\bibinfo{date}{Jun} \bibinfo{year}{2022}).
\newblock


\bibitem[Wu et~al\mbox{.}(2022b)]%
        {wu2022partiallyspan}
\bibfield{author}{\bibinfo{person}{Haibin Wu}, \bibinfo{person}{Heng-Cheng Kuo}, \bibinfo{person}{Naijun Zheng}, \bibinfo{person}{Kuo-Hsuan Hung}, \bibinfo{person}{Hung-Yi Lee}, \bibinfo{person}{Yu Tsao}, \bibinfo{person}{Hsin-Min Wang}, {and} \bibinfo{person}{Helen Meng}.} \bibinfo{year}{2022}\natexlab{b}.
\newblock \showarticletitle{Partially fake audio detection by self-attention-based fake span discovery}. In \bibinfo{booktitle}{\emph{ICASSP 2022-2022 IEEE International Conference on Acoustics, Speech and Signal Processing (ICASSP)}}. IEEE, \bibinfo{pages}{9236--9240}.
\newblock


\bibitem[Wu et~al\mbox{.}(2020b)]%
        {wu2020defense}
\bibfield{author}{\bibinfo{person}{Haibin Wu}, \bibinfo{person}{Songxiang Liu}, \bibinfo{person}{Helen Meng}, {and} \bibinfo{person}{Hung-yi Lee}.} \bibinfo{year}{2020}\natexlab{b}.
\newblock \showarticletitle{Defense against adversarial attacks on spoofing countermeasures of asv}. In \bibinfo{booktitle}{\emph{ICASSP 2020-2020 IEEE International Conference on Acoustics, Speech and Signal Processing (ICASSP)}}. IEEE, \bibinfo{pages}{6564--6568}.
\newblock


\bibitem[Wu et~al\mbox{.}(2022c)]%
        {Wu2022levelfusion}
\bibfield{author}{\bibinfo{person}{Haibin Wu}, \bibinfo{person}{Lingwei Meng}, \bibinfo{person}{Jiawen Kang}, \bibinfo{person}{Jinchao Li}, \bibinfo{person}{Xu Li}, \bibinfo{person}{Xixin Wu}, \bibinfo{person}{Hung-yi Lee}, {and} \bibinfo{person}{Helen Meng}.} \bibinfo{year}{2022}\natexlab{c}.
\newblock \showarticletitle{Spoofing-aware speaker verification by multi-level fusion}.
\newblock \bibinfo{journal}{\emph{Interspeech 2022}} (\bibinfo{date}{Sep} \bibinfo{year}{2022}).
\newblock


\bibitem[Wu et~al\mbox{.}(2024)]%
        {wu2024codecfake}
\bibfield{author}{\bibinfo{person}{Haibin Wu}, \bibinfo{person}{Yuan Tseng}, {and} \bibinfo{person}{Hung-yi Lee}.} \bibinfo{year}{2024}\natexlab{}.
\newblock \showarticletitle{CodecFake: Enhancing Anti-Spoofing Models Against Deepfake Audios from Codec-Based Speech Synthesis Systems}.
\newblock \bibinfo{journal}{\emph{arXiv preprint arXiv:2406.07237}} (\bibinfo{year}{2024}).
\newblock


\bibitem[Wu et~al\mbox{.}(2020a)]%
        {Wu2020FeatureGenuinization}
\bibfield{author}{\bibinfo{person}{Zhenzong Wu}, \bibinfo{person}{Rohan~Kumar Das}, \bibinfo{person}{Jichen Yang}, {and} \bibinfo{person}{Haizhou Li}.} \bibinfo{year}{2020}\natexlab{a}.
\newblock \showarticletitle{Light convolutional neural network with feature genuinization for detection of synthetic speech attacks}.
\newblock \bibinfo{journal}{\emph{Interspeech 2020}} (\bibinfo{date}{Oct} \bibinfo{year}{2020}).
\newblock


\bibitem[Wu et~al\mbox{.}(2015)]%
        {wu15e_interspeech}
\bibfield{author}{\bibinfo{person}{Zhizheng Wu}, \bibinfo{person}{Tomi Kinnunen}, \bibinfo{person}{Nicholas Evans}, \bibinfo{person}{Junichi Yamagishi}, \bibinfo{person}{Cemal Hanilçi}, \bibinfo{person}{Md. Sahidullah}, {and} \bibinfo{person}{Aleksandr Sizov}.} \bibinfo{year}{2015}\natexlab{}.
\newblock \showarticletitle{{ASVspoof 2015: the first automatic speaker verification spoofing and countermeasures challenge}}. In \bibinfo{booktitle}{\emph{Proc. Interspeech 2015}}. \bibinfo{pages}{2037--2041}.
\newblock
\showISSN{2958-1796}


\bibitem[Wu et~al\mbox{.}(2013)]%
        {wu2013synthetic}
\bibfield{author}{\bibinfo{person}{Zhizheng Wu}, \bibinfo{person}{Xiong Xiao}, \bibinfo{person}{Eng~Siong Chng}, {and} \bibinfo{person}{Haizhou Li}.} \bibinfo{year}{2013}\natexlab{}.
\newblock \showarticletitle{Synthetic speech detection using temporal modulation feature}. In \bibinfo{booktitle}{\emph{2013 IEEE International Conference on Acoustics, Speech and Signal Processing}}. \bibinfo{pages}{7234--7238}.
\newblock


\bibitem[Xiang et~al\mbox{.}(2023)]%
        {xiang2023extracting}
\bibfield{author}{\bibinfo{person}{Ziyue Xiang}, \bibinfo{person}{Amit Kumar~Singh Yadav}, \bibinfo{person}{Stefano Tubaro}, \bibinfo{person}{Paolo Bestagini}, {and} \bibinfo{person}{Edward~J Delp}.} \bibinfo{year}{2023}\natexlab{}.
\newblock \showarticletitle{Extracting efficient spectrograms from MP3 compressed speech signals for synthetic speech detection}. In \bibinfo{booktitle}{\emph{Proceedings of the 2023 ACM Workshop on Information Hiding and Multimedia Security}}. \bibinfo{pages}{163--168}.
\newblock


\bibitem[Xie et~al\mbox{.}(2023)]%
        {Xie2023generailized}
\bibfield{author}{\bibinfo{person}{Yuankun Xie}, \bibinfo{person}{Haonan Cheng}, \bibinfo{person}{Yutian Wang}, {and} \bibinfo{person}{Long Ye}.} \bibinfo{year}{2023}\natexlab{}.
\newblock \showarticletitle{Learning a self-supervised domain-invariant feature representation for generalized audio deepfake detection}.
\newblock \bibinfo{journal}{\emph{INTERSPEECH 2023}} (\bibinfo{date}{Aug} \bibinfo{year}{2023}).
\newblock


\bibitem[Xie et~al\mbox{.}(2024)]%
        {xie2024codecfake}
\bibfield{author}{\bibinfo{person}{Yuankun Xie}, \bibinfo{person}{Yi Lu}, \bibinfo{person}{Ruibo Fu}, \bibinfo{person}{Zhengqi Wen}, \bibinfo{person}{Zhiyong Wang}, \bibinfo{person}{Jianhua Tao}, \bibinfo{person}{Xin Qi}, \bibinfo{person}{Xiaopeng Wang}, \bibinfo{person}{Yukun Liu}, \bibinfo{person}{Haonan Cheng}, {et~al\mbox{.}}} \bibinfo{year}{2024}\natexlab{}.
\newblock \showarticletitle{The Codecfake Dataset and Countermeasures for the Universally Detection of Deepfake Audio}.
\newblock \bibinfo{journal}{\emph{arXiv preprint arXiv:2405.04880}} (\bibinfo{year}{2024}).
\newblock


\bibitem[Xie et~al\mbox{.}(2021)]%
        {Xie2021SENET}
\bibfield{author}{\bibinfo{person}{Yang Xie}, \bibinfo{person}{Zhenchuan Zhang}, {and} \bibinfo{person}{Yingchun Yang}.} \bibinfo{year}{2021}\natexlab{}.
\newblock \showarticletitle{Siamese network with wav2vec feature for spoofing speech detection}.
\newblock \bibinfo{journal}{\emph{Interspeech 2021}} (\bibinfo{date}{Aug} \bibinfo{year}{2021}).
\newblock


\bibitem[Xue et~al\mbox{.}(2022)]%
        {xue2022audio}
\bibfield{author}{\bibinfo{person}{Jun Xue}, \bibinfo{person}{Cunhang Fan}, \bibinfo{person}{Zhao Lv}, \bibinfo{person}{Jianhua Tao}, \bibinfo{person}{Jiangyan Yi}, \bibinfo{person}{Chengshi Zheng}, \bibinfo{person}{Zhengqi Wen}, \bibinfo{person}{Minmin Yuan}, {and} \bibinfo{person}{Shegang Shao}.} \bibinfo{year}{2022}\natexlab{}.
\newblock \showarticletitle{Audio deepfake detection based on a combination of f0 information and real plus imaginary spectrogram features}. In \bibinfo{booktitle}{\emph{Proceedings of the 1st International Workshop on Deepfake Detection for Audio Multimedia}}. \bibinfo{pages}{19--26}.
\newblock


\bibitem[Xue et~al\mbox{.}(2023a)]%
        {xue2023learning}
\bibfield{author}{\bibinfo{person}{Jun Xue}, \bibinfo{person}{Cunhang Fan}, \bibinfo{person}{Jiangyan Yi}, \bibinfo{person}{Chenglong Wang}, \bibinfo{person}{Zhengqi Wen}, \bibinfo{person}{Dan Zhang}, {and} \bibinfo{person}{Zhao Lv}.} \bibinfo{year}{2023}\natexlab{a}.
\newblock \showarticletitle{Learning from yourself: A self-distillation method for fake speech detection}. In \bibinfo{booktitle}{\emph{ICASSP 2023-2023 IEEE International Conference on Acoustics, Speech and Signal Processing (ICASSP)}}. IEEE, \bibinfo{pages}{1--5}.
\newblock


\bibitem[Xue et~al\mbox{.}(2023b)]%
        {xue2023cross}
\bibfield{author}{\bibinfo{person}{Junxiao Xue}, \bibinfo{person}{Hao Zhou}, \bibinfo{person}{Huawei Song}, \bibinfo{person}{Bin Wu}, {and} \bibinfo{person}{Lei Shi}.} \bibinfo{year}{2023}\natexlab{b}.
\newblock \showarticletitle{Cross-modal information fusion for voice spoofing detection}.
\newblock \bibinfo{journal}{\emph{Speech Communication}}  \bibinfo{volume}{147} (\bibinfo{year}{2023}), \bibinfo{pages}{41--50}.
\newblock


\bibitem[Yadav et~al\mbox{.}(2023)]%
        {yadav2023assd}
\bibfield{author}{\bibinfo{person}{Amit Kumar~Singh Yadav}, \bibinfo{person}{Ziyue Xiang}, \bibinfo{person}{Emily~R Bartusiak}, \bibinfo{person}{Paolo Bestagini}, \bibinfo{person}{Stefano Tubaro}, {and} \bibinfo{person}{Edward~J Delp}.} \bibinfo{year}{2023}\natexlab{}.
\newblock \showarticletitle{ASSD: Synthetic Speech Detection in the AAC Compressed Domain}. In \bibinfo{booktitle}{\emph{ICASSP 2023-2023 IEEE International Conference on Acoustics, Speech and Signal Processing (ICASSP)}}. IEEE, \bibinfo{pages}{1--5}.
\newblock


\bibitem[Yamagishi et~al\mbox{.}(2019a)]%
        {yamagishi2019asvspoof}
\bibfield{author}{\bibinfo{person}{Junichi Yamagishi}, \bibinfo{person}{Massimiliano Todisco}, \bibinfo{person}{Md Sahidullah}, \bibinfo{person}{H{\'e}ctor Delgado}, \bibinfo{person}{Xin Wang}, \bibinfo{person}{Nicolas Evans}, \bibinfo{person}{Tomi Kinnunen}, \bibinfo{person}{Kong~Aik Lee}, \bibinfo{person}{Ville Vestman}, {and} \bibinfo{person}{Andreas Nautsch}.} \bibinfo{year}{2019}\natexlab{a}.
\newblock \showarticletitle{Asvspoof 2019: The 3rd automatic speaker verification spoofing and countermeasures challenge database}.
\newblock  (\bibinfo{year}{2019}).
\newblock


\bibitem[Yamagishi et~al\mbox{.}(2019b)]%
        {yamagishi2019cstr}
\bibfield{author}{\bibinfo{person}{Junichi Yamagishi}, \bibinfo{person}{Christophe Veaux}, \bibinfo{person}{Kirsten MacDonald}, {et~al\mbox{.}}} \bibinfo{year}{2019}\natexlab{b}.
\newblock \showarticletitle{Cstr vctk corpus: English multi-speaker corpus for cstr voice cloning toolkit (version 0.92)}.
\newblock \bibinfo{journal}{\emph{University of Edinburgh. The Centre for Speech Technology Research (CSTR)}} (\bibinfo{year}{2019}).
\newblock


\bibitem[Yan et~al\mbox{.}(2022)]%
        {yan2022initial}
\bibfield{author}{\bibinfo{person}{Xinrui Yan}, \bibinfo{person}{Jiangyan Yi}, \bibinfo{person}{Jianhua Tao}, \bibinfo{person}{Chenglong Wang}, \bibinfo{person}{Haoxin Ma}, \bibinfo{person}{Tao Wang}, \bibinfo{person}{Shiming Wang}, {and} \bibinfo{person}{Ruibo Fu}.} \bibinfo{year}{2022}\natexlab{}.
\newblock \showarticletitle{An initial investigation for detecting vocoder fingerprints of fake audio}. In \bibinfo{booktitle}{\emph{Proceedings of the 1st International Workshop on Deepfake Detection for Audio Multimedia}}. \bibinfo{pages}{61--68}.
\newblock


\bibitem[Yang and Das(2020)]%
        {yang2020long}
\bibfield{author}{\bibinfo{person}{Jichen Yang} {and} \bibinfo{person}{Rohan~Kumar Das}.} \bibinfo{year}{2020}\natexlab{}.
\newblock \showarticletitle{Long-term high frequency features for synthetic speech detection}.
\newblock \bibinfo{journal}{\emph{Digital Signal Processing}}  \bibinfo{volume}{97} (\bibinfo{year}{2020}), \bibinfo{pages}{102622}.
\newblock


\bibitem[Yang et~al\mbox{.}(2018)]%
        {yang2018extended}
\bibfield{author}{\bibinfo{person}{Jichen Yang}, \bibinfo{person}{Rohan~Kumar Das}, {and} \bibinfo{person}{Haizhou Li}.} \bibinfo{year}{2018}\natexlab{}.
\newblock \showarticletitle{Extended constant-Q cepstral coefficients for detection of spoofing attacks}. In \bibinfo{booktitle}{\emph{2018 Asia-Pacific Signal and Information Processing Association Annual Summit and Conference (APSIPA ASC)}}. IEEE, \bibinfo{pages}{1024--1029}.
\newblock


\bibitem[Yang et~al\mbox{.}(2021)]%
        {yang2021modified}
\bibfield{author}{\bibinfo{person}{Jichen Yang}, \bibinfo{person}{Hongji Wang}, \bibinfo{person}{Rohan~Kumar Das}, {and} \bibinfo{person}{Yanmin Qian}.} \bibinfo{year}{2021}\natexlab{}.
\newblock \showarticletitle{Modified magnitude-phase spectrum information for spoofing detection}.
\newblock \bibinfo{journal}{\emph{IEEE/ACM Transactions on Audio, Speech, and Language Processing}}  \bibinfo{volume}{29} (\bibinfo{year}{2021}), \bibinfo{pages}{1065--1078}.
\newblock


\bibitem[Yang et~al\mbox{.}(2023)]%
        {yang2023comparative}
\bibfield{author}{\bibinfo{person}{Minjiao Yang}, \bibinfo{person}{Kangfeng Zheng}, \bibinfo{person}{Xiujuan Wang}, \bibinfo{person}{Yudao Sun}, {and} \bibinfo{person}{Zhe Chen}.} \bibinfo{year}{2023}\natexlab{}.
\newblock \showarticletitle{Comparative analysis of asv spoofing countermeasures: Evaluating res2net-based approaches}.
\newblock \bibinfo{journal}{\emph{IEEE Signal Processing Letters}} (\bibinfo{year}{2023}).
\newblock


\bibitem[Yi et~al\mbox{.}(2021)]%
        {yi21_interspeech}
\bibfield{author}{\bibinfo{person}{Jiangyan Yi}, \bibinfo{person}{Ye Bai}, \bibinfo{person}{Jianhua Tao}, \bibinfo{person}{Haoxin Ma}, \bibinfo{person}{Zhengkun Tian}, \bibinfo{person}{Chenglong Wang}, \bibinfo{person}{Tao Wang}, {and} \bibinfo{person}{Ruibo Fu}.} \bibinfo{year}{2021}\natexlab{}.
\newblock \showarticletitle{Half-Truth: A Partially Fake Audio Detection Dataset}. In \bibinfo{booktitle}{\emph{Interspeech 2021}}. \bibinfo{pages}{1654--1658}.
\newblock
\showISSN{2958-1796}


\bibitem[Yi et~al\mbox{.}(2022)]%
        {yi2022add}
\bibfield{author}{\bibinfo{person}{Jiangyan Yi}, \bibinfo{person}{Ruibo Fu}, \bibinfo{person}{Jianhua Tao}, \bibinfo{person}{Shuai Nie}, \bibinfo{person}{Haoxin Ma}, \bibinfo{person}{Chenglong Wang}, \bibinfo{person}{Tao Wang}, \bibinfo{person}{Zhengkun Tian}, \bibinfo{person}{Ye Bai}, \bibinfo{person}{Cunhang Fan}, {et~al\mbox{.}}} \bibinfo{year}{2022}\natexlab{}.
\newblock \showarticletitle{Add 2022: the first audio deep synthesis detection challenge}. In \bibinfo{booktitle}{\emph{ICASSP 2022-2022 IEEE International Conference on Acoustics, Speech and Signal Processing (ICASSP)}}. IEEE, \bibinfo{pages}{9216--9220}.
\newblock


\bibitem[Yi et~al\mbox{.}(2023a)]%
        {yi2023add}
\bibfield{author}{\bibinfo{person}{Jiangyan Yi}, \bibinfo{person}{Jianhua Tao}, \bibinfo{person}{Ruibo Fu}, \bibinfo{person}{Xinrui Yan}, \bibinfo{person}{Chenglong Wang}, \bibinfo{person}{Tao Wang}, \bibinfo{person}{Chu~Yuan Zhang}, \bibinfo{person}{Xiaohui Zhang}, \bibinfo{person}{Yan Zhao}, \bibinfo{person}{Yong Ren}, {et~al\mbox{.}}} \bibinfo{year}{2023}\natexlab{a}.
\newblock \showarticletitle{ADD 2023: the Second Audio Deepfake Detection Challenge}.
\newblock \bibinfo{journal}{\emph{arXiv preprint arXiv:2305.13774}} (\bibinfo{year}{2023}).
\newblock


\bibitem[Yi et~al\mbox{.}(2023b)]%
        {yi2023audio}
\bibfield{author}{\bibinfo{person}{Jiangyan Yi}, \bibinfo{person}{Chenglong Wang}, \bibinfo{person}{Jianhua Tao}, \bibinfo{person}{Xiaohui Zhang}, \bibinfo{person}{Chu~Yuan Zhang}, {and} \bibinfo{person}{Yan Zhao}.} \bibinfo{year}{2023}\natexlab{b}.
\newblock \showarticletitle{Audio deepfake detection: A survey}.
\newblock \bibinfo{journal}{\emph{arXiv preprint arXiv:2308.14970}} (\bibinfo{year}{2023}).
\newblock


\bibitem[Yi et~al\mbox{.}(2020)]%
        {yi2020voice}
\bibfield{author}{\bibinfo{person}{Zhao Yi}, \bibinfo{person}{Wen-Chin Huang}, \bibinfo{person}{Xiaohai Tian}, \bibinfo{person}{Junichi Yamagishi}, \bibinfo{person}{Rohan~Kumar Das}, \bibinfo{person}{Tomi Kinnunen}, \bibinfo{person}{Zhenhua Ling}, {and} \bibinfo{person}{Tomoki Toda}.} \bibinfo{year}{2020}\natexlab{}.
\newblock \showarticletitle{Voice conversion challenge 2020—intra-lingual semi-parallel and cross-lingual voice conversion—}. In \bibinfo{booktitle}{\emph{Proc. Joint Workshop for the Blizzard Challenge and Voice Conversion Challenge}}, Vol.~\bibinfo{volume}{2020}. \bibinfo{pages}{80--98}.
\newblock


\bibitem[Yu et~al\mbox{.}(2016)]%
        {yu2016effect}
\bibfield{author}{\bibinfo{person}{Hong Yu}, \bibinfo{person}{Achintya Sarkar}, \bibinfo{person}{Dennis Alexander~Lehmann Thomsen}, \bibinfo{person}{Zheng-Hua Tan}, \bibinfo{person}{Zhanyu Ma}, {and} \bibinfo{person}{Jun Guo}.} \bibinfo{year}{2016}\natexlab{}.
\newblock \showarticletitle{Effect of multi-condition training and speech enhancement methods on spoofing detection}. In \bibinfo{booktitle}{\emph{2016 First International Workshop on Sensing, Processing and Learning for Intelligent Machines (SPLINE)}}. IEEE, \bibinfo{pages}{1--5}.
\newblock


\bibitem[Zeinali et~al\mbox{.}(2019)]%
        {Zeinali2019Detecting}
\bibfield{author}{\bibinfo{person}{Hossein Zeinali}, \bibinfo{person}{Themos Stafylakis}, \bibinfo{person}{Georgia Athanasopoulou}, \bibinfo{person}{Johan Rohdin}, \bibinfo{person}{Ioannis Gkinis}, \bibinfo{person}{Lukáš Burget}, {and} \bibinfo{person}{Jan Černocký}.} \bibinfo{year}{2019}\natexlab{}.
\newblock \showarticletitle{Detecting spoofing attacks using VGG and SincNet: But-omilia submission to ASVSPOOF 2019 challenge}.
\newblock \bibinfo{journal}{\emph{Interspeech 2019}} (\bibinfo{date}{Sep} \bibinfo{year}{2019}).
\newblock


\bibitem[Zhang and Sim(2022)]%
        {zhang2022localizing}
\bibfield{author}{\bibinfo{person}{Bowen Zhang} {and} \bibinfo{person}{Terence Sim}.} \bibinfo{year}{2022}\natexlab{}.
\newblock \showarticletitle{Localizing fake segments in speech}. In \bibinfo{booktitle}{\emph{2022 26th International Conference on Pattern Recognition (ICPR)}}. IEEE, \bibinfo{pages}{3224--3230}.
\newblock


\bibitem[Zhang et~al\mbox{.}(2023a)]%
        {zhang2023audio}
\bibfield{author}{\bibinfo{person}{Jiachen Zhang}, \bibinfo{person}{Guoqing Tu}, \bibinfo{person}{Shubo Liu}, {and} \bibinfo{person}{Zhaohui Cai}.} \bibinfo{year}{2023}\natexlab{a}.
\newblock \showarticletitle{Audio Anti-Spoofing Based on Audio Feature Fusion}.
\newblock \bibinfo{journal}{\emph{Algorithms}} \bibinfo{volume}{16}, \bibinfo{number}{7} (\bibinfo{year}{2023}), \bibinfo{pages}{317}.
\newblock


\bibitem[Zhang et~al\mbox{.}(2022b)]%
        {Zhang2022sasvbackend}
\bibfield{author}{\bibinfo{person}{Li Zhang}, \bibinfo{person}{Yue Li}, \bibinfo{person}{Huan Zhao}, \bibinfo{person}{Qing Wang}, {and} \bibinfo{person}{Lei Xie}.} \bibinfo{year}{2022}\natexlab{b}.
\newblock \showarticletitle{Backend ensemble for speaker verification and spoofing countermeasure}.
\newblock \bibinfo{journal}{\emph{Interspeech 2022}} (\bibinfo{date}{Sep} \bibinfo{year}{2022}).
\newblock


\bibitem[Zhang et~al\mbox{.}(2022c)]%
        {zhang2022partialspoof}
\bibfield{author}{\bibinfo{person}{Lin Zhang}, \bibinfo{person}{Xin Wang}, \bibinfo{person}{Erica Cooper}, \bibinfo{person}{Nicholas Evans}, {and} \bibinfo{person}{Junichi Yamagishi}.} \bibinfo{year}{2022}\natexlab{c}.
\newblock \showarticletitle{The partialspoof database and countermeasures for the detection of short fake speech segments embedded in an utterance}.
\newblock \bibinfo{journal}{\emph{IEEE/ACM Transactions on Audio, Speech, and Language Processing}}  \bibinfo{volume}{31} (\bibinfo{year}{2022}), \bibinfo{pages}{813--825}.
\newblock


\bibitem[Zhang et~al\mbox{.}(2023b)]%
        {zhang2023Range}
\bibfield{author}{\bibinfo{person}{Lin Zhang}, \bibinfo{person}{Xin Wang}, \bibinfo{person}{Erica Cooper}, \bibinfo{person}{Nicholas Evans}, {and} \bibinfo{person}{Junichi Yamagishi}.} \bibinfo{year}{2023}\natexlab{b}.
\newblock \showarticletitle{{Range-Based Equal Error Rate for Spoof Localization}}. In \bibinfo{booktitle}{\emph{Proc. INTERSPEECH 2023}}. \bibinfo{pages}{3212--3216}.
\newblock


\bibitem[Zhang et~al\mbox{.}(2021d)]%
        {zhang2021multitask}
\bibfield{author}{\bibinfo{person}{Lin Zhang}, \bibinfo{person}{Xin Wang}, \bibinfo{person}{Erica Cooper}, {and} \bibinfo{person}{Junichi Yamagishi}.} \bibinfo{year}{2021}\natexlab{d}.
\newblock \showarticletitle{Multi-task learning in utterance-level and segmental-level spoof detection}.
\newblock \bibinfo{journal}{\emph{2021 Edition of the Automatic Speaker Verification and Spoofing Countermeasures Challenge}} (\bibinfo{date}{Sep} \bibinfo{year}{2021}).
\newblock


\bibitem[Zhang et~al\mbox{.}(2021e)]%
        {zhang2021initialpartial}
\bibfield{author}{\bibinfo{person}{Lin Zhang}, \bibinfo{person}{Xin Wang}, \bibinfo{person}{Erica Cooper}, \bibinfo{person}{Junichi Yamagishi}, \bibinfo{person}{Jose Patino}, {and} \bibinfo{person}{Nicholas Evans}.} \bibinfo{year}{2021}\natexlab{e}.
\newblock \showarticletitle{An initial investigation for detecting partially spoofed audio}.
\newblock \bibinfo{journal}{\emph{Interspeech 2021}} (\bibinfo{date}{Aug} \bibinfo{year}{2021}).
\newblock


\bibitem[Zhang et~al\mbox{.}(2022a)]%
        {Zhang2022sasvscorefusion}
\bibfield{author}{\bibinfo{person}{Peng Zhang}, \bibinfo{person}{Peng Hu}, {and} \bibinfo{person}{Xueliang Zhang}.} \bibinfo{year}{2022}\natexlab{a}.
\newblock \showarticletitle{Norm-constrained score-level ensemble for spoofing aware speaker verification}.
\newblock \bibinfo{journal}{\emph{Interspeech 2022}} (\bibinfo{date}{Sep} \bibinfo{year}{2022}).
\newblock


\bibitem[Zhang(2022)]%
        {zhang2022deepfake}
\bibfield{author}{\bibinfo{person}{Tao Zhang}.} \bibinfo{year}{2022}\natexlab{}.
\newblock \showarticletitle{Deepfake generation and detection, a survey}.
\newblock \bibinfo{journal}{\emph{Multimedia Tools and Applications}} \bibinfo{volume}{81}, \bibinfo{number}{5} (\bibinfo{year}{2022}), \bibinfo{pages}{6259--6276}.
\newblock


\bibitem[Zhang et~al\mbox{.}(2023d)]%
        {zhang2023adaptive}
\bibfield{author}{\bibinfo{person}{Xiaohui Zhang}, \bibinfo{person}{Jiangyan Yi}, \bibinfo{person}{Jianhua Tao}, \bibinfo{person}{Chenlong Wang}, \bibinfo{person}{Le Xu}, {and} \bibinfo{person}{Ruibo Fu}.} \bibinfo{year}{2023}\natexlab{d}.
\newblock \showarticletitle{Adaptive Fake Audio Detection with Low-Rank Model Squeezing}.
\newblock \bibinfo{journal}{\emph{Proceedings of IJCAI 2023 Workshop on Deepfake Audio Detection and Analysis (DADA 2023)}} (\bibinfo{year}{2023}).
\newblock


\bibitem[Zhang et~al\mbox{.}(2023c)]%
        {zhang2023you}
\bibfield{author}{\bibinfo{person}{Xiaohui Zhang}, \bibinfo{person}{Jiangyan Yi}, \bibinfo{person}{Jianhua Tao}, \bibinfo{person}{Chenglong Wang}, {and} \bibinfo{person}{Chu~Yuan Zhang}.} \bibinfo{year}{2023}\natexlab{c}.
\newblock \showarticletitle{Do you remember? Overcoming catastrophic forgetting for fake audio detection}. In \bibinfo{booktitle}{\emph{International Conference on Machine Learning}}. PMLR, \bibinfo{pages}{41819--41831}.
\newblock


\bibitem[Zhang et~al\mbox{.}(2024b)]%
        {zhang2024remember}
\bibfield{author}{\bibinfo{person}{Xiaohui Zhang}, \bibinfo{person}{Jiangyan Yi}, \bibinfo{person}{Chenglong Wang}, \bibinfo{person}{Chu~Yuan Zhang}, \bibinfo{person}{Siding Zeng}, {and} \bibinfo{person}{Jianhua Tao}.} \bibinfo{year}{2024}\natexlab{b}.
\newblock \showarticletitle{What to remember: Self-adaptive continual learning for audio deepfake detection}. In \bibinfo{booktitle}{\emph{Proceedings of the AAAI Conference on Artificial Intelligence}}, Vol.~\bibinfo{volume}{38}. \bibinfo{pages}{19569--19577}.
\newblock


\bibitem[Zhang et~al\mbox{.}(2021b)]%
        {Zhang2021oneclass}
\bibfield{author}{\bibinfo{person}{You Zhang}, \bibinfo{person}{Fei Jiang}, {and} \bibinfo{person}{Zhiyao Duan}.} \bibinfo{year}{2021}\natexlab{b}.
\newblock \showarticletitle{One-class learning towards synthetic voice spoofing detection}.
\newblock \bibinfo{journal}{\emph{IEEE Signal Processing Letters}}  \bibinfo{volume}{28} (\bibinfo{year}{2021}), \bibinfo{pages}{937–941}.
\newblock


\bibitem[Zhang et~al\mbox{.}(2020)]%
        {zhang2020black}
\bibfield{author}{\bibinfo{person}{Yuekai Zhang}, \bibinfo{person}{Ziyan Jiang}, \bibinfo{person}{Jes{\'u}s Villalba}, {and} \bibinfo{person}{Najim Dehak}.} \bibinfo{year}{2020}\natexlab{}.
\newblock \showarticletitle{Black-Box Attacks on Spoofing Countermeasures Using Transferability of Adversarial Examples.}. In \bibinfo{booktitle}{\emph{Interspeech}}. \bibinfo{pages}{4238--4242}.
\newblock


\bibitem[Zhang et~al\mbox{.}(2024a)]%
        {Zhang2024temporal}
\bibfield{author}{\bibinfo{person}{Yuxiang Zhang}, \bibinfo{person}{Zhuo Li}, \bibinfo{person}{Jingze Lu}, \bibinfo{person}{Wenchao Wang}, {and} \bibinfo{person}{Pengyuan Zhang}.} \bibinfo{year}{2024}\natexlab{a}.
\newblock \showarticletitle{Synthetic speech detection based on the temporal consistency of speaker features}.
\newblock \bibinfo{journal}{\emph{IEEE Signal Processing Letters}}  \bibinfo{volume}{31} (\bibinfo{year}{2024}), \bibinfo{pages}{944–948}.
\newblock


\bibitem[Zhang et~al\mbox{.}(2021c)]%
        {Zhang2021silence}
\bibfield{author}{\bibinfo{person}{Yuxiang Zhang}, \bibinfo{person}{Wenchao Wang}, {and} \bibinfo{person}{Pengyuan Zhang}.} \bibinfo{year}{2021}\natexlab{c}.
\newblock \showarticletitle{The effect of silence and dual-band fusion in anti-spoofing system}.
\newblock \bibinfo{journal}{\emph{Interspeech 2021}} (\bibinfo{date}{Aug} \bibinfo{year}{2021}).
\newblock


\bibitem[Zhang et~al\mbox{.}(2022d)]%
        {Zhang2022sasvproba}
\bibfield{author}{\bibinfo{person}{You Zhang}, \bibinfo{person}{Ge Zhu}, {and} \bibinfo{person}{Zhiyao Duan}.} \bibinfo{year}{2022}\natexlab{d}.
\newblock \showarticletitle{A probabilistic fusion framework for spoofing aware speaker verification}.
\newblock \bibinfo{journal}{\emph{The Speaker and Language Recognition Workshop (Odyssey 2022)}} (\bibinfo{date}{Jun} \bibinfo{year}{2022}).
\newblock


\bibitem[Zhang et~al\mbox{.}(2021g)]%
        {Zhang2021channeleffect}
\bibfield{author}{\bibinfo{person}{You Zhang}, \bibinfo{person}{Ge Zhu}, \bibinfo{person}{Fei Jiang}, {and} \bibinfo{person}{Zhiyao Duan}.} \bibinfo{year}{2021}\natexlab{g}.
\newblock \showarticletitle{An empirical study on channel effects for synthetic voice spoofing countermeasure systems}.
\newblock \bibinfo{journal}{\emph{Interspeech 2021}} (\bibinfo{date}{Aug} \bibinfo{year}{2021}).
\newblock


\bibitem[Zhang et~al\mbox{.}(2021a)]%
        {zhang2021fmfcc}
\bibfield{author}{\bibinfo{person}{Zhenyu Zhang}, \bibinfo{person}{Yewei Gu}, \bibinfo{person}{Xiaowei Yi}, {and} \bibinfo{person}{Xianfeng Zhao}.} \bibinfo{year}{2021}\natexlab{a}.
\newblock \showarticletitle{FMFCC-a: a challenging Mandarin dataset for synthetic speech detection}. In \bibinfo{booktitle}{\emph{International Workshop on Digital Watermarking}}. Springer, \bibinfo{pages}{117--131}.
\newblock


\bibitem[Zhang et~al\mbox{.}(2021f)]%
        {zhang2021fake}
\bibfield{author}{\bibinfo{person}{Zhenyu Zhang}, \bibinfo{person}{Xiaowei Yi}, {and} \bibinfo{person}{Xianfeng Zhao}.} \bibinfo{year}{2021}\natexlab{f}.
\newblock \showarticletitle{Fake speech detection using residual network with transformer encoder}. In \bibinfo{booktitle}{\emph{Proceedings of the 2021 ACM workshop on information hiding and multimedia security}}. \bibinfo{pages}{13--22}.
\newblock


\bibitem[Zhou et~al\mbox{.}(2022)]%
        {zhou2022spoof}
\bibfield{author}{\bibinfo{person}{Ye Zhou}, \bibinfo{person}{Jianwu Zhang}, {and} \bibinfo{person}{Pengguo Zhang}.} \bibinfo{year}{2022}\natexlab{}.
\newblock \showarticletitle{Spoof speech detection based on raw cross-dimension interaction attention network}. In \bibinfo{booktitle}{\emph{Chinese Conference on Biometric Recognition}}. Springer, \bibinfo{pages}{621--629}.
\newblock


\bibitem[Zhu et~al\mbox{.}(2023)]%
        {zhu2023characterizing}
\bibfield{author}{\bibinfo{person}{Yi Zhu}, \bibinfo{person}{Saurabh Powar}, {and} \bibinfo{person}{Tiago~H Falk}.} \bibinfo{year}{2023}\natexlab{}.
\newblock \showarticletitle{Characterizing the temporal dynamics of universal speech representations for generalizable deepfake detection}.
\newblock \bibinfo{journal}{\emph{arXiv preprint arXiv:2309.08099}} (\bibinfo{year}{2023}).
\newblock


\end{thebibliography}
\end{document}